\documentclass[prd,singlecolumn,amsmath,nobibnotes,nofootinbib,superscriptaddress,preprintnumbers]{revtex4}

\usepackage{bm} \usepackage{graphicx}
\usepackage{epstopdf}
\def\ba{\begin{eqnarray}}
\def\ea{\end{eqnarray}}
\def\be{\begin{equation}}
\def\ee{\end{equation}}
\def\({\left(}
\def\){\right)}
\def\[{\left[}
\def\]{\right]}
\def\<{\left<}
\def\>{\right>}

\begin{document}

\title{Dynamical compactification from de Sitter space}
\date{\today}
\author{Sean M. Carroll}
\affiliation{California Institute of Technology, Pasadena, CA 91125, USA}
\author{Matthew C. Johnson}
\affiliation{California Institute of Technology, Pasadena, CA 91125, USA}
\author{Lisa Randall}
\affiliation{Harvard University, Cambridge, MA 02138, USA}

\begin{abstract}
We show that $D$-dimensional de Sitter space is unstable to the
nucleation of non-singular geometries containing spacetime regions
with different numbers of macroscopic dimensions, leading to a
dynamical mechanism of compactification. These and other solutions
to Einstein gravity with flux and a cosmological constant are
constructed by performing a dimensional reduction under the
assumption of $q$-dimensional spherical symmetry in the full
$D$-dimensional geometry. In addition to the familiar black holes,
black branes, and compactification solutions we identify a number of new geometries,
some of which are completely non-singular. The dynamical
compactification mechanism populates lower-dimensional vacua very
differently from false vacuum eternal inflation, which occurs
entirely within the context of four-dimensions. We outline the
phenomenology of the nucleation rates, finding that the dimensionality of the vacuum plays a key role and that
 among vacua of the same dimensionality, the rate is highest for smaller values of the cosmological
 constant. We consider the cosmological constant problem and propose
  a novel model of slow-roll inflation that is triggered by the compactification process.
\end{abstract}

\preprint{CALT-68.2727}

\maketitle

\section{introduction}

Strong theoretical reasons for considering extra dimensions arise in many theories of physics beyond the Standard Model:
the potential role of string theory as a theory of quantum gravity,
the possible role of extra dimensions in addressing the hierarchy
 and flavor problems, and the ability of Kaluza-Klein theory
to unify forces. Key to this program is
identifying ways in which extra dimensions can be hidden from our
macroscopic view. The dominant paradigm of compactification has been
greatly successful in producing reasonable four-dimensional
effective theories, especially with recent developments in moduli
stabilization in string theory. Indeed, there appears to be an
embarrassment of riches, with compactifications using fluxes and
branes leading to an enormous ``landscape'' of possible
four-dimensional solutions~\cite{Susskind:2003kw}.

Determining the existence of phenomenologically
acceptable four-dimensional vacuum states is important. Equally important is
addressing the intrinsically cosmological question of the evolution
into these and other vacua. How did our observable universe come to
be effectively four-dimensional? If the laws of physics allow for
more than one long-lived vacuum state, why do we find ourselves in a
vacuum with a small cosmological constant and the particle spectrum
of the Standard Model? These and other fundamental questions require
an understanding of  transitions into different vacua, including
states of different macroscopic dimensionality.

In this paper we explore a new possibility for dynamical
compactification: a spontaneous transition from a higher-dimensional
de Sitter background to  lower-dimensional spacetimes with  vacuum
energy that depends on details of the solution. The only ingredients
in the simple models we consider are a positive cosmological
constant in the original $D$-dimensional background space and a
$q$-form gauge field strength. We show that this theory allows for
the nucleation of regions containing a macroscopic $D-q$-dimensional
spacetime with compactified $q$-spheres stabilized by flux. This
process is the inverse of the ``spontaneous decompactification"
studied by Giddings and Myers~\cite{Giddings:2004vr}, and was first
suggested therein. We explore the cosmology of these models,
including a post-transition inflationary phase that could lead to a
realistic universe.

Other attempts to dynamically account for our observed
four-dimensional world in a higher-dimensional theory include the
suggestion of Brandenberger and Vafa~\cite{Brandenberger:1988aj},
who used the properties of intersecting strings to suggest that
three spatial dimensions could naturally expand while others
remained compact. Karch and Randall~\cite{Karch:2005yz} argued that
a gas of branes with various dimensionalities would relax to one
dominated by three-branes. Our proposal starts from a dramatically
different initial state: empty de~Sitter space in higher dimensions.
Of course this is not necessarily a natural starting point from the
perspective of string theory, which seems to favor Minkowski or
anti-de~Sitter vacua. Nevertheless, if some other process formed
(for example) a six-dimensional de~Sitter vacuum in string theory,
our proposal could explain the transition from six to four
dimensions and given the paucity of dynamical explanations for
compactification, all such theories are potentially of interest.

The post-transition solutions are derived from spherically symmetric
black brane solutions of Einstein-Maxwell theory, and so share some
properties with charged black hole solutions in four dimensions. In
fact, we can explain most of the major elements of the
$D$-dimensional solutions and their nucleation by analogy with black
holes.

To begin, the interior of a black hole can be interpreted as a
two-dimensional ``big-crunch" cosmological spacetime. The static
metric of the Schwarzschild black hole inside of the horizon
\begin{equation}\label{eq:schwbh}
ds^2 =  - \left(\frac{2M}{R}-1\right)^{-1} dR^2 + \left(\frac{2M}{R}-1\right) dt^2+ R^2 d\Omega_2^2,
\end{equation}
can be written in a suggestive form by making the
change of
coordinates 
\begin{equation}
x=t/4M, \ \ \tau = \left( 16M^2 - 8M R \right)^{1/2},
\end{equation}
and looking at the region just inside of the event horizon $R<2 M$
(see also Ref.~\cite{Larsen:1996pb})
\begin{equation}
ds^2 \simeq -d\tau^2 + \tau^2 dx^2 + 4 M^2 d\Omega_2^2.
\end{equation}
The $\tau$ and $x$ coordinates describe
a two-dimensional cosmology,
with the ``extra dimensions" compactified on a two-sphere of
momentarily constant radius. At $\tau=0$ the scale factor $a = \tau$
goes to zero. In FRW cosmology this corresponds to the ``big-bang,"
but in this case it is clear that the $a=0$ surface is merely a
coordinate singularity coinciding with the location of the event
horizon of the Schwarzschild black hole. The solution just outside
the horizon can be obtained by analytically continuing $\tau
\rightarrow i \tau$, in which case $\tau$ and $x$ switch roles as
the timelike or spacelike coordinates. In the region outside the
horizon the radius of the two-spheres of course grows,
``decompactifying" the extra dimensions as asymptotically flat
four-dimensional space is reached far from the black hole.

Exploring this scenario in more detail, we can write the metric Eq.~\ref{eq:schwbh} in the form
\begin{equation}
ds^2 = - d\tau^2 + a^2(\tau) dx^2 + R^2(\tau) d\Omega_2^2,
\end{equation}
which in the $\tau$, $x$ plane describes a two-dimensional FRW
universe in which the radius of the two-sphere $R$ evolves with
proper time $\tau$. The Einstein equations yield (see
Ref.~\cite{Carroll:2009maa} for more details)
\begin{equation}
\ddot{R} + \frac{{\dot{R}}^2}{2 R} = - \frac{1}{2R} = - \frac{dV_{eff}}{dR}, \ \ a = \dot{R},
\end{equation}
where it can be seen that the system of equations is that of a field
$R$ subject to the potential $V_{eff} = \frac{1}{2} \log R$ evolving
in a two-dimensional FRW universe. Again, the ``big-bang" at $a=0$
corresponds to an event horizon in the black hole geometry. The
radius at which this occurs is in one-to-one correspondence with the
mass of the black hole, and we can therefore classify solutions by
stationary points in the motion of $R$ (where, by the equations of
motion, $a=0$). As expected, the potential drives $R \rightarrow 0$,
corresponding to the singularity of the black hole. In addition to the 
non-singular big-bang surface, a non-singular big-crunch can be obtained 
by time reversing this solution.

\begin{figure*}
\begin{center}
\includegraphics[height=7cm]{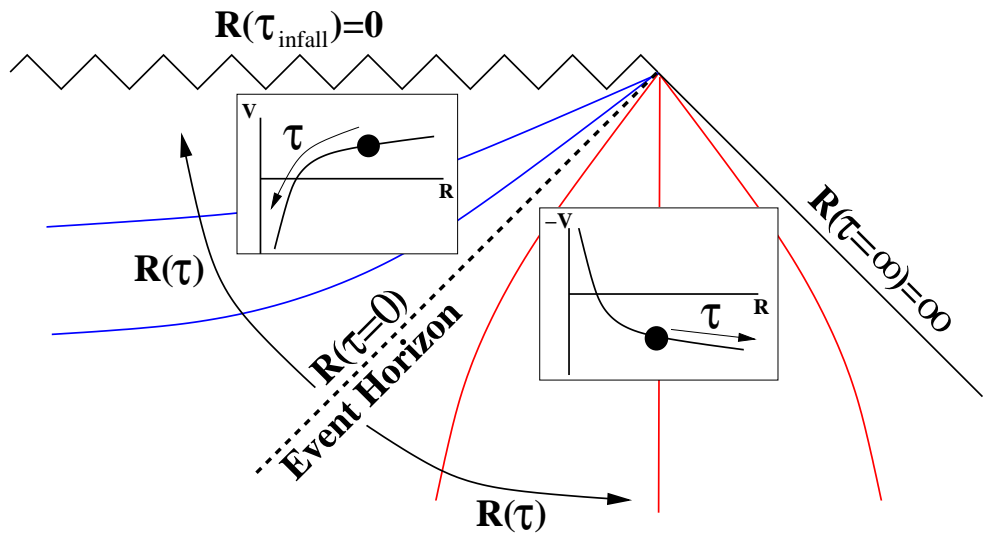}
\end{center}
\caption{A portion of the causal diagram for a Schwarzschild black
hole. To the future of the event horizon, $\tau$ is a timelike
coordinate, and the $\tau$, $x$ plane (which is shown here)
resembles a big-crunch cosmology that begins from a completely
non-singular ``big-bang" at the event horizon. The value of $R$ at
the horizon is determined by the stationary point shown on the
potential (filled circle).  On subsequent surfaces of constant
$\tau$, $R$ evolves in the sketched potential $V$, eventually
reaching a singularity as $R$ is pushed to zero. To the past of the
event horizon, $\tau \rightarrow i\tau$, and $\tau$ becomes a
spacelike coordinate. The value of $R$ on
surfaces of constant $\tau$ is determined by its motion in the
upside-down potential $-V$. This pushes $R \rightarrow \infty$ as
the asymptotically flat region far from the black hole is
approached.} \label{fig:bhdiagram}
\end{figure*}

To generate the region of
the black hole outside of the horizon, we
use the intuition developed above, and analytically continue $\tau
\rightarrow i \tau$. This takes $a \rightarrow i a$ and $V_{eff}
\rightarrow - V_{eff}$ in the equations of motion. Now, starting
from a stationary point, $R$ is pushed to infinity, corresponding to
the asymptotically flat region far from the black hole. In this
piecewise manner, it is possible to sew together the regions inside
and outside the horizon by looking at the motion of $R$ away from a
stationary point in either the potential $V_{eff}$ or $-V_{eff}$ as
shown in Fig,~\ref{fig:bhdiagram}.

Within the context of black hole physics, we can find an
example of the dynamical compactification mechanism mentioned above.
In de Sitter space there is a non-zero probability to pair-produce
charged black holes from the
vacuum~\cite{Bousso:1995cc,Bousso:1996au,Bousso:1996pn,Bousso:1999iq,Hawking:1995ap,Mann:1995vb,Ginsparg:1982rs}.
The two-dimensional FRW regions in the interior of each black hole
are produced dynamically, and so black hole nucleation can be
regarded as the dynamical compactification of two extra dimensions.
The nucleation of many black holes with different charges leads to a
fractally bifurcated structure of future
infinity~\cite{Bousso:1999ms} reminiscent of eternal inflation (a
process in which pocket universes of diverse properties are
nucleated out of a background de Sitter space, dividing future
infinity among many different vacua). However, since we are
discussing black holes, the two-dimensional cosmologies always end
in a big-crunch.

\begin{figure*}
\begin{center}
\includegraphics[width=13cm]{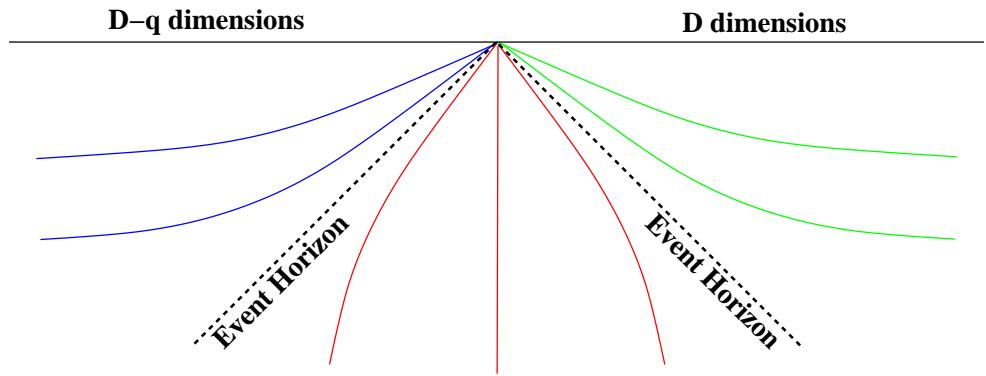}
\end{center}
\caption{A sketch of the causal structure of a solution that
interpolates
between asymptotically $D-q$ and $D$-dimensional regions
across event horizons. On the left, a $D-q$ dimensional non-singular
cosmological spacetime is located behind an event horizon. Outside
this horizon is an interpolating region to another event horizon,
that encloses an asymptotically $D$-dimensional de Sitter region.
The nucleation of such solutions from empty $D$-dimensional de
Sitter space represents the dynamical compactification of some
number of extra dimensions.} \label{fig:interpdiagram}
\end{figure*}

A crucial new feature that arises when considering higher
dimensional gravity is the existence of black brane solutions that
are completely non-singular~\cite{Gibbons:1994vm}. In contrast to
the black hole solutions, this raises the possibility that there are
true interpolating solutions between a region with $D$ large
dimensions and a lower dimensional non-singular cosmology.
Semi-classical processes akin to black hole nucleation can then
produce such regions dynamically.  A sketch of the causal structure
of the types of solutions we will discuss is shown in
Fig.~\ref{fig:interpdiagram}. The lower dimensional cosmology is
contained within an event horizon, just as it was for the black hole
solutions. Outside the horizon is an interpolating region to another
event horizon that encloses an asymptotically $D$-dimensional de
Sitter region~\footnote{Because infinity is spacelike in de Sitter
space, the approach to the asymptotically $D$-dimensional region is
in this case in a timelike direction. The same will be true for the
black hole geometries in de Sitter space, in the region outside of the cosmological horizon.}. In this geometry, either
the $D$ or $p+2$-dimensional spaces can be accessed from the
interpolating region. Similar proposals for embedding a lower
dimensional cosmology within a higher dimensional space have been
made in~\cite{Gibbons:1986wg,Larsen:1996pb,Lu:1996er,Behrndt:1994ev,Bergshoeff:2005bt}. 

The lower-dimensional FRW universe can evolve to a variety of vacua
in which the radion field is stabilized. Our model therefore
describes a landscape that is simpler than string theory
constructions, but nonetheless involves the existence of extra dimensions in a
more fundamental way through the dynamical compactification
mechanism. Using this mechanism, our model allows one to scan not only over
different vacuum energies, but over fundamentally different
spacetimes with different numbers of dimensions. This method of populating the states of a landscape is
fundamentally different than the standard paradigm of
four-dimensional eternal inflation. Finally, we note that the existence of landscapes should in
fact be viewed as a generic consequence of theories with extra
dimensions, underlying the importance of studying a number of
different models.

Summarizing the major elements of this picture, and their description in the remainder of the paper:
\begin{itemize}
\item It is possible to construct $D$-dimensional geometries piecewise by dimensionally
reducing to a theory of lower-dimensional gravity with a radion field encoding the radius of some assumed
number of spherically symmetric dimensions. The set of radion potentials define a landscape
of lower-dimensional effective theories. The properties of the dimensionally reduced theory
are outlined in Sec.~\ref{sec:actionandmethod}.
\item A lower dimensional cosmological spacetime can be embedded behind an event
horizon located in a higher dimensional geometry. The event horizons correspond
to non-singular big-bang or big-crunch surfaces. We classify the possible $D$-dimensional
geometries under the assumption of an FRW metric ansatz in
Sections~\ref{sec:FRWsolutions},~\ref{sec:solutionsL0}, and~\ref{sec:FRWLg0}.
We examine the case where the cosmology is homogenous but anisotropic in
Appendix~\ref{sec:bianchisolutions}. Throughout we consider a zero or positive
cosmological constant $\Lambda \geq 0$. A negative cosmological constant will
yield solutions qualitatively similar to the $\Lambda = 0$ case, and so we do
not consider it here.
\item In a $D$-dimensional de Sitter space, there is a non-zero probability to
nucleate solutions containing a non-singular lower-dimensional cosmological spacetime.
We calculate the probabilities and describe the global structure of the $D$-dimensional
spacetime after many nucleation events in Sec.~\ref{sec:dynamicalcompactification}.
\item In any theory that can access a number of sufficiently finely-spaced vacua, it is possible to address
 the cosmological constant problem. In Sec.~\ref{sec:ccproblem}, we briefly
 discuss the ability to obtain the observed value of the cosmological constant
 naturally from the landscape of radion potentials.
\item It is possible to embed an epoch of inflation inside of the lower-dimensional
cosmological spacetime by adding a scalar field that obtains a negative mass squared
inside an event horizon. This can be accomplished by coupling the scalar field to the
$D$-dimensional curvature and flux, as we describe in Sec.~\ref{sec:inflation}.
\end{itemize}

\section{Dimensional reduction}\label{sec:actionandmethod}

The theory that we consider throughout this paper is $D$-dimensional
Einstein gravity (with and without a cosmological constant) coupled
to a $q$-form field strength. This system is described by the action
\begin{equation}\label{eq:Ddimaction}
S =  \frac{M_D^{D-2}}{2} \int d^D x \sqrt{-\tilde{g}^{(D)}} \left( \tilde{\mathcal{R}}^{(D)} - 2 \Lambda - \frac{1}{2 q!} \tilde{F}_q^2  \right),
\end{equation}
where $M_D$ is the D--dimensional planck mass (with $8 \pi G_D =
M_D^{2-D}$), and
the tilde denotes that each quantity is evaluated in
the $D$-dimensional Einstein frame. In each $D$-dimensional theory,
there can be multiple $q$-form field strengths ranging from $q=2,3
\ldots, D-2$.

Assuming $q$-dimensional spherical symmetry, we can write the metric in the product form
\begin{equation}\label{eq:Ddimeinsteinmetric}
d\tilde{s}^2 = \tilde{g}^{p+2}_{\mu \nu} ( {\bf x} ) dx^{\mu} dx^{\nu} + R^2( {\bf x} ) d\Omega_q^2,
\end{equation}
where
\begin{equation}
D = p + 2 + q,
\end{equation}
$d\Omega_q^2 = d\theta_1^2 + \sin^2 \theta_1 d\theta_2^2 + \ldots + \sin^2 \theta_1 \ldots \sin^2 \theta_{q-1} d\theta_q$, and $\mu, \ \nu = 0, 1 , \ldots p+1$. We employ this decomposition into $q$ and $p+2$ because many of the objects we study are of codimension $p$ (i.e. $p$-branes). The radius of the $q$-sphere $R({\bf x})$ and $p+2$-dimensional metric $\tilde{g}^{p+2}_{\mu \nu} ( {\bf x} )$ will in general be a function of the $p+2$-dimensional coordinates ${\bf x}$. Magnetic $q$-form field strengths solving Maxwell's equations (trivially) and respecting the $q$-dimensional spherical symmetry are given by
\begin{equation}
\tilde{F}_q = Q \sin^{q-1} \theta_1 \ldots \sin \theta_{q-1} d\theta_1 \ldots \wedge d\theta_q.
\end{equation}
For simplicity we do not consider electric fluxes in the following
and construct solutions with a single fixed $q$ (i.e. multiple $q$-form 
charges are not turned on simultaneously).

Under our assumption of $q$-dimensional spherical symmetry,
solutions to the full $D$-dimensional theory Eq.~\ref{eq:Ddimaction} can be
described by the behavior of $R({\bf x})$ in a $p+2$-dimensional
spacetime (in analogy with the discussion of black holes in the
Introduction). In other words, we can perform a dimensional
reduction by integrating over the angles of the $q$-sphere. This
yields  a $p+2$-dimensional theory of gravity coupled to a scalar
field, the ``radion," encoding the radius of the $q$-sphere at each
point. This picture is crucial for describing how four-dimensional
vacua and cosmologies can be obtained from a theory with extra
dimensions. More generally, we will find that the dimensionally
reduced theory is a powerful tool for constructing a variety of
$D$-dimensional geometries not necessarily related to the
description of four-dimensional universes. We now explore this
picture in more detail, deriving the dimensionally reduced theory
and describing its properties.

\subsection{Defining the dimensionally reduced theory}

Because we have assumed $q$-dimensional spherical symmetry, it is
possible to recast the $D$-dimensional
theory in the form of a
$p+2$-dimensional gravitational action coupled to the scalar field
$R( {\bf x} )$. We can write the $D$-dimensional Ricci scalar and
metric determinant as
\begin{equation}\label{eq:Ricciddim}
\tilde{\mathcal{R}}^{(D)} = \tilde{\mathcal{R}} + \frac{q
(q-1)}{R^2} - \frac{2 q}{R}
\tilde{g}^{\mu \lambda} \partial_{\mu}
\partial_{\lambda} R -  \frac{q (q-1)}{R^2} \tilde{g}^{\mu \lambda}
(\partial_{\mu} R) (\partial_{\nu} R),
\end{equation}
and
\begin{equation}
\sqrt{- \tilde{g}^{(D)}} = \sqrt{- \tilde{g} } R^{q} \sin^{q-1} \theta_1 \ldots \sin \theta_{q-1}.
\end{equation}
The absence of a superscript on $\mathcal{R}$ and $g$ indicates a
$p+2$-dimensional quantity.
Integrating over the spherical
coordinates and performing an integration by parts to remove the
second derivative on $R$, the action Eq.~\ref{eq:Ddimaction} becomes
\begin{equation}
S = \frac{ M_D^{D-2} {\rm Vol} (S^{q}) }{2} \int d^{p+2} x \sqrt{-
\tilde{g} }
\left[ R^q \tilde{\mathcal{R}}  - q (q-1) R^{q-2}
\tilde{g}^{\mu \lambda} (\partial_{\mu} R) (\partial_{\nu} R) + q
(q-1) R^{q-2} - 2 \Lambda R^q - \frac{M_D^{2 (1-q)} Q^2}{2 R^q}
\right],
\end{equation}
where ${\rm Vol} (S^{q})$ is the volume of a unit $q$-sphere
\begin{equation}
{\rm Vol} (S^{q}) = \frac{2 \pi^{(q+1) / 2}}{\Gamma \left( \frac{q+1}{2} \right)}.
\end{equation}

For $p \geq 1$, it is possible to perform a conformal transformation to the $p+2$-dimensional Einstein frame
\begin{equation}
g_{\mu \nu} = (M_D R)^{2 \frac{q}{p}} \tilde{g}_{\mu \nu}.
\end{equation}
The absence of a tilde indicates that a quantity is evaluated in the $p+2$-dimensional Einstein frame. Defining
\begin{equation}\label{eq:MdMpp2}
M_{p+2} \equiv M_D \left( {\rm Vol} (S^{q}) \right)^{1/p},
\end{equation}
the action becomes
\begin{equation}
S =\int d^{p+2} x \sqrt{- g} \left[ \frac{M_{p+2}^p}{2} \mathcal{R}
- \frac{M_{p+2}^p}{2}
\frac{q (p+q)}{p R^2} g^{\mu \lambda}
(\partial_{\mu} R) (\partial_{\nu} R) - V(R) \right],
\end{equation}
where the potential for $R$ in the Einstein frame is given by
\begin{equation}\label{eq:Rradionpotential}
V(R) = \frac{M_{p+2}^p M_D^2}{2} \left[ -q (q-1) (M_D R)^{-2
\frac{(q+p)}{p}} + \frac{2
\Lambda}{M_D^2} (M_D R)^{-2 \frac{q}{p}} +
\frac{Q^2}{2} (M_D R)^{-2 (p+1) \frac{q}{p}} \right],
\end{equation}
and we recover the standard Einstein-Hilbert action for the gravitational sector.

It is possible to define a canonically normalized field $\phi$ by making the change of variables
\begin{equation}\label{eq:Rdef}
M_D R = \exp \left[ \sqrt{\frac{p}{q (p+q)}} \frac{\phi}{M_{p+2}} \right].
\end{equation}
This yields the action
\begin{equation}\label{eq:einsteinaction}
S = \int d^{p+2} x \sqrt{ - g} \left[ \frac{M_{p+2}^p}{2}
\mathcal{R} -
\frac{M_{p+2}^{p-2}}{2} g^{\mu \nu} (\partial_{\mu}
\phi) (\partial_{\nu} \phi) - V (\phi)  \right],
\end{equation}
with
\begin{eqnarray}\label{eq:radionpotential}
V (\phi) &=& \frac{M_{p+2}^p M_D^2}{2} \left[ - q (q-1) \exp\left( -
2 \sqrt{\frac{p+q}{pq}}
\frac{\phi}{M_{p+2}} \right) + \frac{2
\Lambda}{M_D^2} \exp\left( - 2 \sqrt{\frac{q}{p (p+q)}}
\frac{\phi}{M_{p+2}} \right) \right. \nonumber \\ && \left. \ \ \ \
\ \ \ \ \ \ \ \ \ \ \ \ + \frac{Q^2}{2} \exp\left( - 2 (p+1)
\sqrt{\frac{q}{p (p+q)}} \frac{\phi}{M_{p+2}} \right) \right].
\end{eqnarray}
In the following sections, we will make frequent use of both the
potential Eq.~\ref{eq:radionpotential}
for $\phi$ and the potential
Eq.~\ref{eq:Rradionpotential} for $R$.

\subsection{A landscape from radion potentials}\label{sec:radionpotential}

\begin{figure*}
\begin{center}
\includegraphics[height=6cm]{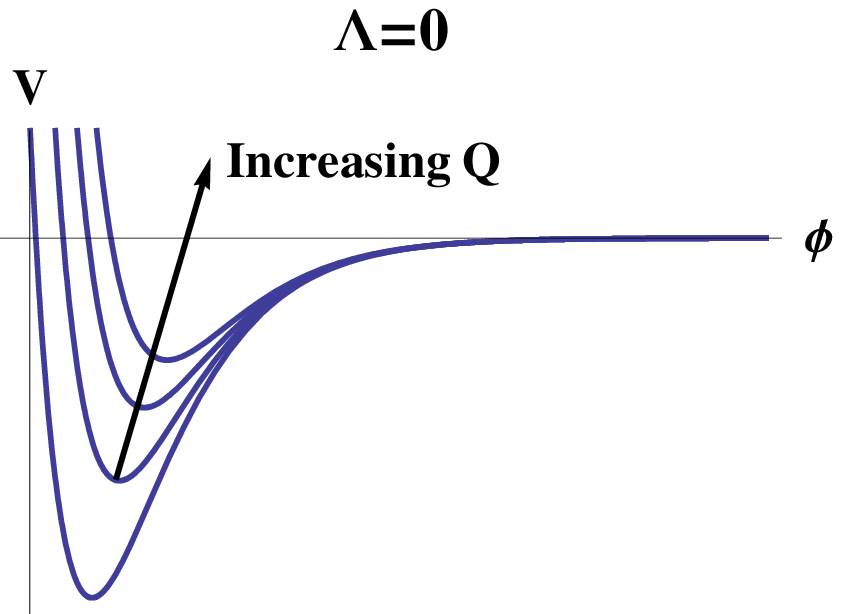}
\includegraphics[height=6cm]{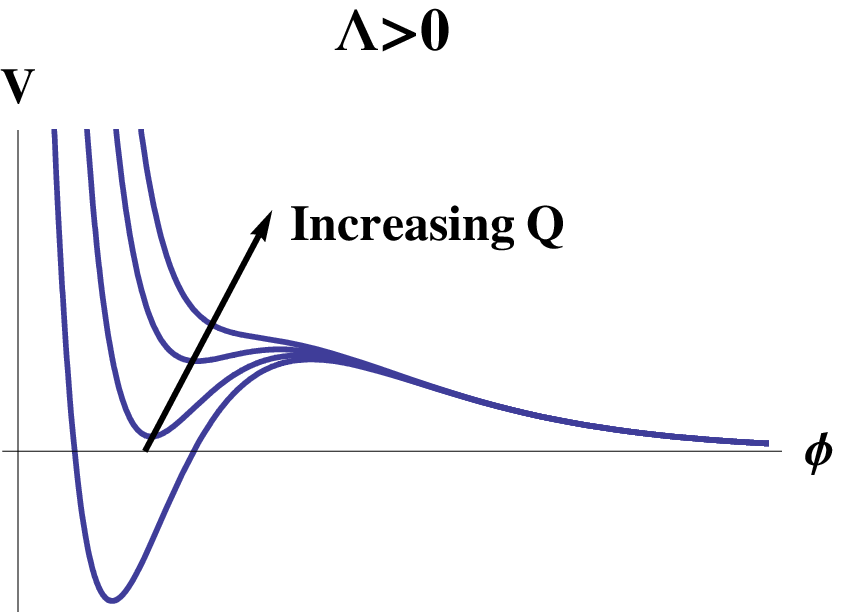}
\end{center}
\caption{The radion potential Eq.~\ref{eq:radionpotential} with
$\Lambda = 0$ (left)
and $\Lambda > 0$ (right). In each plot, a
number of potentials at fixed $\Lambda$ are shown with successively
larger values of $Q$ from bottom to top. For $\Lambda = 0$ the
potential always has a negative minimum and approaches zero from
below as $\phi \rightarrow \infty$, while for $\Lambda > 0$ there
can exist a minimum of negative, zero, or positive energy and the
potential approaches zero from above at large $\phi$.}
\label{fig:effpot}
\end{figure*}

The radion potential Eq.~\ref{eq:radionpotential} is defined in
terms of a number of fixed parameters, such as the total number of
dimensions $D = q+p+2$ and the $D$-dimensional cosmological constant
$\Lambda$. However, it also depends on the charge $Q$ and the number
of compact and non-compact dimensions $q$ and $p+2$ (under the
constraint of fixed $D$), which are variable. This defines an indexed set 
of radion potentials, one for each allowed value of
$Q$, $p$, and $q$. The potential Eq.~\ref{eq:radionpotential} with fixed $p$
and $q$ is sketched in Fig.~\ref{fig:effpot} for $\Lambda = 0$ (left
panel) and $\Lambda > 0$ (right panel), in each case for a number of
values of $Q$. Since it will be qualitatively similar to the
$\Lambda = 0$ case, we will not consider $\Lambda < 0$. The
potential has the same qualitative features for all choices of $p$
and $q$.

Perhaps most interestingly, there are minima where the radius of
the $q$-sphere can be stabilized, yielding $p+2$-dimensional vacua. 
The set of vacua correspond to a ``landscape" of lower dimensional theories~\footnote{The existence of a landscape of vacua
is strongly associated with string theory. However, a landscape of
vacua can be viewed as a generic consequence of theories that have
extra dimensions. There is even a landscape of two-dimensional vacua
in the completely pedestrian four-dimensional Einstein-Maxwell
system.}. For $\Lambda = 0$, the vacua are always negative, decreasing in
depth with increasing charge. For $\Lambda > 0$, the vacua are
negative for small $Q$, but eventually reach zero and then become
positive as $Q$ increases. For large enough $Q$, the minimum
disappears completely. The existence of 4-dimensional vacua with
positive vacuum energy will be important in our discussion of
cosmological solutions.

\begin{figure*}
\begin{center}
\includegraphics[height=6cm]{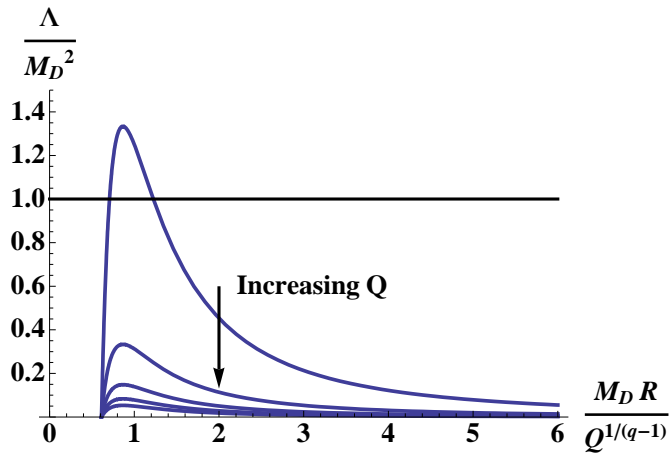}
\end{center}
\caption{Extrema of the radion potential are located at the
intersection of
each curve (Eq.~\ref{eq:lambdaQrel} for $\{p=2, q=2
\}$) with a line of fixed $\Lambda$. There is only one intersection
at $\Lambda = 0$ and zero, one, or two intersections at $\Lambda >
0$. When there are two intersections, the first corresponds to a
minimum (referred to as $R_{-}$ or $\phi_-$) and the second to a
maximum (referred to as $R_{+}$ or $\phi_+$). Note that the location
of the first intersection is relatively independent of $\Lambda$.} \label{fig:zeroes}
\end{figure*}

The locations of potential extrema can be determined most easily
from
Eq.~\ref{eq:Rradionpotential}. We will refer to the location of
potential minima as $R_{-}$ or $\phi_{-}$ and potential maxima as
$R_{+}$ or $\phi_+$. Setting the derivative of the potential with respect to $R$
equal to zero, the extrema are given by the roots of
\begin{equation}\label{eq:lambdaQrel}
\Lambda  = \frac{1}{2} (p+q) (q-1) R^{-2} -\frac{1}{4} (p+1) Q^2 M_D^{-2 (q-1)} R^{-2 q}.
\end{equation}
In Fig.~\ref{fig:zeroes}, the right hand side of this equality is plotted for various $Q$.

When $\Lambda = 0$, there is a single extremum, whose position can be found directly from Eq.~\ref{eq:lambdaQrel}
\begin{equation}\label{eq:nolamroot}
(M_D R_{-})^{2 (q-1)} = \frac{(p+1)}{2 (p+q) (q-1)} Q^2.
\end{equation}
This is a global minimum of the potential. Inserting this back into Eq.~\ref{eq:Rradionpotential}, the value of the potential is
\begin{equation}\label{eq:vphimin}
V(\phi_{-}) = - M_{p+2}^p M_D^{2} \frac{p}{2} \left(
\frac{(q-1)}{(p+1)} \right)^{\frac{q(p+1)}{p(q-1)}}
\left( \frac{2
(p+q)}{Q^2} \right)^{\frac{(q+p)}{p(q-1)}}.
\end{equation}
The depth of the potential minimum decreases with increasing $Q$, as can also be seen in Fig.~\ref{fig:effpot}. 

For fixed $\Lambda > 0$, there are zero, one, or two roots depending
on the value of $Q$.
This is seen in Fig~\ref{fig:zeroes} by locating
the intersections of the curve with lines of fixed $\Lambda$. The
maximum value of $Q$ for which a root exists at fixed $\Lambda$ can
be found by determining the maximum of Eq.~\ref{eq:lambdaQrel}.
There are two roots as long as
\begin{equation}\label{eq:Qlimit}
Q < \frac{2^{1-q/2} (q-1)^{q-1/2} (p+q)^{q/2}}{q^{q/2} (1+p)^{1/2}} \left( \frac{M_D^2}{\Lambda} \right)^{(q-1)/2},
\end{equation}
which grows for decreasing $\Lambda$ and increasing $p$ and $q$.

When it exists, the location of the smaller root is not strongly
affected by the value of $\Lambda$,
and given approximately by
Eq.~\ref{eq:nolamroot}. This is a local minimum of the potential.
The location of the second root, when it exists, can be approximated
by neglecting the $Q$-dependent term in Eq.~\ref{eq:nolamroot}
(since the second root is at relatively large $R$, this term is
subdominant), yielding
\begin{equation}\label{eq:R0maximum}
R_{+} \simeq \sqrt{\frac{(p+q) (q-1)}{2 \Lambda}}.
\end{equation}
Inserting this into Eq.~\ref{eq:Rradionpotential}, the height of the potential maximum is given by
\begin{equation}\label{eq:vbarrier}
V(\phi_{+}) \simeq M_{p+2}^p M_D^2 \frac{p}{2 (p+q)^{\frac{q}{p} +1}  (q-1)^{\frac{q}{p}} } \left(2 \frac{\Lambda}{M_D^2} \right)^{\frac{q}{p} + 1}.
\end{equation}
In addition, this roughly bounds the possible vacuum energy. As we will discuss in Sec.~\ref{sec:dynamicalcompactification}, this bound ensures that the entropy associated with solutions at the positive potential minima is always greater than the entropy associated with a $D$-dimensional de Sitter space with cosmological constant $\Lambda$.

Another property of note is the asymptotic behavior of the
potentials as $\phi \rightarrow \infty$. For $\Lambda = 0$ the
negative term in the potential dominates at large $\phi$, and the
potential approaches zero from below. In the solutions of the following
sections, we will see that this implies that the radius of the $q$-spheres
increase in a spacelike direction, consistent with the assumption
of asymptotically flat space.

For $\Lambda > 0$, the term in the potential proportional to $\Lambda$
dominates at large $\phi$, and the potential approaches zero from above.
We will find that this implies the radius of $q$-spheres increase in a timelike
direction, consistent with a $D$-dimensional de Sitter space.

Therefore, both the vacuum structure and asymptotic properties of the potentials will play
prominent roles in the construction of solutions in the
dimensionally reduced theory, to which we now turn.

\section{Solutions with a $p+2$-dimensional FRW metric ansatz}\label{sec:FRWsolutions}

We generate solutions by assuming that the $p+2$-dimensional
spacetime can be foliated into either spacelike or timelike
$p+1$-dimensional surfaces of homogeneity. A full classification of
homogenous solutions in arbitrary dimensions is beyond the scope of
the present paper and we will restrict ourselves to two different
classes of metrics. In the following three sections, we will treat
the FRW case, classifying all possible solutions for open, closed,
and flat spatial sections. In Appendix~\ref{sec:bianchisolutions},
we treat the case of an anisotropic metric, assuming that the
spatial sections are flat (Bianchi Type I in four dimensions). The
set of solutions generated from these two different forms of the
metric will include both known extremal and non-extremal charged
$p$-brane solutions, as well as a number of new solutions for
$\Lambda = 0$ and $\Lambda > 0$ (again, the solutions for $\Lambda <
0$ will be qualitatively similar to those for $\Lambda = 0$ and we
do not consider them in detail here).

The evolution of the radion field on successive
$p+1$-dimensional
surfaces specifies the location and properties of singularities,
event horizons, and asymptotic infinity. Specializing to the FRW
metric ansatz, in addition to the radion field, there is one scale
factor describing the geometry. We define the relevant equations of
motion in Sec.~\ref{sec:equationsofmotion}. 

As illustrated in the introduction,
regions with spacelike and
timelike surfaces of homogeneity can be matched across event
horizons in the full geometry. From the perspective of the
dimensionally reduced theory, these are big-bang and big-crunch
surfaces of $p+2$-dimensional cosmological spacetimes. We describe
the general circumstances in which this correspondence is valid in
Sec.~\ref{sec:nonsingularsurfaces}.

With the relevant formalism in hand, we will move on to discuss the
particulars of the solutions in Sec.~\ref{sec:solutionsL0}
and~\ref{sec:FRWLg0}. Although it is in most cases impossible to
find a closed analytic form for the metric, we numerically generate
examples of various classes of solutions. See Ref.~\cite{Krishnan:2005su} for a similar 
numerical study of the evolution of the radion in the presence of an FRW metric ansatz. 
The solutions relevant for
the dynamical compactification mechanism are discussed in
Sec.~\ref{sec:FRWLg0}, although the more simple geometries of
Sec.~\ref{sec:solutionsL0} provide a nice illustration of our method
of finding solutions.

In Sec.~\ref{sec:solutionsL0}, we catalog
solutions with $\Lambda =
0$. These include the $AdS_{p+2} \times S^q$ compactification
solutions (corresponding to sitting at minima of the radion
potential) as well as various extremal and non-extremal $p$-brane
solutions in asymptotically flat $D$-dimensional space. We
illustrate in detail the construction of the $p$-brane solutions
from the dimensionally reduced theory. In addition to the known
solutions, we find a novel non-singular geometry in which two
identical asymptotically flat regions are connected across a pair of
event horizons.

Sec.~\ref{sec:FRWLg0} treats the case of a positive $D$-dimensional
cosmological
constant $\Lambda$. The compactification solutions in
this case can be either $AdS_{p+2} \times S^q$,  ${\mathcal M}_{p+2} \times
S^q$, or $dS_{p+2} \times S^q$, corresponding to the negative, zero,
or positive vacuum energy minima and the positive energy maximum of the
radion potential. Among the geometries that contain an
asymptotically $D$-dimensional de Sitter region will be the
``interpolating" solutions mentioned in the introduction. These
completely non-singular geometries interpolate across event horizons
between an asymptotically $D$-dimensional de Sitter space and a
$p+2$-dimensional open FRW universe that can evolve to $dS_{p+2}
\times S^q$. This geometry will form the basis of the dynamical
compactification mechanism described in
Sec.~\ref{sec:dynamicalcompactification}.

\subsection{Equations of motion}\label{sec:equationsofmotion}
For spacelike surfaces of homogeneity, the metric is that of an FRW universe
\begin{equation}
ds^2 = - d \tau^2 + a(\tau)^2 \left[ d \chi^2 + S_k^2 (\chi) d\Omega_{p}^2 \right],
\end{equation}
where
\begin{equation}
S_k^2 = \{ \sin \chi , \ \chi, \ \sinh \chi \},
\end{equation}
for closed, flat, or open spatial slices respectively. The resulting
equations of motion
for the field $\phi(\tau)$ and the scale factor
$a(\tau)$ are given by
\begin{equation}\label{eq:phifrweom}
\ddot{\phi} + (p + 1) \frac{\dot{a}}{a} \dot{\phi} = - M_{p+2}^{2-p} V',
\end{equation}
\begin{equation}\label{eq:ddimenfeqn}
\left( \frac{\dot{a}}{a} \right)^2 = \frac{2}{ M_{p+2}^2 p (p + 1)} \left( \frac{\dot{\phi}^2}{2} + M_{p+2}^{2- p} V (\phi) \right) - \frac{k}{a^2},
\end{equation}
where $k = \{ -1,0,+1 \}$ for open, flat, and closed slicing respectively and $V (\phi)$ is given by Eq.~\ref{eq:radionpotential}.
We will also have occasion to use the second order equation for $a$, given by
\begin{equation}\label{eq:ddimauxfeqn}
\frac{\ddot{a}}{a} = \frac{1}{M_{p+2}^2 p (p+1) } \left( - p \dot{\phi}^2 + 2 M_{p+2}^{2- p} V \right).
\end{equation}
We refer to surfaces where $a=0$ as big-bang or big-crunch surfaces depending on their orientation in time.

When the surfaces of homogeneity are timelike, $\tau$ becomes a
spacelike variable. The metric in these case can be obtained by
analytic continuation. Closed FRW universes always yield spaces with
closed timelike curves upon analytic continuation, and we do
not
consider them here. Starting with an open FRW universe, and taking
\begin{equation}
\tau \rightarrow i \tau, \ \ \chi \rightarrow \chi + i \frac{3 \pi}{2},
\end{equation}
yields
\begin{equation}\label{eq:wallmetric}
ds^2 = - a(\tau)^2 d\chi^2 + d\tau^2 + a(\tau)^2 \cosh^2 \chi d\Omega_p^2,
\end{equation}
where the surfaces of homogeneity are hyperboloids with spacelike
norm. The
equations of motion for the scale factor $a(\tau)$ and
$\phi(\tau)$ are given by Eqns.~\ref{eq:phifrweom},
~\ref{eq:ddimenfeqn}, and~\ref{eq:ddimauxfeqn} with $V \rightarrow
-V$. Recall that a similar flip of the potential occurred in the
black hole example. We will refer to this as the ``upside-down" or
``Euclidean" potential.

In the case of a flat universe,
the analytic continuation $\tau
\rightarrow i \tau$ also takes $V \rightarrow -V$ in the equations
of motion. Continuing one of the cartesian coordinates $z
\rightarrow i z$ as well, the metric becomes
\begin{equation}
ds^2 = - a(\tau)^2 d\chi^2 + d\tau^2 + a(\tau)^2  \chi^2 \left( d\theta_1^2 + \sinh^2 \theta_1 d\Omega_{p-1} \right).
\end{equation}
The surfaces of homogeneity in this case are planes with spacelike norm.

Solutions can be generated by numerically evolving the equations of motion. To do so, it is useful to define the dimensionless variables
\begin{equation}\label{eq:dimlessvariables}
\bar{\phi} = \frac{\phi}{M_{p+2}}, \ \ \  \bar{V} =
\frac{V}{\mu^{p+2}}, \ \ \ \bar{\tau} = \frac{\mu^{(p+2)/2}
}{M_{p+2}^{p/2} }
\tau, \ \ \ \bar{a} =
\frac{\mu^{(p+2)/2}}{M_{p+2}^{p/2} } a,
\end{equation}
in terms of which the equations of motion become
\begin{equation}
\ddot{\bar{\phi}} + ( p+ 1) \frac{\dot{\bar{a}}}{\bar{a}} \dot{\bar{\phi}} = - \bar{V}',
\end{equation}
\begin{equation}
\left( \frac{\dot{\bar{a}}}{\bar{a}} \right)^2 = \frac{2}{ p (p +
1)} \left( \frac{\dot{\bar{\phi}}^2}{2} +  \bar{V}
(\bar{\phi})
\right) - \frac{k}{\bar{a}^2},
\end{equation}
\begin{equation}
\frac{\ddot{\bar{a}}}{\bar{a}} = \frac{1}{ p (p+1) } \left( - p \dot{\bar{\phi}}^2 + 2 \bar{V} \right).
\end{equation}

\subsection{Non-singular big-bang and big-crunch surfaces}\label{sec:nonsingularsurfaces}

Just as for the black hole, we can match different regions across
non-singular big-bang or big-crunch surfaces, which from the
perspective of the full $D$-dimensional spacetime correspond to
event horizons. However, unlike the example of the black hole, the
equations of motion no longer guarantee that stationary points in
the radion field occur when $a=0$, and we must worry about the
existence of curvature singularities. We will again find that a
non-singular initial value surface is only possible by requiring a
stationary point $\dot{\phi} = 0$ at $a=0$. In addition, the scale
factor must have the universal behavior $a = \tau$ in the case of an
open FRW cosmology or $a = \exp(H \tau)$ in the case of a flat FRW
cosmology. In each of these cases, the initial value surface is null
and corresponds to an event horizon when the solutions are lifted to
$D$-dimensions.

To determine the location of a singularity, we can write the Ricci scalar in terms of the field $\phi$ as
\begin{equation}\label{eq:FRWricci}
\mathcal{R} = - \frac{\dot{\phi}^2}{M_{p+2}^2} + \frac{2 (p+2)}{p} \frac{V(\phi)}{M_{p+2}^p}.
\end{equation}
Therefore, as long as the kinetic and potential energy of the field remain bounded, there is no curvature singularity.

A trivial way to satisfy this condition is for the field to sit at a
critical
point of the potential for all times, yielding a
$p+2$-dimensional de Sitter, Minkowski, or Anti-de Sitter space
depending on the sign of the potential in regions where $\tau$ is
timelike. Such solutions were first discussed in
Ref.~\cite{Freund:1980xh}, and are sometimes referred to as
Freund-Rubin compactifications. These spaces are homogenous in time
as well as along the spatial sections, and so admit multiple time
slicings, corresponding to the choices for spatial curvature in the
FRW metric ansatz. de Sitter space admits a closed, flat, or open
foliation while Minkowsi and Anti-de Sitter admit a flat or open
foliation only. In each of these cases, the $a = 0$ surface is
simply a coordinate singularity which can be continued across.

There are also non-singular solutions when the field has non-trivial
dynamics. It is necessary to examine each ansatz for spatial
curvature separately when evaluating these solutions, since the
curvature term in Eq.~\ref{eq:ddimenfeqn} typically dominates as $a
\rightarrow 0$.

For a closed universe ($k=+1$), there are solutions where $a
\rightarrow 0$ only if the kinetic or potential energy of the field
diverge faster than $ + 1 / a^2$ (otherwise $\dot{a} / a$ becomes
imaginary). Since the energy density in the field becomes unbounded,
such solutions produce a curvature singularity at $a = 0$.
Otherwise, the scale factor has a non-zero minimal value, going from
contracting to expanding and producing a ``bouncing" cosmology with
no $a=0$ surfaces. We will not consider this case further.

In an open universe, the kinetic and potential energy of the field
can remain finite as $a \rightarrow 0$. In fact, it is possible to
have curvature alone, in which case the Milne patch of Minkowski
space is generated. Here, the $a = 0$ surface is obviously
non-singular, and simply corresponds to a boundary of the coordinate
patch. Putting $\phi$ back in, the absence of a curvature
singularity requires an everywhere-finite energy density in the
field. This in turn guarantees that the curvature term in
Eq.~\ref{eq:ddimenfeqn} will always dominate as $a \rightarrow 0$,
yielding a universal behavior for the scale factor $a = \tau$ as
$\tau \rightarrow 0$. Inserting this into the equation of motion
Eq.~\ref{eq:phifrweom} for $\phi$, consistency requires that
$\dot{\phi} (\tau = 0) = 0$ due the divergent friction/anti-friction
coefficient at $a = 0$. Solutions with these boundary conditions
have a completely non-singular $a = 0$ surface.

In a flat universe, it is again necessary to set $\dot{\phi} = 0$
when $a = 0$ to avoid a divergent friction/anti-friction
coefficient. From Eq.~\ref{eq:ddimenfeqn}, this requires that $V >
0$ to avoid an imaginary Hubble parameter. These conditions imply a
universal behavior near the big bang surface since the positive
vacuum energy comes to dominate at early enough times. In this case,
one obtains the flat slicing metric coefficient $a = \exp(H \tau)$
as $\tau \rightarrow - \infty$. This can be recognized as the metric
coefficient for de Sitter space in the flat slicing. The $\tau
\rightarrow - \infty$ surface is merely a boundary of the
coordinates, but in contrast to the case of negative spatial
curvature, the initial conditions surface is pushed infinitely far
to the past $\tau \rightarrow -\infty$ along the worldlines of
comoving observers (although boosted observers will encounter the
initial value surface in finite proper time).

This is an exhaustive list of the possible non-singular initial
value surfaces
under the FRW assumption for the $p+2$-dimensional
metric.

\section{Solutions for $\Lambda = 0$}\label{sec:solutionsL0}

\begin{figure*}
\begin{center}
\includegraphics[height=6cm]{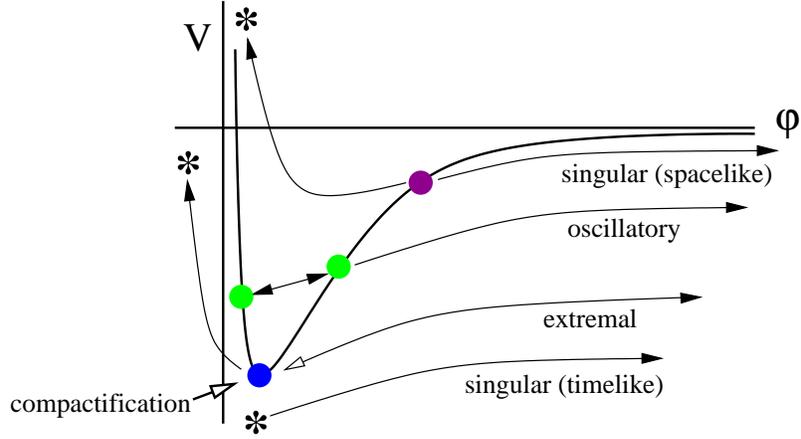}
\end{center}
\caption{A sketch of the solutions for $\Lambda = 0$ with a flat
or
open FRW ansatz for the $p+2$-dimensional metric. Beginning from
stationary points $\dot{\phi}=0$ (filled circles), the evolution is
in the potential $V$ for $\tau$ timelike (trajectories above the
potential) or $- V$ for $\tau$ spacelike (trajectories below the
potential). Singularities along a trajectory are denoted by an
asterisk. The compactification solution sits at the potential
minimum for all $\tau$. The singular (timelike and spacelike),
extremal, and oscillatory solutions all have a region where
$\tau$ is spacelike and $\phi \rightarrow \infty$, universally
approaching the attractor solution Eq.~\ref{eq:L0asymptoticmetric}.
Continuing across an arbitrary stationary point to a region where
$\tau$ is timelike will result in a spacelike singularity. There are
special stationary points from which $\phi$ oscillates about the
potential minimum (indicated by the double arrows on the oscillatory
trajectory). For a flat FRW metric ansatz, there are solutions where
$\tau$ is everywhere spacelike, and the field interpolates between
the potential minimum and $\phi \rightarrow \pm \infty$. Solutions
without a stationary point are always singular (timelike). }
\label{fig:Lzerosolutions}
\end{figure*}

\begin{figure*}
\begin{center}
\includegraphics[width=10cm]{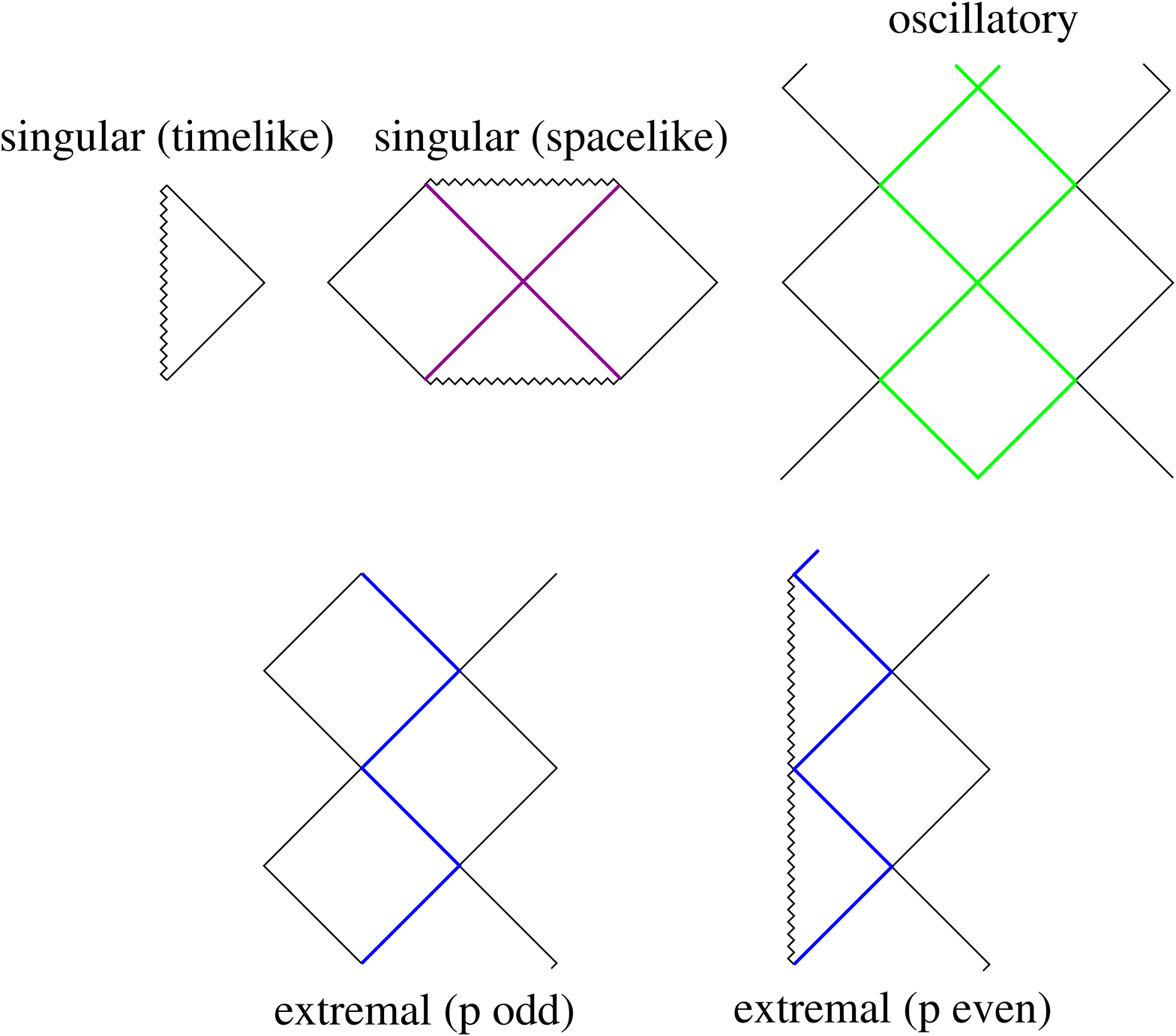}
\end{center}
\caption{The $D$-dimensional causal structure of the solutions
discussed
in the text and shown in Fig.~\ref{fig:Lzerosolutions}. In
the top row, we show the timelike singular (left), spacelike
singular (center), and oscillatory (right) solutions generated from
an open FRW metric ansatz. In the bottom row are the extremal
solutions generated from a flat FRW metric ansatz for $p$ odd (left)
and $p$ even (right). Event horizons correspond to the non-singular
big-bang and big-crunch surfaces at the stationary points of the
$p+2$ dimensional cosmological solutions, and are color-coded to
correspond to the stationary points in
Fig.~\ref{fig:Lzerosolutions}. Singularities occur when the energy
density in $\phi$ diverges at $a=0$, while the asymptotically flat
$D$-dimensional regions are described by the attractor solutions at
large $\phi$.} \label{fig:oscillconf}
\end{figure*}

In this section, we focus on geometries in the absence of a
$D$-dimensional cosmological constant. In addition to the $AdS_{p+2}
\times S^q$ compactification solutions, there are three types of
geometries that include an asymptotically flat $D$-dimensional
region: singular solutions (timelike and spacelike), oscillatory
solutions, and extremal solutions. We will construct examples of
each of these geometries in a piecewise manner by matching solutions
for $a(\tau)$ and $\phi(\tau)$ across non-singular big-bang and
big-crunch surfaces (event horizons) where $\dot{\phi}=0$ and $a=0$.
A representative example of the trajectory $\phi(\tau)$ in the
presence of the potential Eq.~\ref{eq:radionpotential} is sketched
in Fig.~\ref{fig:Lzerosolutions} for each type of solution. The
causal structure of the corresponding $D$-dimensional spacetimes are
shown in Fig.~\ref{fig:oscillconf}. Each of these solutions is an
extended object of codimension $p$ (i.e. a $p$-brane) embedded in
$D$-dimensional flat space.

A few properties of these solutions can be understood without a
detailed analysis.
First, as described in
Sec.~\ref{sec:radionpotential}, when $\Lambda = 0$, $V(\phi)$ always
has a negative global minimum and approaches zero from below as
$\phi \rightarrow \infty$. Recall that the motion is in the
potential $V (\phi)$ in regions where $\tau$ is timelike and $-
V(\phi)$ in regions where $\tau$ is spacelike. Therefore, the radius
of the $q$-sphere will always approach $D$-dimensional space in a
spacelike manner, consistent with an asymptotically flat solution.
If the field energy diverges along the trajectory of $\phi$, the
orientation of the potential also determines if the resulting
curvature singularity is spacelike or timelike. In geometries
without an event horizon (where there is no stationary point in
$\phi$), the naked singularities are timelike, while singularities
cloaked behind an event horizon are necessarily  spacelike.

We begin our detailed discussion of the various geometries with
the
compactification solution, where the field sits at the potential
minimum for all $\tau$. Since the minimum is negative, the full
solution is a $p+2$-dimensional Anti-de Sitter space with a
stabilized $q$-sphere of constant radius: $AdS_{p+2} \times S^q$.
The scale factor can be found from Eq.~\ref{eq:ddimauxfeqn}, which
if $\tau$ is timelike, becomes
\begin{equation}
\ddot{a} = - \omega_{a}^2 a,
\end{equation}
where
\begin{equation}
\omega_{a}^2 \equiv \frac{2 |V(\phi_{-}| )}{M_{p+2}^p p (p+1) }.
\end{equation}
For an open FRW metric ansatz, the solution is
\begin{equation}\label{eq:aoftau}
a (\tau) = \omega_{a}^{-1} \sin \left( \omega_a \tau \right).
\end{equation}
There is no solution for a flat or closed FRW ansatz with $\tau$
timelike
(the Hubble constant would be imaginary). However, for
spacelike $\tau$ there are solutions, corresponding to different
time slicings of the $AdS$ space.

In all other cases, the field has dynamics. As discussed
in
Sec.~\ref{sec:nonsingularsurfaces}, the behavior of the field will
be quite different  for a closed, flat, or open FRW metric ansatz.
We will only consider the open and flat cases in detail since they
allow for the possibility of non-singular big-bang and/or big-crunch
surfaces.

\subsection{Open FRW metric ansatz}

We first specialize to the case of open surfaces of homogeneity.
At
large $\phi$, far from the potential minimum, consider evolution
from a generic set of initial conditions $\{ a (\tau_i), \phi
(\tau_i), \dot{\phi} (\tau_i) \}$. To obtain asymptotically flat
$D$-dimensional space, $\tau$ must be a spacelike variable, and the
evolution is in the inverted potential. Evolving to either earlier
or later $\tau$ from $\tau_i$, the field will overshoot the
inverted-potential maximum if $\dot{\phi}$ is to large. This causes
the field to run off to $\phi \rightarrow -\infty$, resulting in a
curvature singularity. An example of such a ``singular (timelike)"
solution is shown in Fig.~\ref{fig:Lzerosolutions}, and the
resulting $D$-dimensional spacetime is shown in the top left panel
of Fig.~\ref{fig:oscillconf} (which, as could be ascertained from
the naming scheme, has a naked timelike singularity).

However, as described in Sec.~\ref{sec:nonsingularsurfaces},
these
singularities can be avoided by ensuring that $\dot{\phi} = 0$ when
$a = 0$. Specifying boundary conditions for $\phi$ and $a$ at the
position of a stationary point in the field ($\dot{\phi} = 0$), we can
pick out the non-singular solutions by setting $a=0$. If the
stationary point occurs at some $\phi > \phi_{-}$, then the field
will evolve to $\phi \rightarrow \infty$, and therefore all such
solutions have an asymptotically $D$-dimensional region containing
an event horizon. At the event horizon, it is possible to continue
to a region where $\tau$ is timelike, and the motion is in the
original potential. By analyzing the motion away from stationary
points at various $\phi$ for both orientations of the
potential, it is possible to build up the full $D$-dimensional
geometry piece by piece. A similar classification of solutions by
their stationary points was made in the black hole example described
in the Introduction.

All of the solutions will have a region with spacelike $\tau$
that
evolves towards $\phi \rightarrow \infty$. At large $\phi$, the
radion potential Eq.~\ref{eq:radionpotential} becomes dominated by a
single term
\begin{equation}
V (\phi) \simeq  \frac{M_{p+2}^p M_D^2 q (q-1)}{2}  \exp\left( - 2 \sqrt{\frac{p+q}{pq}} \frac{\phi}{M_{p+2}} \right).
\end{equation}
It is well known that there are attractor solutions for the
evolution of a scalar field in
the presence of an exponential
potential \cite{Halliwell:1986ja,Ratra:1987rm}. Generic initial
conditions will approach these fixed points in phase space, and the
asymptotics can be determined. Each geometry should therefore have
the same behavior in the asymptotically flat region.

We will find it convenient to work in terms of the dimensionless set
of FRW variables
defined in Eq.~\ref{eq:dimlessvariables}, and begin
by making the ansatz that the scale factor has power law behavior
(see e.g. \cite{Liddle:1988tb,Giddings:2004vr})
\begin{equation}\label{eq:aansatz}
\bar{a} = \bar{a}_0 (c \bar{\tau})^{\delta}.
\end{equation}
For an exponential potential of the form we have been studying
\begin{equation}
\bar{V}(\bar{\phi}) = A e^{- \gamma \bar{\phi}},
\end{equation}
the field equation is satisfied for
\begin{equation}
\bar{\phi} = \frac{2}{\gamma} \log (c \bar{\tau}),
\end{equation}
with the constant $c$ fixed to be
\begin{equation}\label{eq:c}
c = \frac{\gamma}{\sqrt{2}} \sqrt{ \frac{A}{(1+p) \delta - 1} }.
\end{equation}
Since there is both curvature and field energy, it is necessary to
determine
which will dominate the dynamics of the scale factor.
Substituting for $\bar{\phi}$ and $\bar{a}$ from above, we can write
Eq.~\ref{eq:ddimenfeqn} as
\begin{equation}
\left( \frac{\dot{\bar{a}}}{\bar{a}} \right)^2 = \frac{4 \delta}{p \gamma^2 \bar{\tau}^2 } + \frac{1}{\bar{a}_0 c^{2 \delta} \bar{\tau}^{2 \delta}},
\end{equation}
where the first term represents the contribution from the field and
the second from the curvature term.
It can be seen that the field
dominates the dynamics of the scale factor for $\delta > 1$. If this
is the case, using Eq.~\ref{eq:aansatz}, $\delta$ is found to be
\begin{equation}
\delta = \frac{4}{(p \gamma^2)}.
\end{equation}
If this is less than one for a given potential, then neglecting the
curvature term is not
self consistent, and the equations of motion
imply that $\delta = 1$. This analysis was performed by
Ref.~\cite{Giddings:2004vr}, who also showed numerically that these
two behaviors are attractors in phase space.

Setting $\mu^{p+2} =  M_{p+2}^p M_D^2 / 2$ and comparing with the analysis above, we have
\begin{equation}
A = q (q-1), \ \ \ \gamma = 2 \sqrt{\frac{p+q}{pq}}, \ \ \  \frac{4}{(p \gamma^2)} = \frac{q}{p+q} < 1,
\end{equation}
implying that the attractor solution has $\delta = 1$. The constant $c$ is given by
\begin{equation}
c = \sqrt{\frac{2 (p+q) (q-1)}{p^2}}.
\end{equation}
From Eq.~\ref{eq:Rdef}, we can find $R (\tau)$
\begin{equation}
M_D R(\tau) = \left( c \bar{\tau} \right)^{\frac{p}{p+q}}.
\end{equation}
Using this, we can find the full metric Eq.~\ref{eq:Ddimeinsteinmetric} in the $D$-dimensional Einstein frame
\begin{equation}\label{eq:L0asymptoticmetric}
M_D^2 d \tilde{s}^2 = (c \bar{\tau})^{-2 \frac{q}{p+q}} \left[ d
\bar{\tau}^2 +
\bar{a}_0^2 c^2 \bar{\tau}^2 \left( - d\chi^2 +
\cosh^2 \chi d\Omega_{p}^2 \right) \right] + (c \bar{\tau})^{2
\frac{p}{q+p}} d\Omega_q^2.
\end{equation}
Calculating the $D$-dimensional Ricci scalar, in all cases it is proportional to
\begin{equation}
\tilde{\mathcal{R}}^{(D)} \propto \bar{\tau}^{-\frac{2 p}{p+q}} \propto \frac{1}{R^2},
\end{equation}
and the metric is asymptotically flat as $\bar{\tau} \rightarrow \infty$ as expected.

Returning to the stationary point $\dot{\phi}=0$, $a=0$ of the
non-singular solutions
(i.e. the event horizon in the $D$-dimensional
geometry), we can analytically continue to timelike $\tau$. Here,
there is no universal behavior, and different stationary points will
produce qualitatively different solutions. For most choices, the
solutions become singular. These portions of the ``singular
(spacelike)" solutions describe a  region containing a spacelike
singularity separated from asymptotically flat $D$-dimensional space
by an event horizon. For some special choices of the stationary
point, completely non-singular solutions can be generated. These
``oscillatory" solutions describe a region between two event
horizons in which the radius of the $q$-sphere oscillates. Across
each event horizon is a different region of asymptotically flat
$D$-dimensional space. We now analyze these cases in more detail.

For stationary points in the near vicinity of the negative potential
minimum,
Eq.~\ref{eq:ddimauxfeqn} implies that the scale factor $a$
is  bounded, and possess two zeroes. For small amplitude motion of
$\phi$, then $|V| \gg \dot{\phi}^2$, and the solution is given
approximately by Eq.~\ref{eq:aoftau}. Substituting
Eq.~\ref{eq:aoftau} into the field equation Eq.~\ref{eq:phifrweom}
and Taylor expanding the potential about the minimum we obtain
\begin{equation}
\ddot{\phi} + (p+1) \cot (\omega_a \tau) \dot{\phi} = - M_{p+2}^{2-p} |V'' (\phi_{-})| (\phi - \phi_{-}).
\end{equation}
This has an exact solution in terms of Gegenbauer polynomials
\begin{equation}
\phi - \phi_{-} = A C_{\sigma}^{3/2} \left[ \cos \left( \omega_a \tau \right) \right],
\end{equation}
where $A$ is the amplitude of oscillations and the index $\sigma$ is determined by the positive root of
\begin{equation}
\sigma^2 + (p+1) \sigma = M_{p+2}^{2-p} \frac{|V'' (\phi_{-})|}{\omega_a^2} = \frac{p (p+1)}{2} \frac{|V'' (\phi_{-})|}{|V(\phi_{-})|}.
\end{equation}
The ratio of the potential to its second derivative can be evaluated using the results of Sec.~\ref{sec:radionpotential}, yielding
\begin{equation}\label{eq:VppoverV}
\frac{|V'' (\phi_{-})|}{|V (\phi_{-})|} = \frac{4(p+1)}{p},
\end{equation}
Finally, solving for $\sigma$, we find
\begin{equation}
\sigma = p+1.
\end{equation}
Amazingly, for the radion potentials $\sigma$ is always an integer.
For $\sigma$ integer,
the Gegenbauer polynomial has $\sigma$ nodes
and $\dot{\phi}$ is zero at the endpoints of the motion in $a$. Such
solutions are completely non-singular.

Considering finite size excursions away from the minimum, the
corrections to the frequency
of oscillations of $\phi$ become
important. As $a$ approaches its second zero, $\dot{\phi}$ is no
longer zero, and the $\dot{\phi}$ term in the field equation
Eq.~\ref{eq:phifrweom} yields a divergent anti-friction. This
corresponds to the appearance of a spacelike curvature singularity
(since we are evolving in a region where $\tau$ is timelike) in the
geometry. A non-singular stationary point to the right of the
potential minimum  therefore generically connects the attractor
solutions discussed above to a region containing a spacelike
singularity. The $D$-dimensional causal structure of such singular
(spacelike) solutions is shown in the top center panel of
Fig.~\ref{fig:oscillconf}.

Although a generic choice of initial conditions leads to a
singularity, there are
special choices that lead to a completely
non-singular evolution of the field. For an initial condition to the
right of the minimum, the period of oscillation in $\phi$ increases
monotonically with increasing amplitude (the potential becomes more
gradual at large $\phi$). By increasing the period of oscillation,
eventually another harmonic of the period in $a$ will be reached.
The asymmetry of the potential forbids solutions with an odd number
of nodes, and it follows that there will be initial conditions
leading to non-singular solutions with $\sigma = 2, 4, \ldots p-1$
for $p$ odd and $\sigma = 2, 4, \ldots p$ for $p$ even. These
non-singular ``oscillatory" solutions connect two asymptotic regions
where $\phi \rightarrow \infty$ across non-singular big bang and
big-crunch surfaces. The $D$-dimensional causal structure is shown
in the top right panel of Fig.~\ref{fig:oscillconf}.

\begin{figure*}
\begin{center}
\includegraphics[height=4cm]{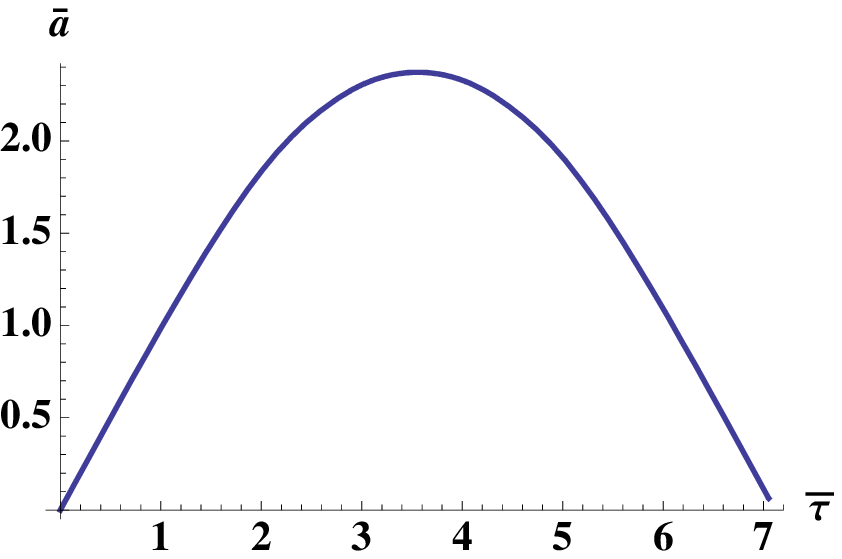}
\includegraphics[height=4cm]{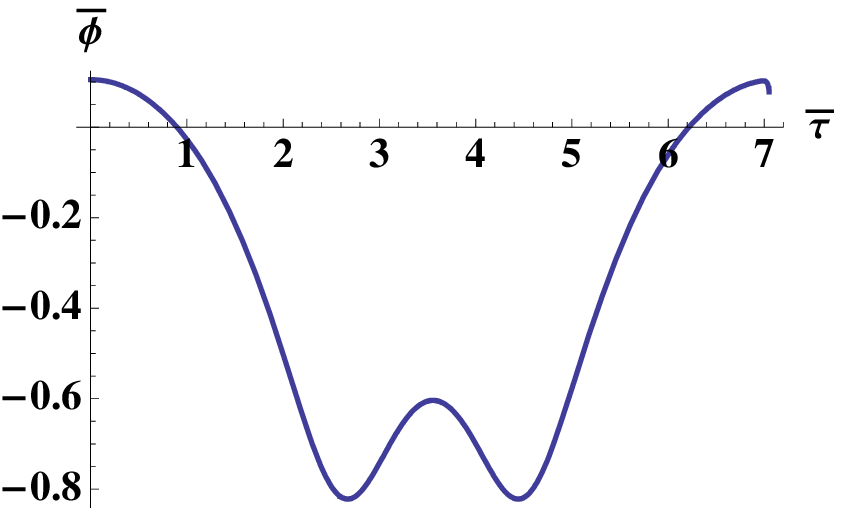}
\includegraphics[height=4cm]{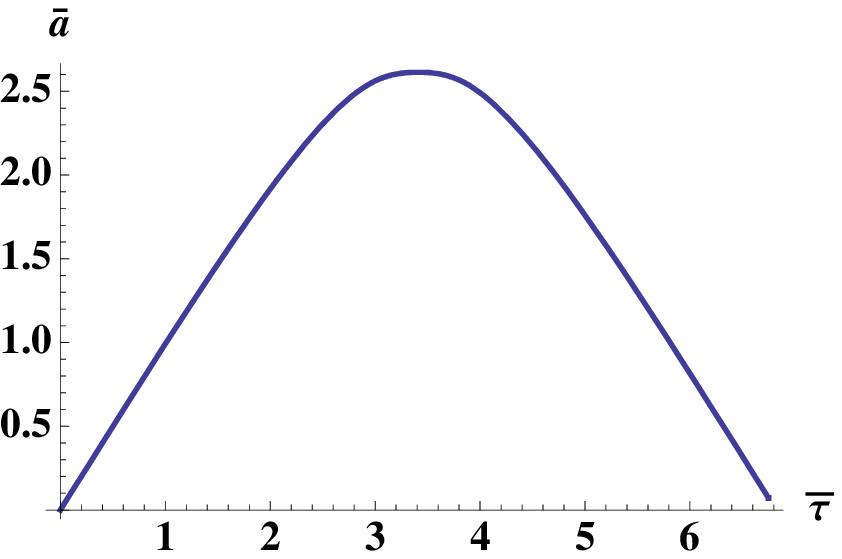}
\includegraphics[height=4cm]{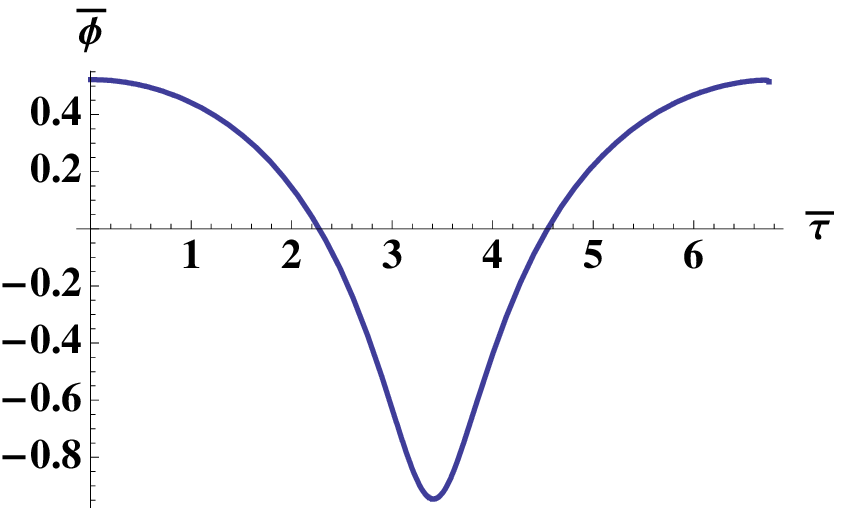}
\end{center}
\caption{The oscillatory solutions for timelike $\tau$ with $\Lambda
= 0$,
$Q=1$, $p=5$, and $q=2$. The solutions for
$\bar{a}(\bar{\tau})$ (left in each row) and $\bar{\phi}
(\bar{\tau})$ (right) describe the spacetime between two event
horizons of the $D$-dimensional geometry (see
Fig.~\ref{fig:oscillconf}). Event horizons are located at the zeros
in $a$. The top panel shows the oscillatory  solutions with $4$
nodes, generated from the initial condition $\phi_0 \simeq .105$.
The bottom panel shows the oscillatory solution with $2$ nodes
generated from the initial condition $\phi_0 \simeq .523$. A scan
over initial conditions reveals that these are the unique
non-singular solutions.} \label{fig:oscill}
\end{figure*}

We have confirmed the existence of such solutions
numerically.
Solutions for $p=5$ are shown in Fig.~\ref{fig:oscill},
where we have set $Q = 1$ and $\mu^{p+2} = M_{p+2}^p M_D^2 / 2$. As
expected, there are solutions with $4$ (top panel) and $2$ (bottom
panel) nodes for initial conditions at increasingly large $\phi$.

\subsection{Flat FRW metric ansatz}

\begin{figure*}
\begin{center}
\includegraphics[height=4.2cm]{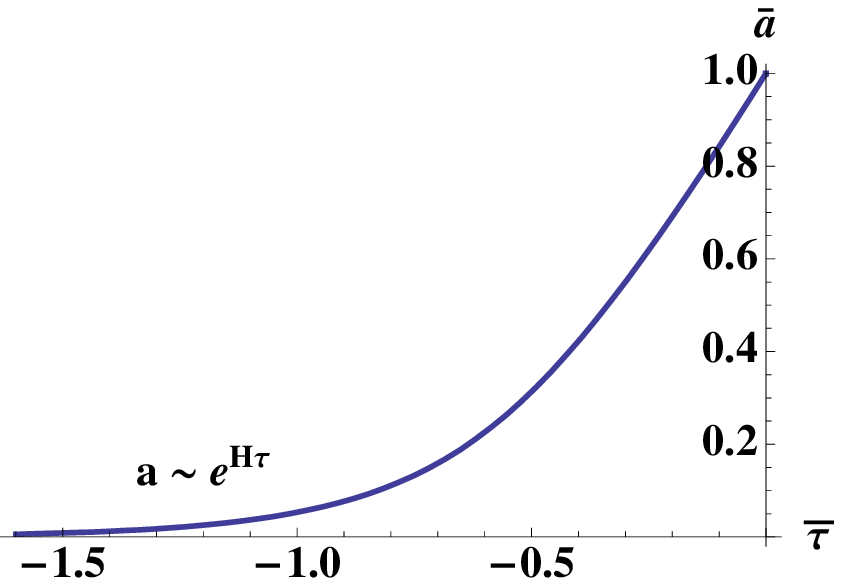}
\includegraphics[height=4.2cm]{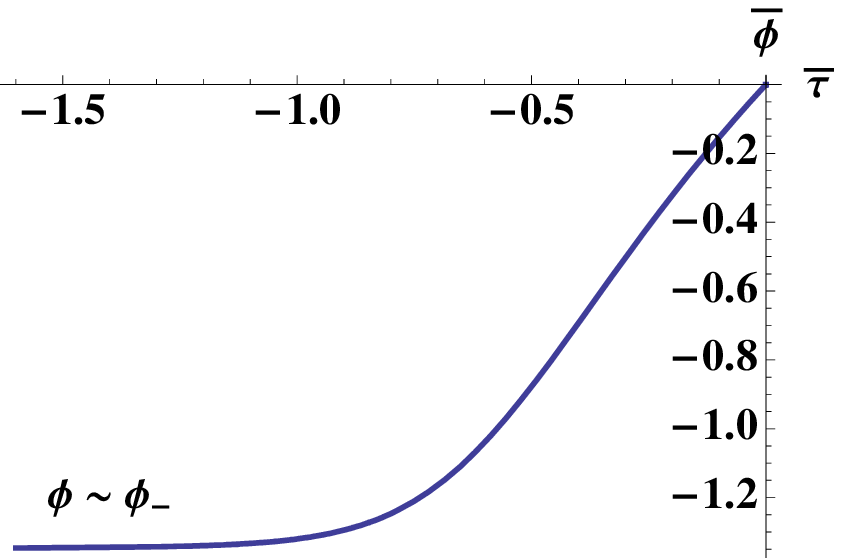}
\includegraphics[height=4.2cm]{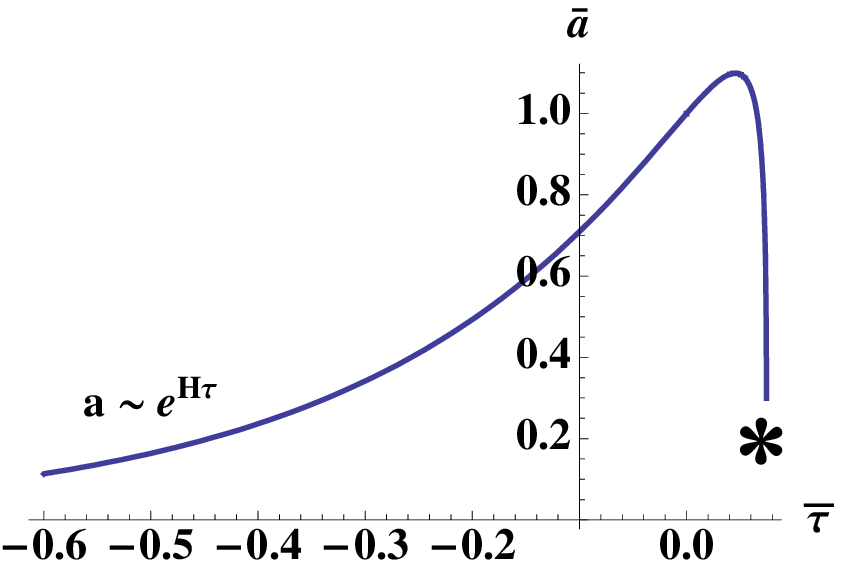}
\includegraphics[height=4.2cm]{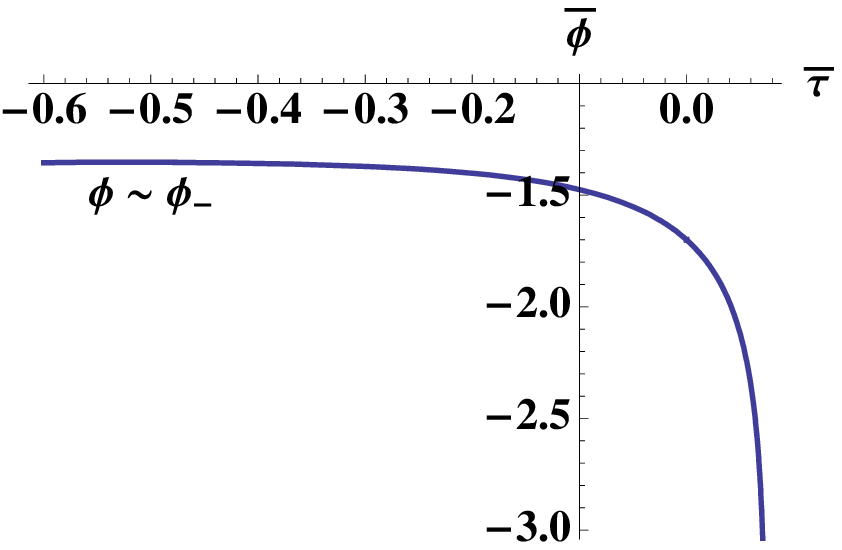}
\end{center}
\caption{The extremal solutions connecting $\phi_{-}$ and $\phi
\rightarrow \infty$ (top panel)
and  $\phi_{-}$ and $\phi \rightarrow
- \infty$ (bottom panel). In the bottom panel, there is a
singularity at finite $\tau$ when $a$ reaches a second zero, denoted
by the asterisk. Near $\tau \rightarrow -\infty$, the field
approaches the potential maximum, and $a$ approaches an exponential.
Event horizons of the $D$-dimensional geometry are located at $\tau
= - \infty$, where the two types of solutions in the top and bottom
panels (going to $\phi = \pm \infty$, i.e. $R=0$ or $R=\infty$) are
sewn together. For $p$ odd, two regions of asymptotically flat
$D$-dimensional space are matched across the horizon (two copies of
the solution in the top panel), while for $p$ even, a region of
asymptotically flat $D$-dimensional space is matched across the
event horizon to a region containing a timelike singularity
(corresponding to the top and bottom panel). The causal structure is
shown in Fig.~\ref{fig:oscillconf}. These are the non-dilatonic
extremal $p$-branes of Ref.~\cite{Gibbons:1994vm}, and the infinite
range in $\tau$ corresponds to the infinite proper distance
separating any point from the event horizon.} \label{fig:extremal}
\end{figure*}

We now discuss the extremal solution, which is obtained for a flat
FRW metric ansatz. Because the potential is negative in most regions
of interest,
the field cannot have a stationary point $\dot{\phi} =
0$ in a region with timelike $\tau$ (from Eq.~\ref{eq:ddimenfeqn}
this would cause the Hubble constant to become imaginary). This,
together with the condition $\ddot{a} < 0$ from
Eq.~\ref{eq:ddimauxfeqn}, implies that such spacetimes always begin
with a singular big-bang and end in a singular big-crunch. If
instead $\tau$ is a spacelike variable, the evolution is in the
upside-down potential $-V$, and it is possible to have a non-singular stationary
point $\dot{\phi} = 0$. As described in
Sec.~\ref{sec:nonsingularsurfaces}, nonsingular solutions must be
potential dominated as $a \rightarrow 0$ and the metric coefficient
$a$ must behave asymptotically like $a = \exp (H \tau)$ as $\tau
\rightarrow - \infty$, with $\dot{\phi} (\tau \rightarrow -\infty) =
0$. The only two solutions that can meet these requirements
interpolate between the inverted-potential maximum and $\phi
\rightarrow \pm \infty$, as shown in Fig.~\ref{fig:extremal}. These
``extremal" solutions are the only trajectories where $\phi$ can
loiter for an infinite range in $\tau$ near a stationary point. All
other solutions have a naked timelike singularity (the causal
structure resembles the top left panel of
Fig.~\ref{fig:oscillconf}).

The non-singular surface at $\tau = -\infty$ corresponds to an event
horizon in the
$D$-dimensional geometry. The infinite range in $\tau$
between any point $\phi$ and the potential maximum corresponds to an
infinite proper distance between any point in the $D$-dimensional
geometry and the event horizon. The individual solutions
interpolating between the potential maximum and $\phi = \pm \infty$
can be matched at the event horizon, corresponding to sewing
together regions of asymptotically flat $D$-dimensional space and
regions containing a timelike singularity. These full solutions are
none other than the non-dilatonic extremal $p$-branes described
in~\cite{Gibbons:1994vm}. These authors found that when $p$ is odd,
two identical regions of asymptotically flat space are connected
across the event horizon, while for $p$ even, a region of
asymptotically flat space is connected to a region containing a
timelike singularity across the event horizon.

In order to determine which of our solutions can be pasted together,
we
can take the limit of the open FRW solutions as the endpoint of
the motion approaches the inverted-potential maximum. At the
stationary point in the region where $\tau$ is spacelike, it is
possible to analytically continue to a region where $\tau$ becomes
timelike, and the motion is about the potential minimum. For motion
very near the minimum, we found above that the solution for $\phi$
was approximately a Gegenbauer polynomial with $p+1$ nodes. For any
finite displacement from the potential minimum, non-linearities
cause the solution to become singular near the second stationary
point in $a$, preventing a second analytic continuation back to a
region where $\tau$ is spacelike. For $p$ even, the singularity
occurs to the left of the minimum ($\phi$ has an odd number of
nodes), and for $p$ odd, to the right of the minimum (since $\phi$
has an even number of nodes). Taking the limit where the stationary
point of the field reaches the potential maximum, from the spacelike
side, the fact that we started with an open universe is irrelevant
since an infinite amount of proper time elapses before leaving the
potential maximum. However, the properties of the timelike solution
survive, and determine if the metric is invariant under reflection
of $\phi$ about the maximum. Thus, for $p$ odd, the spacetimes are
completely non-singular, connecting two identical asymptotically
flat regions of the spacetime across an event horizon. For $p$ even,
there is a timelike singularity behind the event horizon. This
agrees with the analysis of Ref.~\cite{Gibbons:1994vm}.

\subsection{Closed FRW metric ansatz}

Finally, we can briefly visit the closed FRW ansatz. Just as for the
flat case, solutions with
$\tau$ timelike always begin with a
singular big-crunch and end in a singular big-bang. For spacelike
$\tau$ the discussion in Sec.~\ref{sec:nonsingularsurfaces}
indicated that there is always a singularity at $a = 0$ (which in
this case will be a naked timelike singularity). This singularity
need not be present if there are bouncing solutions where $\ddot{a}$
is positive definite. From Eq.~\ref{eq:ddimauxfeqn}, this requires
that the field remain potential dominated over its entire evolution,
which simply does not occur in the presence of the radion potentials
(see Ref.~\cite{Giddings:2004vr} for further discussion). We
therefore conclude that all solutions with a closed FRW ansatz are
singular.

In conclusion, each of the solutions that we have found correspond to extended
 objects embedded in asymptotically flat $D$-dimensional space. The singular and 
 oscillatory solutions are new classes of solutions to higher dimensional Einstein gravity, 
 whereas the extremal solutions for $p$ even or odd are identified as the non-dilatonic $p$-brane solutions
 of Ref.~\cite{Gibbons:1994vm}. We now turn to the case with a positive $D$-dimensional 
 cosmological constant, $\Lambda > 0$, where again we will find new classes of solutions. 

\section{Solutions for $\Lambda > 0$}\label{sec:FRWLg0}

We now construct solutions where there is a positive cosmological
constant $\Lambda$ in the $D$-dimensional theory. The methods used
are identical to those in the previous section, but the character of
the solutions will be very different due to a number of qualitative
changes in the radion potential. First, for any non-zero $\Lambda$,
over some range of charge, the radion potential possesses two
extrema (as outlined in Sec.~\ref{sec:radionpotential}). One is a
maximum located at $\phi = \phi_+$ and the other a minimum at $\phi
= \phi_-$. Unlike the case where $\Lambda = 0$, the minimum can be
either positive, zero, or negative energy depending on the charge
$Q$. In addition, the potential in the case where $\Lambda>0$
approaches zero from above as $\phi \rightarrow \infty$.

The extrema of the potential give rise to
$AdS_{p+2} \times S^q$, ${\mathcal M}_{p+2} \times S^q$, and $dS_{p+2} \times S^q$ compactification
solutions. There are also three qualitatively different classes of
solutions that contain an asymptotically $D$-dimensional de Sitter
region: the singular, interpolating, and Nariai solutions. The
singular and interpolating solutions are generated from an open FRW
metric ansatz, and the Nariai solution from a flat FRW ansatz. There
are important qualitative differences between the solutions in the
case where the minimum of the radion potential is negative as
opposed to zero or positive, and so we treat each of these cases
separately.

\begin{figure*}
\begin{center}
\includegraphics[width=10cm]{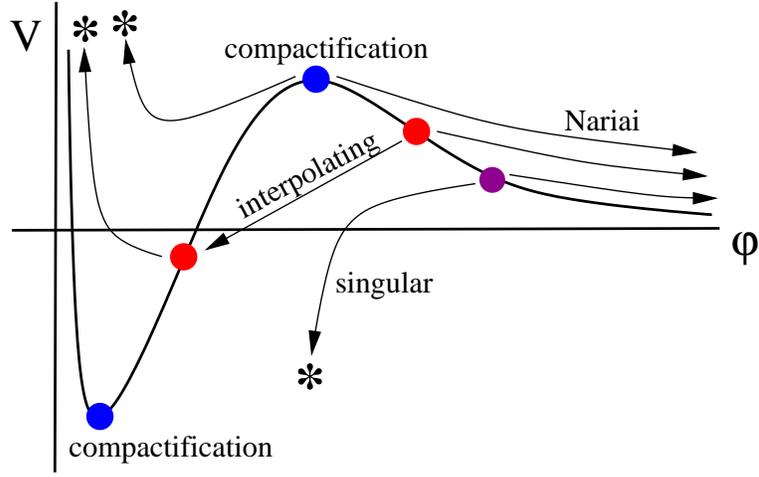}
\end{center}
\caption{A sketch of the solutions for $\Lambda > 0$ when
the radion
potential has a negative energy minimum. Motion above the potential
occurs in a region where $\tau$ is a timelike variable, and motion
below in regions where $\tau$ is a spacelike variable. An asterisk
denotes the occurrence of a curvature singularity along the
trajectory. Non-singular stationary points are denoted by the filled
circles. All of the solutions contain a singularity except for the
compactification solutions that sit at the critical points of the
potential for all $\tau$. At the minimum, an $AdS_{p+2} \times S^q$
geometry results, and at the maximum a $dS_{p+2} \times S^q$
geometry results. For an open FRW ansatz, the interpolating solution
has two non-singular stationary points and connects $\phi
\rightarrow \infty$ to a region containing a spacelike singularity.
All other choices for the location of the stationary points will
lead to a timelike singularity. In either of these cases, the
$D$-dimensional solution approaches de Sitter as $\tau \rightarrow
\infty$. For a flat FRW ansatz, the Nariai solution connects regions
interpolating between $\phi_{+}$ and $\phi \rightarrow \pm \infty$
over an infinite range in $\tau$. The Nariai solution has a
stationary point at $\phi_+$, but the geometry is distinct from the
compactification solution (although the near-horizon geometry of the
Nariai solution is {\em locally} $dS_{p+2} \times S^q$).}
\label{fig:Lnozeronegsolutions}
\end{figure*}

\begin{figure*}
\begin{center}
\includegraphics[width=12cm]{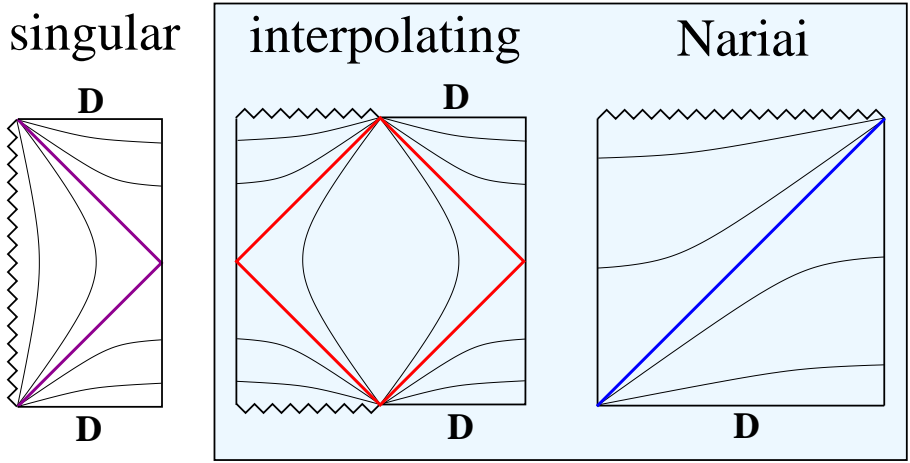}
\end{center}
\caption{Causal diagrams for $\Lambda > 0$ solutions with an
asymptotically
$D$-dimensional region and a radion potential with a
negative minimum. Surfaces of constant $\tau$ are shown as the light
solid lines. Starting from the left, we show the singular,
interpolating, and Nariai solutions. All solutions contain a
timelike or spacelike singularity separated from asymptotically
$D$-dimensional de Sitter space (denoted by the $D$) by event
horizons. The horizons are in each case color coded with the
stationary points of the radion field trajectories in
Fig.~\ref{fig:Lnozeronegsolutions}. We highlight the solutions
relevant for dynamical compactification in the shaded box.}
\label{fig:Lnozeronegconf}
\end{figure*}

\begin{figure*}
\begin{center}
\includegraphics[width=10cm]{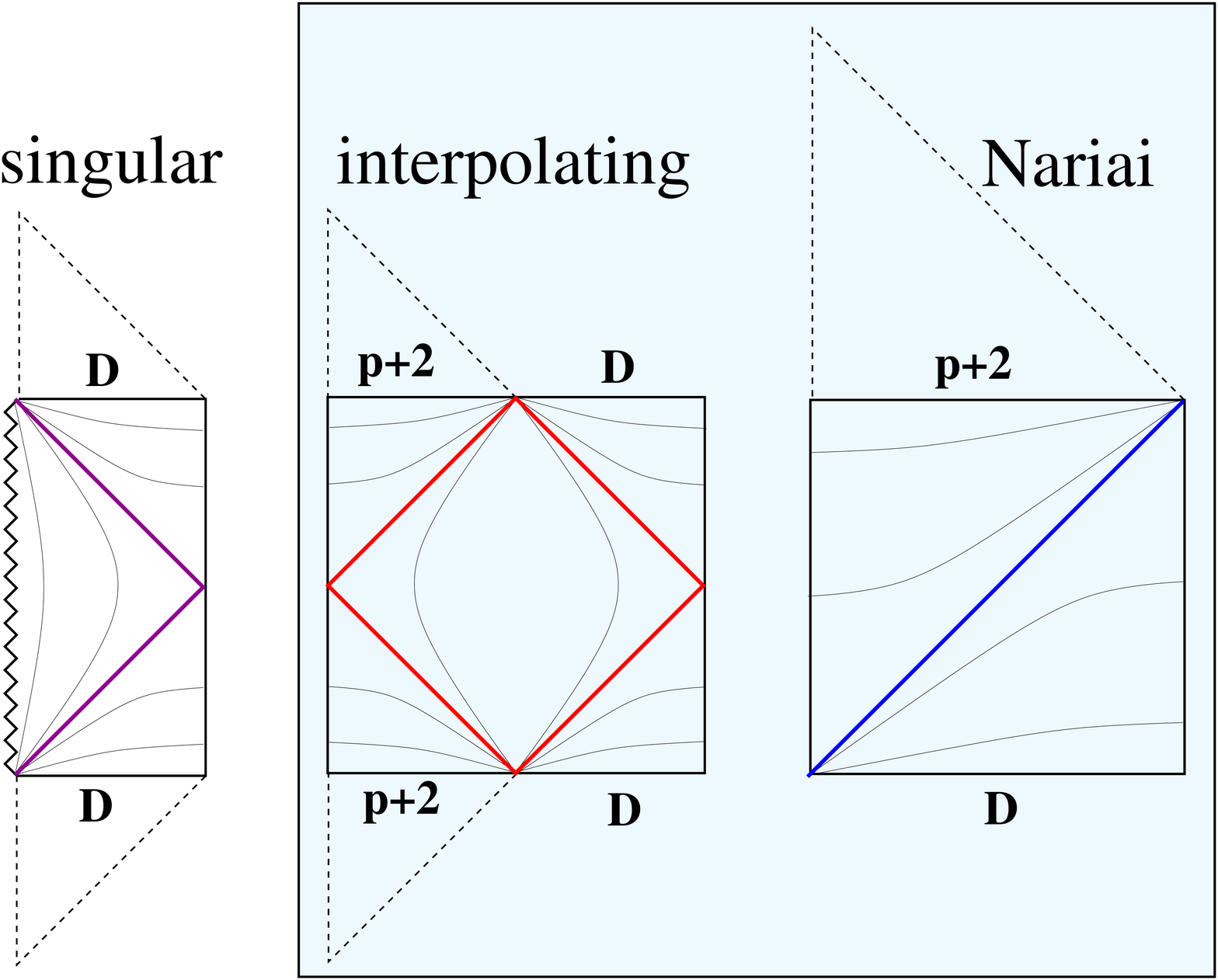}
\end{center}
\caption{Causal diagrams for $\Lambda > 0$ solutions with an
asymptotically
$D$-dimensional region and a radion potential that has
a positive or zero minimum. Surfaces of constant $\tau$ are shown as
the light solid lines. Starting from the left, we show the singular,
interpolating, and Nariai solutions. In the interpolating solution,
future/past infinity is split between non-singular regions of
different effective dimensionality (the dimensionality is labeled
$D$ or $p+2$). When the minimum has zero energy, the
$p+2$-dimensional parts of future infinity is null (denoted by the
dashed ``hat"). The Nariai solution connects the contracting portion
of a $D$-dimensional de Sitter space to a non-singular
$p+2$-dimensional flat FRW universe across an event horizon.  The
horizons are in each case color coded with the stationary points of
the radion field in Fig.~\ref{fig:Lnozeronegsolutions}. We highlight
the solutions relevant for dynamical compactification in the shaded
box.} \label{fig:Lnozeroposconf}
\end{figure*}

The possible trajectories when the radion
potential has a negative
minimum are shown in Fig.~\ref{fig:Lnozeronegsolutions}, the
possible solutions for a positive or zero minimum are shown in
Fig.~\ref{fig:Lzerosolutions}, and the causal structure of the
corresponding $D$-dimensional spacetimes is shown in
Figs.~\ref{fig:Lnozeronegconf} and~\ref{fig:Lnozeroposconf}. Notable
among these geometries are the interpolating and Nariai solutions,
which have the remarkable property of interpolating between
$D$-dimensional and effectively $p + 2$-dimensional regions across
event horizons. We summarize the cosmological significance of these
solutions at the end of this section.

\begin{figure*}
\begin{center}
\includegraphics[width=10cm]{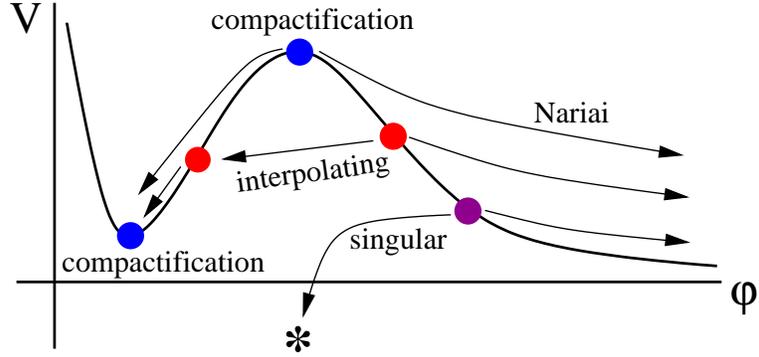}
\end{center}
\caption{A sketch of the solutions for $\Lambda > 0$ when the
radion potential has a zero or positive energy minimum. There is one
de Sitter and one de Sitter or Minkowski compactification solution at the
critical points of the potential. As for the case where the radion
potential has a negative minimum, for an open FRW ansatz there is an
interpolating solution that has two non-singular stationary points
connecting the asymptotic $\phi \rightarrow \infty$ region to the
basin of attraction of the potential minimum. Motion away from the
second stationary point will lead to a completely non-singular
evolution towards $\phi_{-}$, asymptotically producing a region of
$p+2$ dimensional Minkowski or de Sitter space. Solutions with any
other stationary points contain a timelike singularity. For a flat
FRW ansatz, the Nariai solution connects regions interpolating
between $\phi_{+}$ and $\phi_{-}$ or $\phi \rightarrow \infty$ over
an infinite range in $\tau$. The Nariai solution is in this case
completely non-singular.} \label{fig:Lnozeropossolutions}
\end{figure*}

For fixed charge, there is a compactification solution for each of
the critical points of the potential (0, 1, or 2 critical points are
possible). Positive, zero, and negative energy minima of the radion
potential yield a $dS_{p+2} \times S^q$, ${\mathcal M}_{p+2} \times S^q$, and
$AdS_{p+2} \times S^q$ geometry respectively. Since these solutions
sit at a potential minimum, they are stable to homogenous
perturbations in the radion field. The compactification solution
produced by sitting at the radion potential maximum will always be
$dS_{p+2} \times S^q$. This solution is unstable to homogenous
perturbations in the radion field, which would cause spacetime
regions to evolve away from the maximum at $\phi_+$ either towards
smaller or larger $\phi$.

The construction of the Anti-de Sitter compactification solutions will proceed as
described above. For positive or zero energy minima, the behavior of
the scale factor is different.
Minkowski space can be foliated into
flat or open spatial sections, and de Sitter space flat, open, or
closed spatial sections. In any case, the equation of motion for the
scale factor in regions where $\tau$ is timelike is again  given by
\begin{equation}
\ddot{a} = \omega_{a}^2 a,
\end{equation}
where
\begin{equation}\label{eq:afreq}
\omega_{a}^2 \equiv \frac{2 V(\phi_{crit} )}{M_{p+2}^p p (p+1) }.
\end{equation}
The initial conditions are set by the Friedmann equation Eq.~\ref{eq:ddimenfeqn}, and depend on the spatial curvature.

\subsection{Open FRW metric ansatz}

Beginning our study of dynamical solutions with an open FRW metric
ansatz, we need to consider the evolution of the field in three
separate regions of the potential: the asymptotic $\phi \rightarrow
\infty$ region, the region under the potential barrier, and the
region in the basin of attraction of the minimum. For $\Lambda > 0$
the potential approaches zero from above as $\phi \rightarrow
\infty$, implying that $\tau$ is a timelike variable in the
asymptotic region. If there is evolution under the barrier (this
requires that the energy density in the field is small enough),
$\tau$ must be a spacelike variable. Finally, for evolution in the
basin of attraction of the minimum, $\tau$ is again a timelike
variable. The evolution in each of these three regions can be sewn
together across non-singular big-bang and big-crunch surfaces where
$\dot{\phi} =0$ and $a=0$, just as in the previous section.

In the asymptotic region $\phi \rightarrow \infty$, where $\tau$
is timelike, we can again look for attractor solutions that are
approached starting from generic initial conditions. The dominant
term in the potential at large $\phi$ is
\begin{equation}
\bar{V} (\phi) \simeq  \frac{\Lambda}{M_D^2} \exp\left( - 2 \sqrt{\frac{q}{p (p+q)}} \frac{\phi}{M_{p+2}} \right).
\end{equation}
Again, we will find it convenient to work in terms of the
dimensionless variables defined in
Eq.~\ref{eq:dimlessvariables}.
From the asymptotic form of the
potential, we can identify the constants
\begin{equation}
A = \frac{\Lambda}{M_D^2} , \ \ \ \gamma = 2 \sqrt{\frac{q}{p (p+q)}}, \ \ \  \frac{4}{(p \gamma^2)} =
\delta = \frac{p+q}{q} > 1, \ \ \ c = \frac{q}{p} \sqrt{\frac{2 \Lambda}{M_D^2 (p+q) (p+q+1)}}.
\end{equation}
In this case, we have $\delta > 1$, and the field dynamics dominate
the contribution from curvature.
The metric in the $D$-dimensional
Einstein frame is given by
\begin{equation}
M_D^2 d\tilde{s}^2 = (c \bar{\tau})^{-2} \left[ - d \bar{\tau}^2 +
\bar{a}_0^2 (c \bar{\tau})^{2(p+q)/q}
\left( d\chi^2 + \sinh^2 \chi
d\Omega_{p}^2 \right) \right] + (c \bar{\tau})^{2 p / q}
d\Omega_q^2.
\end{equation}
We can define a new time coordinate
\begin{equation}
c M_D T = \log (c \bar{\tau}),
\end{equation}
in terms of which (after substituting for the appropriate values of $c$ from Eq.~\ref{eq:c}) the metric becomes
\begin{equation}
d\tilde{s}^2 =- d T^2 + a_0^2 \exp \left( 2 \sqrt{\frac{2
\Lambda}{(p+q) (p+q+1)} } T  \right)
\left[  d\chi^2 + \sinh^2 \chi
d\Omega_{p}^2  + d\Omega_q^2 \right].
\end{equation}
This is the metric of $D = p+q+2$ dimensional de Sitter space with
cosmological constant
$\Lambda$ (this analysis was first performed in
Ref.~\cite{Giddings:2004vr}).

As in the previous section, if $a(\tau = 0) = 0$ at a stationary
point $\dot{\phi} (\tau = 0) = 0$,
then there is a corresponding
event horizon in the asymptotically $D$-dimensional region.
Otherwise, there is a naked spacelike singularity corresponding to a
singular big-bang or big-crunch surface. In the non-singular case,
we can continue across the stationary point to a region where the
field evolves in the inverted potential and $\tau$ is spacelike.
Considering motion about the near-vicinity of the inverted maximum,
the scale factor will have a sinusoidal behavior, and the field
equation has a solution in terms of Gegenbauer polynomials (as was
the case for motion about the potential minimum for $\Lambda = 0$).
The number of nodes is given by the positive root of
\begin{equation}\label{eq:alphainterpolating}
\sigma^2 + (p+1) \sigma = \frac{p (p+1)}{2} \frac{|V'' (\phi_{+})|}{|V(\phi_{+})|}.
\end{equation}
In the limit of small $Q$, we can use Eqs.~\ref{eq:R0maximum} and~\ref{eq:vbarrier} to find the ratio
\begin{equation}
\frac{|V'' (\phi_{+})|}{|V(\phi_{+})|} = \frac{4}{p}
\end{equation}
Solving for $\sigma$,
\begin{equation}
\sigma = \frac{1}{2} \left[ - (p+1) + \sqrt{8 (p+1) + p^2} \right].
\end{equation}
This ranges between $1 < \sigma < 2$ for all $p$ and is never integer.

Starting from an arbitrary stationary point to the right of the
potential maximum yields a singular solution, since $\sigma$ is not
integer. The $D$-dimensional causal structure of such solutions is
shown in the left panel of Figs.~\ref{fig:Lnozeronegconf}
and~\ref{fig:Lnozeroposconf}. However, this singularity is
potentially avoidable since the field has a sufficiently long time
to cross the potential maximum at least once before $a$ reaches its
second zero. Increasing the initial displacement from the maximum,
the frequency of oscillations in the field decreases, and there will
always be a critical point at which the traversal time of $\phi$
matches the period of the scale factor. This is essentially
Coleman's undershoot/overshoot argument for finding instantons in a
scalar field potential~\cite{Coleman:1977py}. For different values
of the charge, there is a unique stationary point that yields such a
non-singular ``interpolating" solution. Although the requisite set
of initial conditions seems rather special, we will find in
Sec.~\ref{sec:dynamicalcompactification} that there are instanton
solutions that pick them out uniquely.

\begin{figure*}
\begin{center}
\includegraphics[height=4cm]{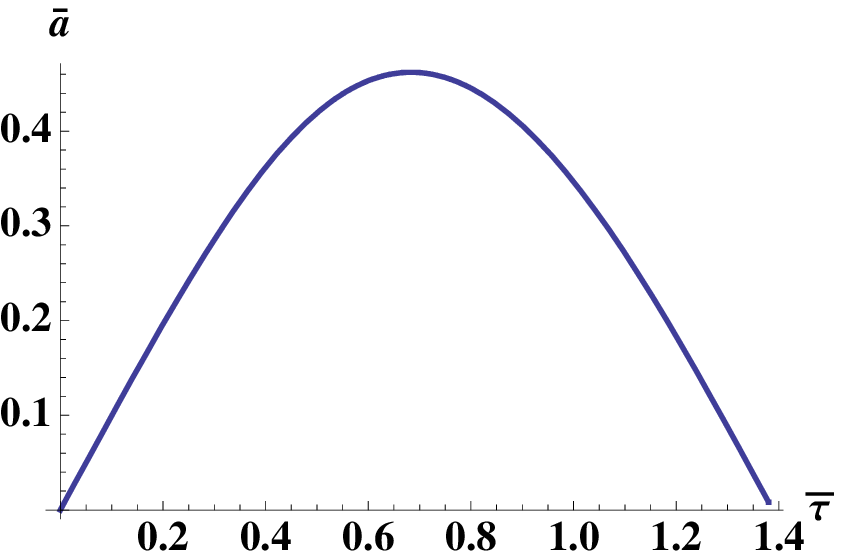}
\includegraphics[height=4cm]{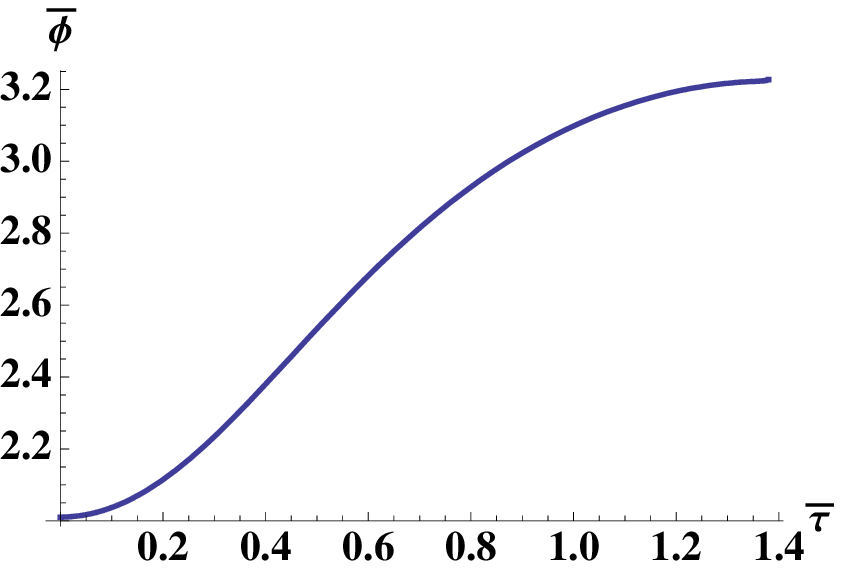}
\end{center}
\caption{The interpolating solution for $p=2$, $q=2$, $Q = 3.2$,
and
$\Lambda / M_{D}^2 = .1$ generated from the initial condition
$\bar{\phi}_0 \simeq 2.01 $. The field interpolates between either
side of the potential barrier while the scale factor evolves between
its two zeros.} \label{fig:interpolating}
\end{figure*}

We have confirmed the existence of such solutions numerically.
An example for the parameters $p=2$, $q=2$, $Q = 3.2$, and $\Lambda /
M_{D}^2 = .1$ is shown in Fig.~\ref{fig:interpolating}. Here, it can
be seen that the radion field interpolates between either side of
the barrier as the scale factor evolves between its two zeroes. In
the $D$-dimensional geometry, this describes the behavior of the
$q$-sphere between two event horizons. As the value of the charge
increases for fixed $p$, $q$, and $\Lambda$, the potential maximum
broadens (recall that the maximum and minimum of the potential merge
at a finite charge Eq.~\ref{eq:Qlimit}). At a large enough value for
the charge, we have $\sigma < 1$, implying that the field does not
have sufficient time to cross the barrier before the scale factor
reaches its second zero. In these cases, the interpolating solution
does not exist.

At the second non-singular stationary point ($\phi < \phi_{+}$)
of the interpolating solution, we can continue to a region where
$\tau$ is again a timelike variable. The properties of the solution
in this region depend on the sign of the potential minimum. For
small charge (much less than the the limit Eq.~\ref{eq:Qlimit}), the
potential near the minimum is negative and approximately the same as
for $\Lambda = 0$. Evolution away from the stationary point of the
interpolating solution always results in a singularity as the
non-zero kinetic energy of the field blue-shifts in the negative
vacuum energy background (this is in exact analogy with the
big-crunch that occurs inside of negative energy vacuum
bubbles~\cite{Abbott:1985kr}). If instead $V (\phi_{-}) > 0$, the
field will simply experience damped oscillation around the positive
minimum while the scale factor grows monotonically. The solution in
this case is completely non-singular.

Solutions of this type interpolate between the asymptotic
$D$-dimensional region $\phi \rightarrow \infty$ and the basin of
attraction of the radion potential minimum where the spacetime is
effectively $p+2$-dimensional. The causal structure of these
solutions is shown in the center panel of
Figs.~\ref{fig:Lnozeronegconf} (for a negative minimum)
and~\ref{fig:Lnozeroposconf} (for a zero or positive minimum). In
each diagram, there are three qualitatively different regions
corresponding to the three segments of the field evolution described
above. In each case, there is a patch in which the field evolves to
$\phi \rightarrow \infty$, asymptotically reaching $D$-dimensional
de Sitter space (the right portions of past and future infinity
marked "D" on the causal diagrams) and a region where the field
interpolates between non-singular stationary points under the
potential maximum (the central diamond on the causal diagrams).
There is then a region that evolves in the basin of attraction of
the radion potential, ending in a singularity if the minimum is of
negative energy (as in Fig.~\ref{fig:Lnozeronegconf}) or in a
$p+2$-dimensional asymptotically flat or de Sitter region (the left
portions of past and future infinity marked "p+2") if the minimum is
zero or positive energy (as in Fig.~\ref{fig:Lnozeroposconf}). At
the boundary of each segment, there is an event horizon.

The regions evolving in the basin of attraction of the
radion potential are $p+2$-dimensional open FRW universes with a
non-singular big-bang surface. This is extremely interesting from
the standpoint of cosmology, since there exists a pre-big bang epoch
in which the effective dimensionality of spacetime changes.

What happens to the interpolating solutions when the charge goes to zero? In this limit, the minimum of the potential goes to $\phi \rightarrow - \infty$. At $Q = 0$, the minimum disappears, since the positive contribution to the potential Eq.~\ref{eq:radionpotential} is no longer present. Studying a few numerically generated examples, we find that the stationary point to the left of the potential maximum follows the location of the minimum towards $\phi \rightarrow -\infty$. At $Q=0$, the location of the stationary point will be at $R = 0$, which is just the non-singular origin of coordinates. The spacetime has a single horizon, corresponding to the stationary point to the right of the maximum. This is nothing other than empty $D$-dimensional de Sitter space. 

Finally, we note that in regions with timelike $\tau$ it is possible to find a portion of the
oscillatory solutions described in the previous section. For small
charge, the potential in the vicinity of the minimum is largely
unaffected by the positive cosmological constant, and so the
analysis will proceed exactly as before. As the charge increases,
the depth of the minimum is decreased, and the term in the potential
proportional to $\Lambda$ begins to contribute to the second
derivative. This causes the ratio Eq.~\ref{eq:VppoverV} to diverge
as $V(\phi_{-}) \rightarrow 0$. As this occurs, it becomes possible
to find an increasing number of oscillatory solutions since the
frequency of the scale factor oscillations goes to zero while the
frequency of field oscillations stays roughly constant. The number
of solutions diverges as $V \rightarrow 0$. However, the
non-singular stationary point of the interpolating solution never
match the non-singular stationary point of the oscillatory solution.
This means, in contrast to the solutions with $\Lambda = 0$, that it
is impossible to match these oscillatory solutions to an
asymptotically $D$-dimensional region.

\subsection{Flat FRW metric ansatz}

For a flat FRW ansatz, the extremal solution evolves away from the
potential minimum in a
region where $\tau$ is spacelike, and can only
exist when $V(\phi_{-}) < 0$. However, unlike the case where
$\Lambda = 0$, these solutions cannot evolve to $\phi \rightarrow
\infty$ (recall that the interpolating solution is the only
non-singular solution in the region under the potential barrier).
Instead, these solutions have two timelike singularities separated
by an event horizon.

\begin{figure*}
\begin{center}
\includegraphics[height=4cm]{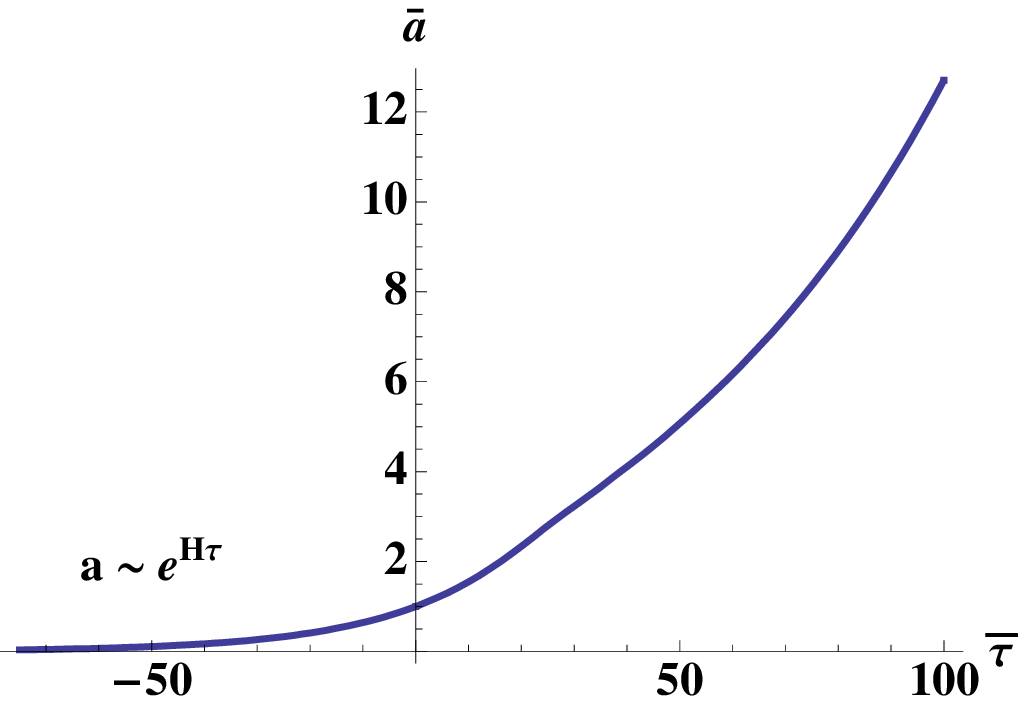}
\includegraphics[height=4cm]{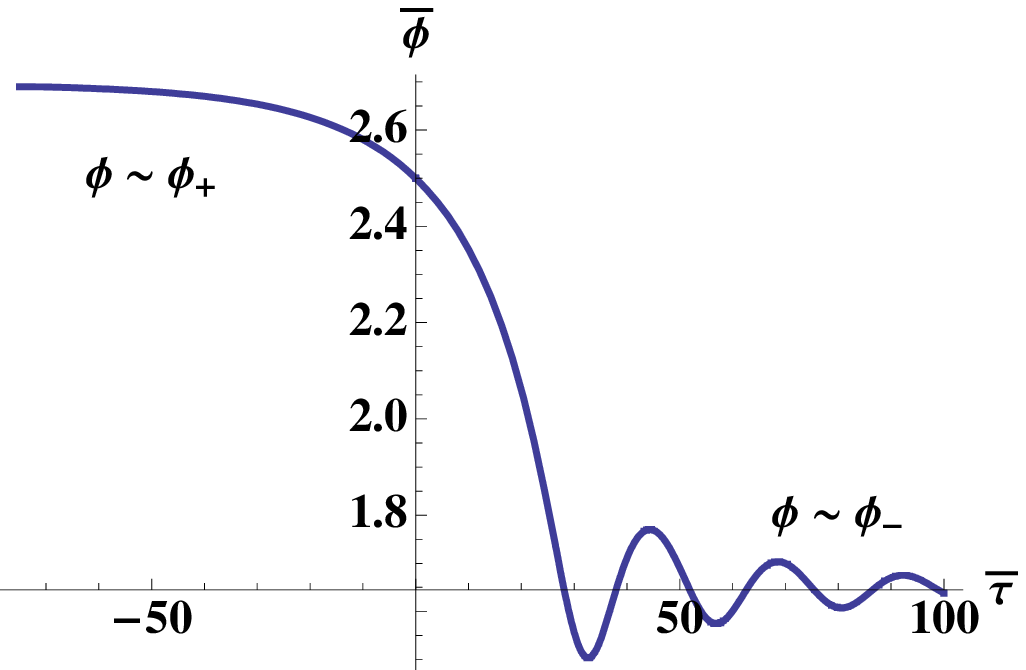}
\includegraphics[height=4cm]{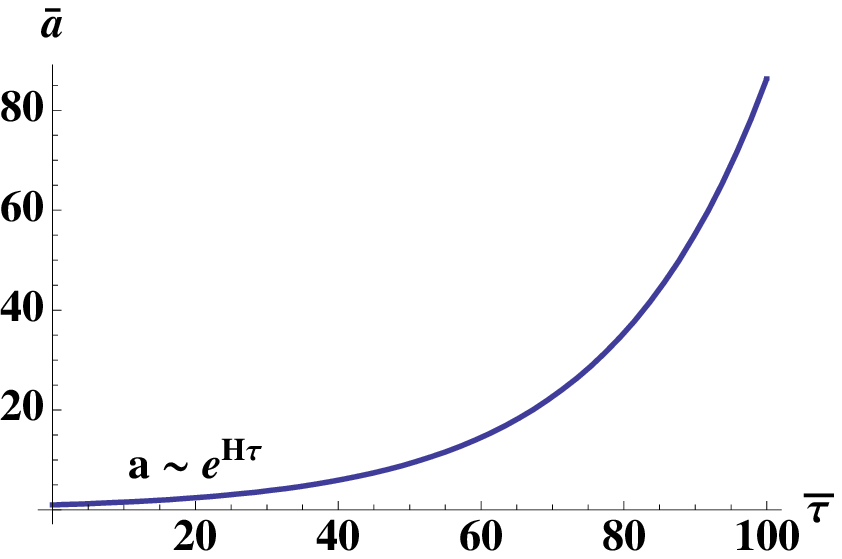}
\includegraphics[height=4cm]{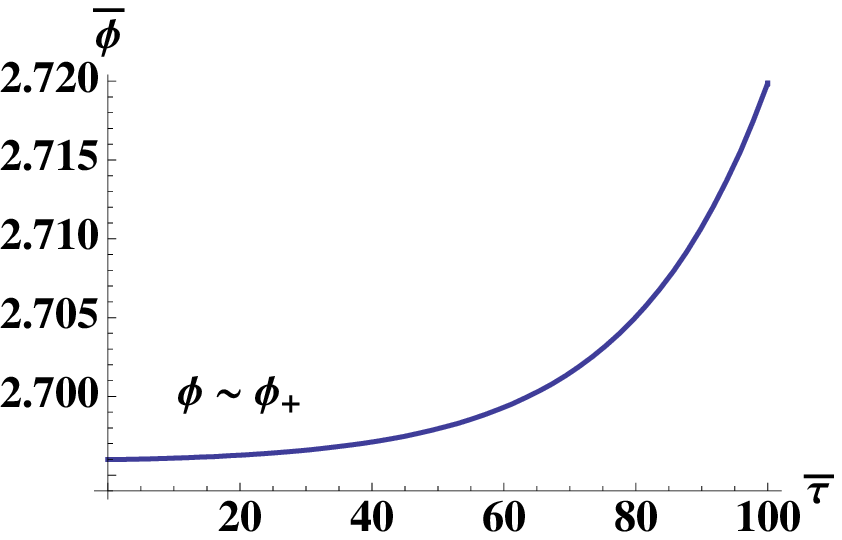}
\end{center}
\caption{An example of the Nariai solution for $p=2$, $q=2$,
$Q=3.2$, and $\Lambda / M_{D}^2 = .1$.
The field evolves from the
potential maximum to either $\phi_{-}$ (top) or $\phi \rightarrow
\infty$ (bottom) while the scale factor $a$ monotonically increases.
The two solutions are matched across the event horizon of the
$D$-dimensional geometry located at $\tau = - \infty$. This connects
the contracting portion of a $D$-dimensional de Sitter space to a
flat $p+2$-dimensional FRW universe that asymptotically has positive
vacuum energy.} \label{fig:nariai}
\end{figure*}

In the case of the Nariai solution, the field evolves away from the
potential maximum
towards either $\phi_{-}$ or $\phi \rightarrow
\infty$ over an infinite range in $\tau$. It is similar to the
extremal solutions found for $\Lambda = 0$, except that here $\tau$
is everywhere timelike. A numerically generated example for a radion
potential with positive minimum is shown in Fig.~\ref{fig:nariai}.
This solution is completely non-singular, and corresponds to a
contracting $D$-dimensional de Sitter space separated by an event
horizon from an asymptotically $p+2$-dimensional de Sitter space
with a $q$-sphere of constant size. Unlike the extremal solutions of
the previous section, an asymptotically $D$-dimensional region where
$\phi \rightarrow \infty$ is always matched across the horizon to an
asymptotically $p+2$-dimensional region where  $\phi \rightarrow
\phi_{-}$ (if the radion potential is zero or positive) or a
singularity at $\phi \rightarrow -\infty$ (if the radion potential
minimum is negative). This is due to the fact that the Gegenbauer
index $\sigma$ for small-amplitude solutions is always less  than 2,
and therefore motion across the potential maximum never has two
nodes.

Like the interpolating solutions, the Nariai solutions
interpolate
between an asymptotically $D$-dimensional de Sitter space
and a $p+2$-dimensional region that is either singular (when the
minimum of the radion potential is negative) or asymptotically de
Sitter/Minkowksi (when the minimum is positive or zero energy). The
causal structure is shown in the right panel of
Figs.~\ref{fig:Lnozeronegconf} and~\ref{fig:Lnozeroposconf}. Unlike
the interpolating solutions, future and past infinity are not
bifurcated into separate regions.

\subsection{Summary of the cosmologically interesting solutions}

The compactification, interpolating, and Nariai solutions all
have
spacetime regions that contain a $p+2$-dimensional FRW universe
with a stabilized $q$-sphere. This allows for the intriguing
possibility that our observable universe could actually be embedded
in one of these geometries.

Compactification solutions generated at the minimum of the
radion
potential are equivalent to one of the vacuum geometries $AdS_{p+2}
\times S^q$, ${\mathcal M}_{p+2} \times S^q$, or $dS_{p+2} \times S^q$. These
solutions are stable vacua, and the existence of de Sitter solutions
is compatible with the observed cosmological acceleration in our
universe. The compactification solution generated at the radion
potential maximum is also a $dS_{p+2} \times S^q$ geometry, but it
is in unstable equilibrium. Perturbations will cause the radion
field to evolve away from the potential maximum in different
spacetime regions. Some such regions can evolve towards the minimum
of the radion potential, resulting (at least locally) at late times
in a $p+2$-dimensional de Sitter vacuum.

The Nariai solutions (which are generated from a flat FRW metric
ansatz) interpolate in a timelike manner from a contracting
$D$-dimensional de Sitter space to an expanding infinite
$p+2$-dimensional flat FRW universe that can evolve towards zero or
positive vacuum energy. These regions are again separated by an
event horizon, although in this case it extends across the entire
universe. The radion field is located at the potential maximum on
the event horizon, and so the near-horizon geometry will be {\em
locally}  equivalent to the compactification solution $dS_{p+2}
\times S^q$. Since the radion field is dynamical, there will also be
portions of the Nariai solution that are locally equivalent to the
spacetime that results from perturbing the compactification solution
away from the potential maximum.

The interpolating solutions (which are generated from an open FRW
metric ansatz) contain an infinite $p+2$-dimensional open FRW
universe that can evolve towards positive vacuum energy at late
times. The big bang of the FRW universe is non-singular, and
corresponds to an event horizon in the $D$-dimensional geometry. On
the other side of the event horizon, there is a region in which the
radius of the $q$-sphere grows in a spacelike manner. As the radius
of the $q$-sphere grows, another horizon is eventually reached,
across which is an asymptotically $D$-dimensional de Sitter region.

\section{Dynamical Compactification}\label{sec:dynamicalcompactification}
Thus far, we have discussed the classical solutions of $D$-dimensional Einstein gravity with a cosmological constant and flux. We have identified a number of solutions that contain a lower-dimensional cosmological spacetime, but these are full time-symmetric solutions, and do not represent a transition. We now examine the semi-classical properties of the theory, where transitions to these classical solutions can occur from empty de Sitter space.

This process has an analog in black hole physics. Because of its thermal character, it has long been appreciated that
four-dimensional de Sitter space is unstable to the nucleation of
both charged and uncharged black holes. Such nucleation events can
be described in the semi-classical approximation by Euclidean
instantons, with the probability per unit four-volume given by
\begin{equation}\label{eq:nucrate}
\Gamma = A e^{-(S_{inst} - S_{dS})},
\end{equation}
where $A$ is a prefactor, $S_{inst}$ is the Euclidean action of the
instanton, and
$S_{dS}$ is the action of Euclideanized de Sitter
space. In higher dimensional de Sitter space with $q$-form field
strength, given the proliferation of classical solutions discussed
in previous sections, one might expect that there is a finite
probability for their nucleation.

\begin{figure*}
\begin{center}
\includegraphics[height=7cm]{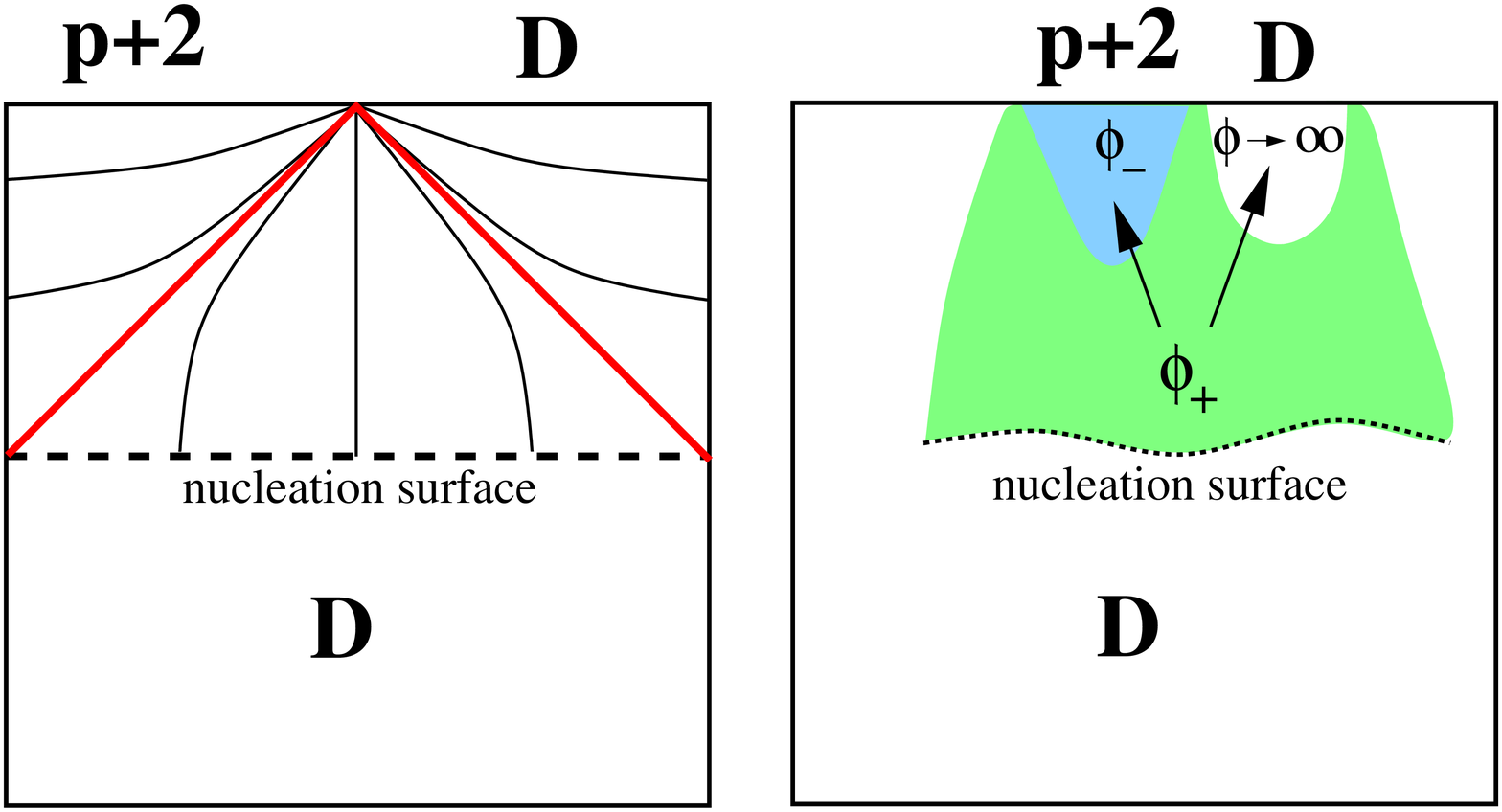}
\end{center}
\caption{The spacetime picture of dynamical
compactification when it
is mediated by the Euclidean interpolating (left) and
compactification (right) solutions. On the left, the initial
$D$-dimensional de Sitter space is matched across the nucleation
surface to the interpolating solution of Sec.~\ref{sec:FRWLg0}. The
surfaces of constant $\phi$ are shown as solid lines, and
interpolate across the barrier of the radion potential along the
nucleation surface. On the right, roughly a horizon-volume of the
initial $D$-dimensional de Sitter space fluctuates to the
compactification solution at the maximum of the radion potential.
After the fluctuation, the radion falls from its unstable
equilibrium in different spacetime regions, producing pockets where
the field either evolves to the minimum of the radion potential or
off to infinity where the $q$-sphere decompactifies.}
\label{fig:instanton}
\end{figure*}

Indeed, we will find that this is the case.
Examining the solutions
for $\Lambda > 0$ presented in Sec.~\ref{sec:FRWLg0}, we must
identify those solutions that when Euclideanized yield a finite
action. A sufficient condition for finite action is that the
Euclidean manifold is compact, and so the most obvious candidates
are the $dS_{p+2} \times S^q$ compactification solutions. Euclidean
de Sitter space of dimension $p+2$ is a $p+2$-sphere, which yields
the compact Euclidean manifold $S^{p+2} \times S^q$. Moving to the
solutions with an open metric ansatz, we note that they possess an
$SO(p+1, 1)$ symmetry. The obvious analytic continuation is to a
metric with $O(p+2)$ symmetry, which can be related to the metric
for spacelike $\tau$ Eq.~\ref{eq:wallmetric} by taking $\chi
\rightarrow i (\chi + 3\pi / 2)$, yielding
\begin{equation}\label{eq:eucmetric}
ds_E^2 = d\tau^2 + a(\tau)^2 d\Omega_{p+1}^2.
\end{equation}
This manifold is also generally compact when
the scale factor
$a(\tau)$ is bounded. Because the radial coordinate $\chi$ does not
appear in the equations of motion for $a$ and $\phi$
(Eqns.~\ref{eq:ddimenfeqn} and~\ref{eq:phifrweom}), they remain
unaltered. The Euclidean profiles for $a$ and $\phi$ are then
identical to those presented in Sec.~\ref{sec:FRWLg0} for regions
with spacelike $\tau$ (the region under the potential barrier). 
Therefore, another class of finite action
instantons can be obtained from the Euclideanized interpolating
solutions in the region underneath the potential barrier, where 
$a(\tau)$ possesses two zeroes (an example is the
profiles shown in Fig.~\ref{fig:interpolating}). Finally, we can
examine the Nariai solution for a flat FRW metric ansatz. In this
case, there is no analytic continuation to a compact manifold, and
so no finite action action instantons will result.

The finite action instantons will mediate a
transition from empty
$D$-dimensional de Sitter space to either the compactification or
interpolating spacetimes. The spacetime picture of these two
processes are shown in Fig.~\ref{fig:instanton}. Because in each
case there is a region after the transition that relaxes to a
$p+2$-dimensional vacuum, this can rightly be viewed as a dynamical
mechanism for the compactification of $q$ dimensions.

When dynamical compactification is mediated by the
Euclideanized
interpolating solution, the spacetime resembles the left panel of
Fig.~\ref{fig:instanton}. A $D$-dimensional de Sitter space is
matched onto the Lorentzian interpolating solution across a
nucleation surface, which by time translation, can be placed at the
throat of the background de Sitter space. Evolving, there are two
infinite open FRW universes separated by event horizons, one of
which is $p+2$-dimensional. This process splits future infinity
between a $p+2$ and $D$ dimensional region.

If the Euclidean compactification solution is the relevant
instanton,
then the spacetime picture is sketched in the right panel of
Fig.~\ref{fig:instanton}. The interpretation is very similar to that
of the Hawking-Moss instanton for a scalar field coupled to
gravity~\cite{Hawking:1981fz}, or the fluctuation of a Nariai black
hole~\cite{Bousso:1995cc}. In particular, a roughly horizon-sized
patch of the $D$-dimensional de Sitter space fluctuates into a
solution very close to the Lorentzian compactification solution at
the maximum of the radion potential.~\footnote{It is also  possible
to obtain a finite Euclidean action from the compactification
solution at a positive {\em minimum} of the radion potential.
However, since there are no negative modes, this cannot be
interpreted as mediating a transition. For a discussion of this
point, see e.g. Ref.~\cite{Batra:2006rz}.}  However, this solution
is unstable, and will fall either into the asymptotically
$D$-dimensional region at $\phi \rightarrow \infty$ or towards the
$p+2$-dimensional vacuum at $\phi_{-}$. This evolution will locally
be similar to the Nariai solution described in the previous section,
although the global causal structure will be different.

In addition to dynamical compactification,
the inverse process can
also occur, whereby a $p+2$ dimensional vacuum can undergo a
``decompactification transition" to a region that empties out into
$D$-dimensions~\cite{Giddings:2003zw,Giddings:2004vr}. This is
mediated by the same instantons responsible for dynamical
compactification~\cite{Giddings:2004vr}. The spacetime picture for
spontaneous decompactification is similar to
Fig.~\ref{fig:instanton}, with the region of $D$-dimensional de
Sitter space to the past of the nucleation surfaces replaced by a
$p+2$-dimensional de Sitter.

In the next subsection, we study the nucleation rates in a general
radion potential. We then briefly describe the global spacetime
picture that emerges from considering multiple transitions between a
$D$-dimensional de Sitter space and the great variety of possible
$p+2$-dimensional vacua. We largely avoid  interpreting our results
in terms of statistical predictions for physical quantities,  but
simply present the possibilities and in some cases relative
likelihoods of populating different vacua.

\subsection{Populating a landscape of vacua}
The many radion potentials for fixed $p$, $q$, and $\Lambda$
that exist for different values of the charge define a landscape of
lower dimensional vacua. Each of the potential minima have a
different $p+2$-dimensional vacuum energy, ranging from arbitrarily
negative values at $Q \rightarrow 0$ either to zero as $Q
\rightarrow \infty$ in the case where $\Lambda=0$ or to a bounded
positive value as $Q \rightarrow Q_{max}$ (by Eq.~\ref{eq:Qlimit})
in the case where $\Lambda >0$. If the charge is quantized, then
there are an infinite number of vacua for $\Lambda = 0$ where $Q$
can become arbitrarily large, but only a finite number for $\Lambda
> 0$ where the charge must be less than $Q_{max}$. Allowing for the
existence of multiple types of $q$-form flux by replacing
\begin{equation}
\frac{1}{2 q!} F_q^2 \rightarrow \sum_{i=2}^{D-3} \frac{1}{2 q_i !} F_{q_i}^2,
\end{equation}
we can also find collections of vacua with different effective
dimensionality.
The dynamical compactification mechanism described
above provides a means to evolve from $D$-dimensional de Sitter
space into each of these vacua. In this section, we will
quantitatively compare the nucleation rates for each transition.

In analogy with the nucleation rate for black holes in four
dimensions
Eq.~\ref{eq:nucrate}, we take the probability per unit
$D$-volume of the background de Sitter space to nucleate a solution
to be
\begin{equation}\label{eq:nucleationprob}
\Gamma = A \exp \left[ - (S_{inst} - S^{(D)}_{dS}) \right].
\end{equation}
The prefactor $A$ is of mass dimension $D$ and represents the
first quantum corrections to the probability. This quantity is in general
difficult to calculate (requiring us to step away from the
minisuperspace of metrics we have considered), but we expect it to
be of order $\Lambda^{D/2}$. The action for the higher dimensional
background dS space is given by
\begin{equation}\label{eq:SdsD}
S^{(D)}_{dS} = - \frac{M_{D}^{D-2}}{2} \int d^D x \sqrt{g} \left( R
- 2 \Lambda \right) = - (p+q+1)^{(p+q+2)/2} (p+q)^{(p+q)/2} {\rm
Vol}
(\Omega_{p+q+2}) \left( \frac{M_D^2}{2 \Lambda}
\right)^{(p+q)/2}.
\end{equation}
Substituting the metric ansatz Eq.~\ref{eq:eucmetric} for
the instanton into the action Eq.~\ref{eq:einsteinaction} and
integrating over the angles of the $p+1$-sphere, the Euclidean
action is given by
\begin{equation}\label{eq:actionwdimensions}
S_{inst} = - 2 {\rm Vol} \left( \Omega_{p+1} \right) \int d \tau a^{p+1} \left[ V + \frac{1}{2} \frac{p(p+1)}{a^2} \right].
\end{equation}

There are two types of instantons that can be responsible for dynamical compactification: the Euclideanized compactification and interpolating solutions. Both types will have an action given by Eq.~\ref{eq:actionwdimensions}. It is possible to find the compactification solutions analytically, but it will be necessary to construct the interpolating solutions numerically, and we will therefore find it convenient to work with the dimensionless variables
defined in Eq.~\ref{eq:dimlessvariables}. A characteristic scale for
the potential is $\omega_a$, which is essentially the barrier
height, and is defined in Eq.~\ref{eq:afreq}. Substituting for the
value of $V(\phi_{+})$, $\omega_a$ is given by
\begin{equation}\label{eq:omegaawvmax}
\omega_a^2 = \frac{M_{p+2}^{2}}{{\rm Vol} \left( \Omega_{q}
\right)^{2/p} (p+1) (p+q)^{q/p+1} (q-1)^{q/p} } \left( \frac{2
\Lambda}{M_D^2} \right)^{q/p + 1}.
\end{equation}
Setting $\mu = M_{p+2}^{p/(p+2)} \omega_a^{2/(p+2)}$,
the
dimensionless FRW variables are given by
\begin{equation}
\bar{\phi} = \frac{\phi}{M_{p+2}}, \ \ \  \bar{V} =
\frac{V}{M_{p+2}^{p} \omega_{a}^2}, \ \ \ \bar{\tau} = \omega_a \tau, \ \ \ \bar{a} = \omega_{a} a,
\end{equation}
Substituting into the action, we obtain
\begin{eqnarray}\label{eq:sinst}
S_{inst} &=& - 2  {\rm Vol} \left( \Omega_{p+1} \right)
\left( \frac{M_{p+2}}{\omega_a}\right)^p \int d \bar{\tau} \bar{a}^{p+1} \left[ \bar{V} + \frac{1}{2} \frac{p(p+1)}{\bar{a}^2} \right] \nonumber \\
&=& - 2 {\rm Vol} \left( \Omega_{p+1} \right) {\rm Vol} \left( \Omega_{q} \right) (p+1)^{p/2} (p+q)^{(p+q)/2} (q-1)^{q/2} \left( \frac{M_D^2}{2 \Lambda} \right)^{(p+q)/2} \int d \bar{\tau} \bar{a}^{p+1} \left[ \bar{V} + \frac{1}{2} \frac{p(p+1)}{\bar{a}^2} \right].
\end{eqnarray}
We use this form of the rescaled action throughout the rest of this section in our computation of the instanton action.

A convenient parametrization of the nucleation rates is in terms of the ratio
\begin{equation}
\alpha \equiv S_{inst} / S^{(D)}_{dS}.
\end{equation}
The nucleation probabilities are then given by
\begin{equation}\label{eq:interpolatingrates}
\Gamma =  A \exp \left[ S^{(D)}_{dS} \left( 1 - \alpha \right) \right],
\end{equation}
which grows with increasing $\alpha$ since $S^{(D)}_{dS} < 0$.
For the semi-classical formalism to accurately describe the transition
process, we must always have suppressed transitions, requiring
$\alpha < 1$. We will find that this is indeed the case for all of
the solutions we construct. Unless $\alpha \simeq 1$, the rates are
set by $S^{(D)}_{dS}$, which is negative and large in magnitude
unless $\Lambda \agt M_{D}^2$. For such a large $\Lambda$, we expect
the semi-classical description of the background $D$-dimensional de Sitter space to break down.

Substituting the instanton action Eq.~\ref{eq:sinst} and the background de Sitter action Eq.~\ref{eq:SdsD}, we find that
\begin{equation}\label{eq:fullalpha}
\alpha = 2 \frac{ {\rm Vol} \left( \Omega_{p+1} \right)
{\rm Vol}
\left( \Omega_{q} \right)}{{\rm Vol} (\Omega_{p+q+2}) }
\frac{(p+1)^{p/2} (q-1)^{q/2}}{(p+q+1)^{(p+q+2)/2} }  \int d
\bar{\tau} \bar{a}^{p+1} \left[ \bar{V} + \frac{1}{2}
\frac{p(p+1)}{\bar{a}^2} \right].
\end{equation}

The first (and simplest) finite-action instanton we consider is the
Euclideanized compactification solution, where the field sits at the
maximum over the entire evolution in $\tau$. The solution for the
scale factor is $a(\tau) = \omega_a^{-1} \sin (\omega_a \tau)$.
Substituting the dimensionless time variable $\bar{\tau} = \omega_a
\tau$, dimensionless scale factor $\bar{a} (\bar{\tau}) = \sin
\bar{\tau}$, and dimensionless potential $\bar{V} =  V / (M_{p+2}{p}
\omega_a^2 ) = - p(p+1) / 2$ into the instanton action, we obtain
\begin{equation}
S_{inst} = - p(p+1) {\rm Vol} \left( \Omega_{p+1} \right)
\left(\frac{M_{p+2}}{\omega_a}\right)^p  \int_0^{\pi} d \bar{\tau}
\sin^{p-1} \bar{\tau} \left[ 1 - \sin^{2} \bar{\tau} \right].
\end{equation}
The integral can be evaluated in terms of $\Gamma$ functions, and after some manipulation we find
\begin{equation}
S_{inst} = - (p+1){\rm Vol} \left( \Omega_{p+2} \right) \left(\frac{M_{p+2}}{\omega_a}\right)^p .
\end{equation}
This is equivalent to the volume of a $p+2$-sphere of radius
$\omega_a^{-1}$,
as expected from the fact that we are simply
Euclideanizing a $p+2$-dimensional de Sitter space. For small $Q$,
we can substitute for $\omega_a$ using Eq.~\ref{eq:omegaawvmax}. The
value of $\alpha$ in this case is
\begin{equation}\label{eq:alphasmallQ}
\alpha = \frac{ {\rm Vol} \left( \Omega_{p+2} \right) {\rm Vol}
\left( \Omega_{q}
\right)}{{\rm Vol} (\Omega_{p+q+2}) }
\frac{(p+1)^{p/2+1} (q-1)^{q/2}}{(p+q+1)^{(p+q+2)/2} }.
\end{equation}

\begin{figure*}
\begin{center}
\includegraphics[width=8cm]{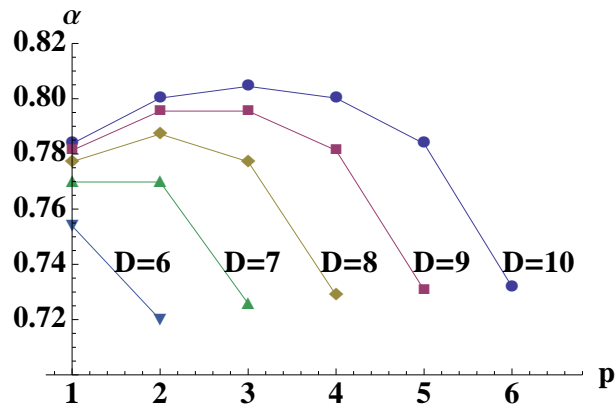}
\end{center}
\caption{The value of $\alpha$ (Eq.~\ref{eq:alphasmallQ}),
the ratio of the instanton action to the $D$-dimensional de Sitter action, for
fixed $D$ and various $p$ in the limit of small $Q$. Larger values
of $\alpha$ correspond to a faster nucleation rate. Circles,
squares, diamonds, triangles, and inverted triangles are for $D=10$,
$9$, $8$, $7$, $6$ respectively. In each case, $p$ can range from to
$1$ to $D-2$. The value of $\alpha$ peaks at a value of $p$
dependent on the total dimensionality.} \label{fig:compaction}
\end{figure*}

In Fig.~\ref{fig:compaction}, we plot the value of $\alpha$
in Eq.~\ref{eq:alphasmallQ} (valid for small $Q$) for fixed $D$ over
a range of $p$. Interestingly, this function peaks at a value of $p$
specified by the total dimensionality. This is due to the relative
volumes of the unit spheres. Relevant for our universe is that for
$D=8$, the nucleation rate for Euclidean compactification solutions
with $p=2$ (and therefore four non-compact dimensions) is largest.
There are also marginal cases: for $D=7$, the rate for $p=1$ and
$p=2$ are identical while for $D=9$, the rate for $p=2$ and $p=3$
are identical. Also of note is the fact that $\alpha$ is in general
larger when the total dimensionality is increased.

\begin{figure*}
\begin{center}
\includegraphics[width=8cm]{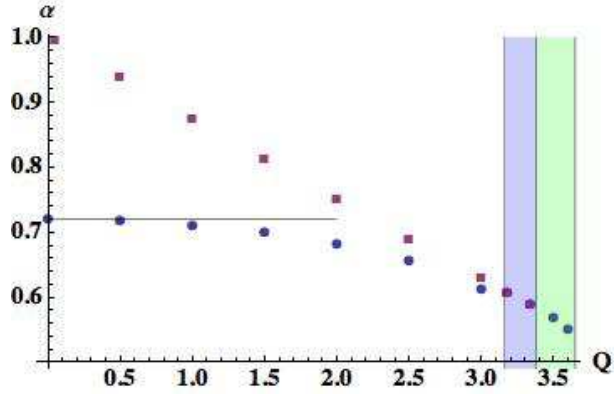}
\end{center}
\caption{A plot of $\alpha$ versus $Q$ for the
Euclidean
interpolating (squares) and compactification (circles)
solutions with $\{ p=2, q=2, \Lambda = .1 \}$. The value of $\alpha$
for the compactification solutions approaches the value defined in
Eq.~\ref{eq:alphasmallQ} (denoted by the horizontal solid line at
$\alpha = .72$) at small $Q$. A few important thresholds in $Q$ are
denoted by the shaded regions. For $3.162 < Q < 3.35$, there exists
a minimum of the radion potential that has positive energy. Between
$3.162 < Q < 3.35$, in the light (blue) shaded region, both the
Euclidean interpolating and compactification solutions are allowed.
For $3.35 < Q < 3.65$, in the dark (green) shaded region, only the
Euclidean compactification solution exists. When it exists, the 
interpolating solution always makes the dominant contribution to the decay rate.}
\label{fig:alphaplotbig}
\end{figure*}

For appreciable values of $Q$, the relative
importance of the various terms in the potential changes, and Eq.~\ref{eq:vbarrier} becomes a
poor estimate for $V (\phi_{+})$. In Fig.~\ref{fig:alphaplotbig}, we
plot the exact value of $\alpha$ versus $Q$ (circles) for $\{ p=2,
q=2, \Lambda = .1 \}$ determined by numerically finding $V
(\phi_{+})$. Here, it can be seen that at small $Q$, $\alpha$
approaches the value determined by Eq.~\ref{eq:alphasmallQ}
(depicted by the solid horizontal line). Further, $\alpha$ decreases
with increasing $Q$. For $Q$ larger than the critical value defined
in Eq.~\ref{eq:Qlimit}, there are no potential extrema, and the
Euclidean compactification solutions do not exist.

Now, we consider the second instanton, found by Euclideanizing the interpolating solution.
 Evolution in the presence of the Euclidean potential was already discussed in
Sec.~\ref{sec:FRWLg0}. This generated the portion of the
interpolating solution in regions with spacelike $\tau$ (underneath the barrier). The
profiles for $a(\tau)$ and $\phi(\tau)$ are qualitatively identical
to those depicted in Fig.~\ref{fig:interpolating}, where we set
$\dot{\phi} = 0$ at the endpoints in $a$, and the field interpolates
across the potential barrier. It is possible to construct such
solutions numerically, and compute the value of $\alpha$ using
Eq.~\ref{eq:fullalpha}.

In Fig.~\ref{fig:alphaplotbig}, we plot the value of $\alpha$ versus
$Q$ for the Euclidean interpolating solutions (squares) using $\{
p=2, q=2, \Lambda = .1 \}$. The values of $\alpha$ at any fixed $Q$
are always larger for the Euclidean interpolating solutions than for
the Euclidean compactification solutions, implying that the
nucleation rate is also larger for the Euclidean interpolating
solutions. As $Q$ increases, $\alpha$ decreases, and the action of
the interpolating solutions approaches that of the compactification
solutions. At small $Q$, $\alpha$ is very close to unity. Because
the potential minimum goes to $V_{-} \rightarrow - \infty$ as $Q
\rightarrow 0$ (see Eq.~\ref{eq:vphimin}), our numerical solutions
become unreliable in this regime, and it is not possible to
determine what the asymptotic value at $Q=0$ is. However, as we discussed in Sec.~\ref{sec:FRWLg0}, the zero charge limit
of the interpolating solutions should be empty $D$-dimensional de Sitter space, which would be consistent with $\alpha \rightarrow 1$. At $Q=0$, the beginning and end states of the ``tunneling" process are identical.

The properties of the potential maximum are altered as $Q$
increases. Eventually the index of the Gegenbauer polynomials
defined in Eq.~\ref{eq:alphainterpolating} becomes less than one,
and it is impossible to find an interpolating solution.
This generally occurs for $Q$ less than the value for which
there are no critical points of the potential (defined in
Eq.~\ref{eq:Qlimit}), and thus there is a range in $Q$ over which
the Euclidean compactification solution is the only viable
transition mechanism. This is depicted by the dark (green) shaded
region in Fig.~\ref{fig:alphaplotbig}. In general (for arbitrary
$p$, $q$, and $\Lambda$), there will be a window of values of $Q$
with a positive cosmological constant minimum and a Euclidean
interpolating solution, depicted by the light (blue) shaded region
of Fig.~\ref{fig:alphaplotbig}. 

\begin{figure*}
\begin{center}
\includegraphics[width=8cm]{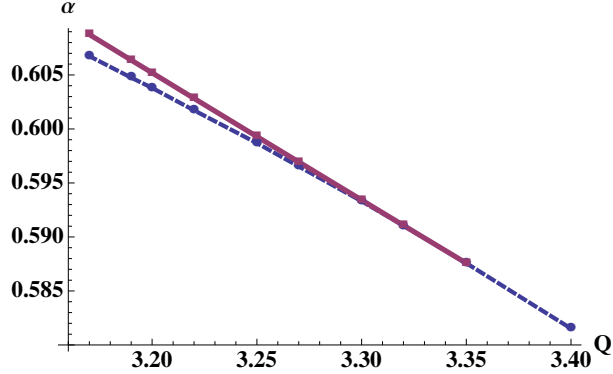}
\end{center}
\caption{The shaded region of the $\alpha$ vs $Q$ plot
of Fig.~\ref{fig:alphaplotbig}, in which the radion potential has a
positive minimum and the Euclidean compactification (dashed line)
and interpolating (solid line) solutions exist. As $Q$ is increased,
the instanton endpoints approach the potential maximum, and the
value of $\alpha$ for the interpolating solution approaches that for
the compactification solutions from above. When it exists, the 
interpolating solution always makes the dominant contribution to the decay rate.}
\label{fig:alphaplotsmall}
\end{figure*}

In Fig.~\ref{fig:alphaplotsmall}, we zoom in on the shaded region of Fig.~\ref{fig:alphaplotbig}. As $Q$ is increased, the endpoints of the interpolating solution approach the potential maximum, eventually merging when it becomes impossible to find the interpolating solution. The action for the interpolating solution approaches that of the compactification solution, dominating the decay rate when it exists.

To the extent that our dimensional reduction captures the relevant physics, the number of negative modes for each type of solution can be determined by  the arguments of Ref.~\cite{Batra:2006rz}. The interpolating solution (similar to the Coleman-de Luccia instanton~\cite{Coleman:1980aw}), when it exists, will always have a single negative mode, and therefore represents a semi-classical instability. The Euclidean compactification solution (similar to the Hawking-Moss instanton~\cite{Hawking:1981fz}) will have a single negative mode only when the interpolating solution does not exist (and it therefore dominates the decay rate as illustrated in Figures~\ref{fig:alphaplotbig}~and~\ref{fig:alphaplotsmall}).

\begin{figure*}
\begin{center}
\includegraphics[width=8cm]{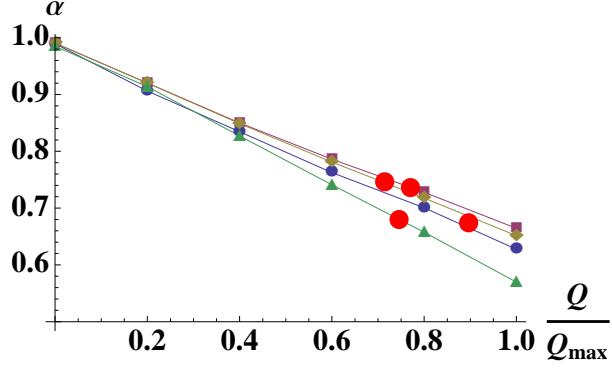}
\end{center}
\caption{The value of $\alpha$ in $D=8$ for $p=1,2,3,4$ (blue
circles, purple squares, yellow diamonds, and green triangles
respectively) versus the normalized charge $Q / Q_{max}$. The red
circle denotes the value of $Q$ yielding an approximately zero
energy $p+2$ dimensional vacuum for each $p$.}
\label{fig:comparealpha}
\end{figure*}

It is also possible to compare the actions for interpolating
solutions with variable $p$ in a fixed total number of dimensions. In
Fig.~\ref{fig:comparealpha} we plot $\alpha$ vs the normalized
charge $Q/Q_{max}$ in $D=8$ for $p=1,2,3,4$ (numerical data points
are shown as circles, squares, diamonds, and triangles
respectively). The value of $Q$ yielding an approximately zero
energy $p+2$ dimensional vacuum in each case is denoted by the large
red circle. At fixed $Q / Q_{max}$, the solutions for $p=2$ always
have the largest value of $\alpha$ (and hence the highest nucleation
rate). It is also of interest to compare the rates to zero energy $p+2$
dimensional vacua, in which case we find that the $p=2$ and $p=3$
solutions have comparable values for $\alpha$ (because of the
numerical error incurred in doing the computation it is unclear
which dominates in this case). In general the comparison of
nucleation rates between solutions is complicated, since there is a
value of $Q$ at each $p$ for which the nucleation rate is the same,
and therefore vacua of different dimensionality and different
cosmological constant will be populated with identical frequency. It
is necessary to determine some physical criterion for making the
comparison, a topic we return to in the next section.

The formation of the interpolating solution brings into being a
region that relaxes to a $p+2$-dimensional big crunch (when the
vacuum energy is negative), or an open universe with zero or
positive vacuum energy. However, when the radion potential has a
positive minimum, the instantons we have found can also mediate a
decompactification
transition~\cite{Giddings:2003zw,Giddings:2004vr}. The late-time
$p+2$-dimensional de Sitter region produced in dynamical
compactification will therefore in some patches undergo transitions
back to $D$-dimensional de Sitter space. The tunneling rate out of a
metastable $p+2$ dimensional de Sitter vacuum is given by
\begin{equation}\label{eq:decompactificationrate}
\Gamma = A \exp \left[ - (S_{inst} - S^{(p+2)}_{dS}) \right] = A \exp \left[ S^{(p+2)}_{dS} \left( 1 - \beta \right) \right],
\end{equation}
where the action of the lower dimensional de Sitter vacuum is
\begin{equation}
S^{(p+2)}_{dS} = \frac{p (p+2)^2 {\rm Vol} \left( \Omega_{p+2} \right)} {2 \omega_a^2 \bar{V} (\phi_{-})},
\end{equation}
and the quantity $\beta$ (a ratio of actions analogous to $\alpha$) is given by
\begin{equation}
\beta \equiv \frac{S_{inst}}{S^{(p+2)}_{dS}} = \frac{4 \bar{V} (\phi_{-}) {\rm Vol} \left( \Omega_{p+1} \right) }{\omega_a^{p-2} p (p+1)^2  {\rm Vol} \left( \Omega_{p+2} \right)}  \int d \bar{\tau} \bar{a}^{p+1} \left[ \bar{V} + \frac{1}{2} \frac{p(p+1)}{\bar{a}^2} \right] .
\end{equation}

\begin{figure*}
\begin{center}
\includegraphics[width=8cm]{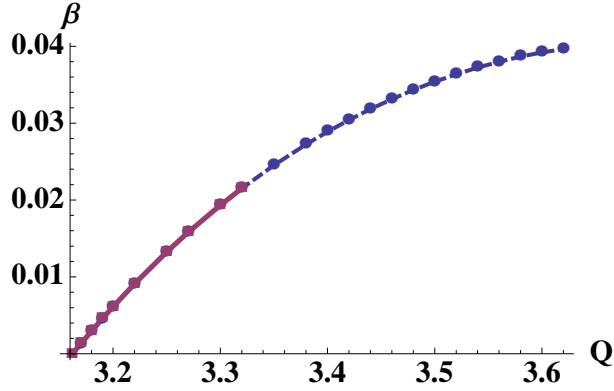}
\end{center}
\caption{A plot of $\beta$, the ratio of the instanton action to
the
$p+2$-dimensional de Sitter action, versus $Q$ for the
interpolating (solid line) and compactification (dashed line)
solutions with $\{ p=2, q=2, \Lambda = .1 \}$. The values of $\beta$
are nearly identical for the two solutions. At $Q \simeq 3.162$, the
minimum of the radion potential goes to zero, sending $\beta
\rightarrow 0$. The interpolating solution no longer exists for $Q >
3.35$, and at the upper range of the plot $Q = 3.65$, there are no
potential extrema.} \label{fig:betaplot}
\end{figure*}

A plot of $\beta$ versus $Q$ for $\{ p=2, q=2, \Lambda = .1 \}$ is shown in Fig.~\ref{fig:betaplot}. As $V(\phi_{-}) \rightarrow 0$, $S^{(p+2)}_{dS} \rightarrow - \infty$. Because the instanton action approaches a constant in this limit, $\beta \rightarrow 0$, and the tunneling rate out of the $p+2$-dimensional dS space is extremely suppressed, with $\Gamma \sim e^{S^{(p+2)}_{dS}}$. In the limit of a vanishing $p+2$-dimensional cosmological constant, the vacuum is completely stable.

In summary,
\begin{itemize}
\item The value of $\alpha$ at a fixed normalized
charge $Q / Q_{max}$ is maximized for a particular $p$ or range of
$p$ depending on the total dimensionality. If $D=7,8,9$ then
solutions with $p=2$ either maximize $\alpha$ (for $D=8$) or are
comparable to solutions with $p=1$ (for $D=7$) or $p=3$ (for $D=9$).
A detailed comparison of nucleation rates for vacua with different
dimensionality and cosmological constant should be possible.
\item For fixed $p$ and $q$,
the rate is highest for small $Q$, and therefore the lowest (negative) energy vacua.
\item When it exists, the rate is higher at fixed $Q$,
$p$, and $q$ for the interpolating solution than the
compactification solution.
There is in general a range of $Q$ for
which there is a positive energy minimum, but no interpolating
solution.
\item Because the instanton action is finite as the vacuum energy of
the minimum goes to zero, the rate into a $p+2$-dimensional vacuum
with a
small (positive) vacuum energy is much higher than the rate
out. $p+2$-dimensional Minkowski vacua are completely stable.
\end{itemize}

\subsection{Global structure of the multiverse}

Given the existence of the instantons described in the previous
section,
empty $D$-dimensional de Sitter space in a theory with
$q$-form field strengths will {\em inevitably} populate all of the
lower-dimensional vacua of the theory. Because the probability per
unit $D$-volume to undergo dynamical compactificaton is much smaller
than a $D$-dimensional Hubble volume $H_D^{-D} \Gamma \ll 1$, it is
very rare to have more than one transition per Hubble volume per
Hubble time. This implies that the $D$-dimensional de Sitter space
can never be totally eaten up by transitions and can populate many
different lower-dimensional vacua, in exact analogy with
four-dimensional eternal inflation. If the lower-dimensional vacua
have positive vacuum energy, they are metastable, with local regions
eventually transitioning back to the $D$-dimensional de Sitter
space. This recycling process continues indefinitely, fragmenting
future infinity into regions with diverse properties.

When a compactification or interpolating solution is nucleated out
of the $D$-dimensional de Sitter space, it occupies nearly a Hubble
volume of the background space. This can be seen by noting that both
the size of the $q$-sphere (through the location of the potential
maximum Eq.~\ref{eq:R0maximum}) and the amplitude of $a$ (through
$\omega_a^{-1}$ Eq.~\ref{eq:omegaawvmax}) is set by $\Lambda$. Thus,
the naive picture that emerges is that a nucleation event
corresponds to taking roughly one horizon volume of the background
dS space and replacing it with the post-nucleation spacetime. In the
asymptotic future, this patch then evolves into two regions: one in
which the spacetime approaches a $p+2$-dimensional vacuum, and the
other in which it returns to the $D$-dimensional de Sitter space.

At the very most, the region containing the $p+2$-dimensional
vacuum can correspond to removing a comoving Hubble volume from the
background dS space. The interpolating solution is anisotropic when
viewed from the full $D$-dimensional theory, which  introduces
additional complications. Leaving this subtlety aside, we can
course-grain, and attempt to determine how the comoving volume of
the original de Sitter space is distributed between the various
vacua.

Consider a fixed parcel of comoving volume that is on some
fixed initial value surface everywhere in the $D$-dimensional de Sitter
space. Extending comoving geodesics from this initial value surface,
we can trace the evolution as a function of proper time of the
fraction of such worldlines that remain in the original vacuum
$\mathcal{P}_{dS} (\tau)$ to those that encounter a nucleation event
$\mathcal{P}_{i} (\tau)$ of type $i$. If we restrict our attention
to the volume cut out by the nucleation events, and do not attempt
to follow geodesics into the interpolating region, this should be a
valid procedure. This type of volume weighting is identical to the
procedures applied in various measures for eternal inflation (see
for example \cite{Garriga:1997ef}, whose conventions we use below,
for further discussion).

Conservation of probability implies that
\begin{equation}
\mathcal{P}_{dS} (\tau) + \sum_{i} \mathcal{P}_{i} (\tau) = 1.
\end{equation}
Disregarding decompactification transitions back to the $D$-dimensional dS space, the rate equations are
\begin{eqnarray}
\frac{d \mathcal{P}_{dS}}{d\tau} &=& - \sum_i \kappa^{dS}_{i} \mathcal{P}_{dS} (\tau), \\
\frac{d \mathcal{P}_{j}}{d\tau} &=& \kappa^{dS}_{j} \mathcal{P}_{dS} (\tau),
\end{eqnarray}
where the $\kappa_i$ are the probability per unit proper time to experience a transition, given by
\begin{equation}
\kappa^{dS}_{i} = {\rm Vol} \left( \Omega_{D-1} \right) H_{dS}^{-(D-1)} \Gamma_i .
\end{equation}
The rates per unit $D$-volume $\Gamma_i$ are given for the various
types of
transitions by Eq.~\ref{eq:interpolatingrates}, and $H_{dS}$
is the $D$-dimensional Hubble constant. On the initial value surface
$\mathcal{P}_{dS} (\tau = 0) = 1$, and solving the rate equations we
obtain
\begin{eqnarray}
\mathcal{P}_{dS} &=& e^{- \sum_i \kappa^{dS}_{i} \tau}, \\
\mathcal{P}_{j} &=& \frac{\kappa^{dS}_j}{\sum_i \kappa^{dS}_{i}} \left( 1- e^{- \sum_i \kappa^{dS}_{i} \tau}  \right).
\end{eqnarray}
The comoving volume fraction remaining in the $D$-dimensional dS
space goes to zero (however,
the physical volume will be
exponentially increasing). At late times, the distribution of
worldlines reaches an equilibrium state, with the relative fraction
distributed between any two vacua $i$ and $j$ given by
\begin{equation}\label{eq:ratioofrates}
\frac{\mathcal{P}_{i}}{\mathcal{P}_{j}} = \frac{\Gamma_i}{\Gamma_j} \sim \exp \left[ -S_{dS}^{(D)} \left( \alpha_i - \alpha_j \right) \right].
\end{equation}
Therefore, under this measure, vacua with the largest possible values of $\alpha$ are exponentially favored.

Taking into account the fact that there are decompactification
transitions back to the
$D$-dimensional dS space requires us to
follow worldlines between the vacua of different dimensionality.
This will undoubtedly involve a number of important subtleties, but
we might hope that a naive addition to the rate equations provides a
rough picture of what happens:
\begin{eqnarray}
\frac{d \mathcal{P}_{dS}}{d\tau} &=& - \sum_i \kappa^{dS}_{i} \mathcal{P}_{dS} (\tau) + \sum_i \kappa^{i}_{dS} \mathcal{P}_{i} (\tau), \\
\frac{d \mathcal{P}_{j}}{d\tau} &=& \kappa^{dS}_{j} \mathcal{P}_{dS} (\tau) - \kappa^{j}_{dS} \mathcal{P}_{j} (\tau).
\end{eqnarray}
Here,
\begin{equation}
\kappa^{i}_{dS} =  {\rm Vol} \left( \Omega_{p+1} \right) H_{p+2}^{-(p+1)} \Gamma_i ,
\end{equation}
where the rates $\Gamma_i$ are given by Eq.~\ref{eq:decompactificationrate}. The full solution to the rate equations can be found using the methods outlined in~\cite{Winitzki:2006rn}, but considering a single transition type, the solution is
\begin{eqnarray}
\mathcal{P}_{dS} &=& \frac{1}{\kappa^{dS}_j + \kappa_{dS}^j} \left(  \kappa_{dS}^j + \kappa^{dS}_j \exp\left[ - \left(  \kappa_{dS}^j + \kappa^{dS}_j \right) \tau \right] \right), \\
\mathcal{P}_{j} &=& \frac{\kappa^{dS}_j}{\kappa^{dS}_j + \kappa_{dS}^j} \left( 1 - \exp\left[ - \left(  \kappa_{dS}^j + \kappa^{dS}_j \right) \tau \right] \right).
\end{eqnarray}
Again, at late times, a steady state behavior is reached,
with the
relative weight of the two vacua given by
\begin{equation}
\frac{\mathcal{P}_{j} }{\mathcal{P}_{dS}} = \frac{\kappa^{dS}_j}{ \kappa_{dS}^j} \sim \exp \left[ |S_{dS}^{p+2}| - |S_{dS}^{D}| \right],
\end{equation}
which is typically much greater than one, since $|S_{dS}^{p+2}| \gg
|S_{dS}^{D}|$
when the scale setting the $p+2$ dimensional vacuum
energy is much smaller than the scale setting the $D$-dimensional
vacuum energy. This result is in accord with the qualitative
expectation that a statistical system spends most of its time in the
state of highest entropy, and was commented on originally in
Refs.~\cite{Bousso:2002fi,Giddings:2004vr}. Here, it comes about
because the rate to evolve back from the $p+2$ dimensional vacuum to
$D$-dimensions becomes arbitrarily small as the $p+2$-dimensional
vacuum energy goes to zero.

This is only one of many possible measures over the vacua produced
from the
background dS space. Connecting such measures to observable
quantities is inherently ambiguous. For example, it is not obvious
how to compare the number of observers in each unit of comoving
volume, both because it is unclear what a unit of comoving volume
has to do with observers, and because we are comparing infinities
and there is no unique regulator (notwithstanding the fact that one
must compare volumes in vacua of different effective
dimensionality). We have little to add to this discussion, but it
should be possible to generalize many of the existing measure
proposals to this scenario. Another possibility is offered by the
no-boundary proposal. Here, each Hubble volume of the background dS
space is treated as an independent system, ``nucleated" with
probability set by the various instantons found above. The relative
probability of obtaining various solutions is proportional to the
ratio of rates, as in Eq.~\ref{eq:ratioofrates}.

We also note that there might exist a progressive relaxation of the
number of
dimensions if the appropriate $q$-form fluxes were
available. In this scenario, after compactifying to ${\bf M}_{p+
q_2 + 2} \times S^{q_1}$, where ${\bf M}_{p+q_2+2}$ is a
$p+q_2+2$-dimensional vacuum solution, a second transition could
produce a product compactification ${\bf M}_{p+2} \times S^{q_1}
\times S^{q_2}$. Including electric flux as well, one could consider
the nucleation of charged membranes in the $p+2$-dimensional vacua,
embedding the standard four-dimensional picture of false-vacuum
eternal inflation inside of the dynamical compactification scenario.

\section{The cosmological constant problem}\label{sec:ccproblem}

Within the landscape of radion potentials, it is possible to
find effectively 4-dimensional vacua with positive vacuum energy. In
order to promote this toy model to a plausible theory of our
universe, it is necessary to ensure that there is at least one
vacuum with sufficiently small positive energy to account for the
observed vacuum energy $\rho_{vac} \sim 1.5 \times
10^{-123}$~\cite{Komatsu:2008hk}. This four-dimensional vacuum
energy is determined by a cancellation between the curvature, flux,
and higher-dimensional cosmological constant contributions to the
radion potential Eq.~\ref{eq:radionpotential}. There is one vacuum
per value of the charge $Q$, and since the flux should obey a
quantization condition $Q=e n$ with $e$ the gauge coupling and $n$
integer, there is a discrete number of vacua.

The question of how miraculous the required cancelation is depends
on how finely the vacuum energy scans with a change in the discrete
values of the charge. If the gap between successive values of the
vacuum energy is large, then the gauge coupling and/or higher
dimensional cosmological constant would have to be extremely
fine-tuned in order to produce the observed value of $\rho_{vac}$.
However, if the gap is quite small, and there are many vacua with
energies in the correct ball-park, then  it is reasonable to believe
that the correct vacua can be populated, even in the absence of
fine-tuning.

We  find two ingredients in our scenario that reduce the amount of
tuning in $\Lambda$ and $e$ necessary to naturally produce the
correct value of the four-dimensional cosmological constant: a large
number of extra dimensions $q$, and the presence of multiple
quantized $q$-forms with incommensurate fundamental charges. In
essence, these are the same methods used by Bousso and
Polchinski~\cite{Bousso:2000xa} to obtain vacua with a naturally
small value of the cosmological constant, although here they arise
in a different context.

If there are multiple copies of a $q$-form, then $Q^2$ in the radion potential Eq.~\ref{eq:radionpotential} is replaced by the sum
\begin{equation}
Q^2 = \sum_{i=1}^J Q_i^2 = \sum_{i=1}^J e_i^2 n_i^2.
\end{equation}
where $e_i$ is the gauge coupling for each copy, and $n_i$ is the
number of units of fundamental charge. Following Bousso and
Polchinski~\cite{Bousso:2000xa}, we can visualize
charge space as a
$J$-dimensional grid of points with a spacing in each direction set
by $e_i$. Inside of this charge space, construct a shell of radius
$Q_0$, and width $\Delta Q$. In the bulk of this shell, the total
charge produces a vacuum energy between approximately zero, which
defines the radius $Q_0$, and the observed value of the vacuum
energy, which defines the width $\Delta Q$. In order for it to be
likely that there is a vacuum with the appropriate value of the
cosmological constant, we require that the volume contained inside
of the shell is greater than the volume of a unit cell in charge
space
\begin{equation}\label{eq:shellcondition}
\prod_{i=1}^{J} e_i < \frac{{\rm Vol} \left( \Omega_{J-1} \right)}{d} Q_0^{J-1} \Delta Q,
\end{equation}
where $d$ is a degeneracy in values of $V_0$. If all the $e_i$ are different, then $d = 2^J$
due to the invariance under $n_i \rightarrow -n_i$ (the total volume in the shell is reduced
to the volume contained in a region with positive $n_i$). It is possible that $d$ can be somewhat larger if the $e_i$ are commensurate.

\begin{figure*}
\begin{center}
\includegraphics[width=8cm]{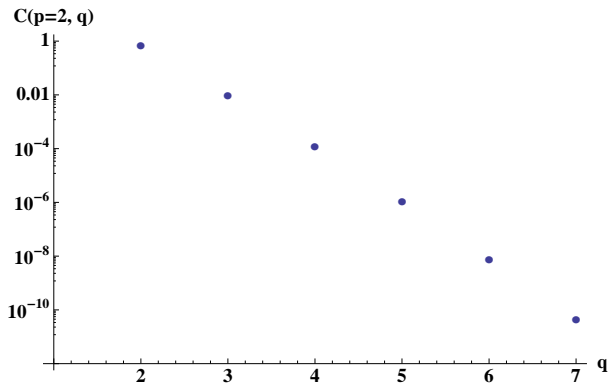}
\end{center}
\caption{A log plot of the function $C(p,q)$ for $p=2$ and various
$q$, obtained
numerically by finding the value of the potential
minimum for values of $Q$ that yield $V_0 \sim 0$, constructing the
function $V ( \phi_{-} [Q] )$, evaluating the derivative, and
factoring out the scaling with $\Lambda$ determined in
Eq.~\ref{eq:dVdQ}. It can be seen that the behavior is approximately
exponential in $q$.} \label{fig:dVdQ}
\end{figure*}

Applying this analysis to the radion potential, the total amount of charge $Q_0$
necessary to produce a zero $p+2$-dimensional vacuum energy can be found by
substituting the approximate location of the minimum Eq.~\ref{eq:nolamroot}
into the radion potential, setting the latter to zero, and solving for $Q$.
The result is an order of magnitude estimate for $Q_0$ given by
\begin{equation}
Q_0 \sim \sqrt{\frac{2 (p+q)^q (q-1)^q}{(p+1)^q}} \left( \frac{M_D^2}{2 \Lambda} \right)^{(q-1)/2}.
\end{equation}
For a given $p+2$-dimensional vacuum energy $\rho_{vac}$, we can define the width $\Delta Q$ by
\begin{equation}
\Delta Q = \rho_{vac} \left( \frac{d V(\phi_{-})}{dQ} \left. \right|_{Q=Q_0} \right)^{-1},
\end{equation}
where the term in parentheses is the derivative of the height of the potential minimum with respect to $Q$
evaluated at $Q_0$. Substituting the approximate position of the potential minimum given by Eq.~\ref{eq:nolamroot}
into the radion potential and evaluating the derivative, we can find the scaling with $\Lambda$
\begin{equation}\label{eq:dVdQ}
 \frac{d V(\phi_{-})}{dQ} \left. \right|_{Q=Q_0} = M_{p+2}^{p+2} C(p,q) \left( \frac{\Lambda}{M_D^2} \right)^{\frac{2q+p+pq}{2p}},
\end{equation}
where $C(p,q)$ is a function of geometrical factors only, and can be
determined numerically.
An example for $p=2$ as a function of $q$ is
shown in Fig.~\ref{fig:dVdQ}, where it can be seen that this factor
decreases exponentially with increasing $q$.

Re-arranging Eq.~\ref{eq:shellcondition}, we require that
\begin{equation}\label{eq:ccinequality}
\frac{{\rm Vol} \left( \Omega_{J-1} \right)}{2^J C(p,q)
\prod_{i=1}^{J} e_i} 
\left( \frac{(p+q)^q (q-1)^q}{2^{q-2} (p+1)^q}
\right)^{(J-1)/2} \left( \frac{M_D^2}{\Lambda}
\right)^{\frac{Jp(q-1) + 2 (p+q)}{2 p}} >
\frac{M_{p+2}^{p+2}}{\rho_{vac}}.
\end{equation}
It is interesting and potentially significant that the effects of
fine-tuning in $D$-dimensions are amplified in the lower-dimensional
theory in a number of different ways. In particular, the
higher-dimensional cosmological constant is raised to a power
dependent upon the number of compact dimensions and the number of
fluxes, the individual gauge coupling appear as a product, and there
are potentially large geometrical factors. With many compact
dimensions and/or fluxes, mildly tuned values $\Lambda < M_D^2$ and
$e_i < 1$ are raised to large powers, vastly increasing the volume
of the shell in charge space.

For example, taking $\{ e=.1, \Lambda = .1 M_D^2 \}$ and $q=7$,
the
inequality is satisfied for $J >18$. With more tuning in
$\Lambda$, it is possible to satisfy the inequality for smaller
values of $J$ and $q$. With $\Lambda= 10^{-6} M_D^2$, the inequality
is satisfied for $\{ q=7, J >5 \}$, $\{q=6 , J > 6 \}$, $\{q=5, J >8
\}$, $\{q=4, J > 10 \}$, $\{q=3, J > 15 \}$, or $\{q=2, J >30\}$.
The inequality can be satisfied for large numbers of extra
dimensions or fluxes individually. For example, with $q \sim 60$,
one requires $\Lambda \sim .1 M_D^2$ and $e \sim .1$, while for
$q=2$, one requires $J>137$ for the same $e$ and $\Lambda$. Even for
the minimal values $\{ q=2 , J=1\}$, after setting $e \sim 1$, the
necessary tuning of $\Lambda = 10^{-48} M_D^2$ is some $75$ orders
of magnitude better than if one had started with a four-dimensional
theory.

Although we will not explore the issue in detail, how finely
the can
lower-dimensional vacuum energy scans will also affect any potential
statistical predictions of the cosmological constant. A key element
in the original arguments of Weinberg~\cite{Weinberg:1987dv} for an
anthropic prediction of the cosmological constant is that there are
many vacua inside of an ``anthropic window" of acceptable vacuum
energies. The validity of this assumption requires that the vacuum
energy scans quite finely, and must be tested on a case-by-case
basis. For example, an analysis of this issue in the context of the
string theory landscape can be found in
Ref.~\cite{SchwartzPerlov:2006hi}. For a modern review of
statistical predictions for the cosmological constant see
e.g.~\cite{Vilenkin:2004fj,Bousso:2007gp}.

\section{Including an epoch of lower dimensional inflation}\label{sec:inflation}

An analysis of the $p+2$-dimensional cosmological constant relies
on
the properties of the $p+2$-dimensional vacua alone. However,
perhaps the most interesting feature of the solutions discussed in
previous sections is the existence of a region containing a
$p+2$-dimensional FRW universe. The radion field dynamically evolves
towards its vacuum, approaching at late times a $p+2$-dimensional de
Sitter, Minkowski, or big-crunch spacetime. In our universe, scalar
field dynamics is thought to drive an epoch of slow-roll inflation,
and it is a natural question to ask if we can embed this into our
model.

We will consider two separate models of inflation in this section,
both of which require the addition of a scalar field $\psi$ into the
$D$-dimensional theory. If this field has a positive mass squared,
then it will be possible to construct the compactification,
interpolating, and Nariai solutions of Sec.~\ref{sec:FRWLg0} exactly
as before. This is not a cosmologically interesting theory, but by
coupling $\psi$ to the $D$-dimensional Ricci scalar and field
strength, it is possible to drive the mass squared negative for
large enough curvature and/or flux. In the dimensionally reduced
theory, this corresponds to couplings between the radion field
$\phi$ and the scalar $\psi$, with the large curvature regime
corresponding to small $\phi$.

\begin{figure*}
\begin{center}
\includegraphics[width=16cm]{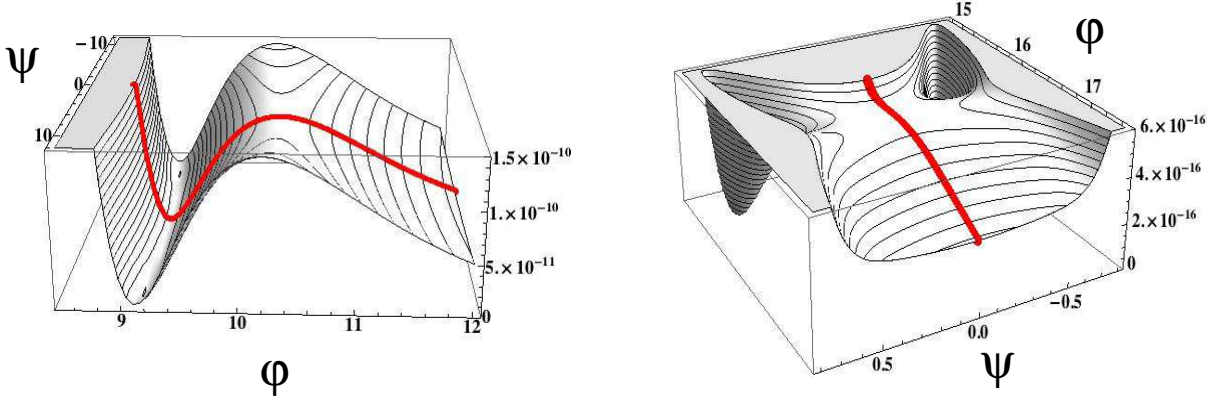}
\end{center}
\caption{Two models of inflation arising from the potential
Eq.~\ref{eq:inflationpotential}.
In both cases, the $\psi$ field
becomes destabilized between the minimum and maximum of the
potential along the $\phi$ direction (the profile along $\psi = 0$
is shown as the solid line). This can either drive (left) or end
(right) an epoch of slow-roll inflation. For the potential on the
left, there exists an interpolating solution, which naturally sets
the initial conditions near $\psi = 0$. Inflation then occurs as the
field rolls in the $\psi$ direction. For the potential on the right,
there is no interpolating solution, but the Euclidean
compactification solution can set the initial conditions for the
field very near the potential maximum. Inflation occurs as the field
rolls in the $\phi$ direction, eventually ending when the field
falls into one of the minima in the $\psi$ direction.}
\label{fig:inflation}
\end{figure*}

In the model shown in the left panel of Fig.~\ref{fig:inflation}, the instability causes $\psi$ to slowly-roll, while $\phi$ is stabilized in a minimum of the potential. In the model shown in the right panel of Fig~\ref{fig:inflation}, $\phi$ is the slowly-rolling inflaton, while $\psi$ is in an unstable equilibrium, eventually rolling off to a global minimum of the potential, and ending inflation. The
slow-roll parameters in each case are satisfied near critical points
of the potential, with slow-roll inflation occurring mostly in the
$\psi$ direction for the potential on the left and mostly in the
$\phi$ direction for the potential on the right.

The coupling of $\psi$ to curvature and flux also allows
for a natural explanation of the initial conditions for inflation. For
example, the potential on the left of Fig.~\ref{fig:inflation}
admits an interpolating solution. It is possible to adjust the
couplings such that the instability in the $\psi$ direction develops
only in the spacetime region containing the $p+2$-dimensional FRW
universe. In such models, slow-roll inflation in the $\psi$
direction is triggered by the compactification of the extra
dimensions. The potential on the right of Fig.~\ref{fig:inflation}
does not admit an interpolating solution, but there is a
compactification solution at the potential maximum along the $\phi$
direction. Evolution away from the maximum in the $\phi$ direction
can drive slow-roll inflation, until the instability in the $\psi$
direction takes over, at which point inflation ends.

We now examine the two models of inflation in detail. In
both cases,
the $D$-dimensional action we will consider is given by
\begin{equation}
S =  \frac{M_D^{q+2}}{2} \int d^{q+4} x
\sqrt{-\tilde{g}^{(q+4)}}
\left( f(\psi) \tilde{\mathcal{R}}^{(q+4)}
- 2 \Lambda - \frac{h(\psi)}{2 q!} \tilde{F}_q^2 \right) + \int
d^{q+4} x \sqrt{-\tilde{g}^{(q+4)}} \left( - M_{\psi}^{q}
\tilde{g}^{\mu \nu} \partial_{\mu} \psi \partial_{\nu} \psi -
v(\psi)  \right).
\end{equation}
The functions $f(\psi)$ and $h(\psi)$ specify the coupling to
gravity
and flux respectively, and $\psi$ has potential $v(\psi)$.
The properties of the $\psi$ field depend on background field and
metric configurations through the couplings $f(\psi)$ and $h(\psi)$.
It will be convenient to go to the Einstein frame in order to
utilize the solutions found previously in Sec.~\ref{sec:FRWLg0}.
After some manipulation, the action becomes
\begin{eqnarray}
S &=&  \frac{M_D^{q+2}}{2} \int d^{q+4} x \sqrt{-g^{(q+4)}} \left( \mathcal{R}^{(q+4)} -  \frac{2}{f(\psi)^{\frac{q+4}{q+2}} } \Lambda - \frac{h(\psi)}{f(\psi)^{\frac{4-q}{q+2}}} \frac{F_q^2}{2 q!} \right) \nonumber \\
&+& \int d^{q+4} x \sqrt{-g^{(q+4)}} \left( - M_{\psi}^{q} \left[ \frac{1}{f(\psi)} + \frac{(q+3)}{f(\psi)^2 (q+2)} \left( \frac{df}{d\psi}\right)^2  \right] g^{\mu \nu} \partial_{\mu} \psi \partial_{\nu} \psi - \frac{v(\psi)}{f(\psi)^{\frac{(q+4)}{q+2}} }  \right).
\end{eqnarray}
Performing the dimensional reduction as before and going to
the
Einstein frame of the $4$-dimensional theory we obtain the action
\begin{eqnarray}
S &=& \int d^{4} x \sqrt{ - g} \left( \frac{M_{4}^p}{2} \mathcal{R} + \frac{M_{4}^2}{4} \frac{q (q+2)}{R^2} g^{\mu \nu } (\partial_{\mu} R) (\partial_{\nu} R)  \right. \nonumber \\
&& \left.
- \frac{M_{\psi}^{q}  {\rm Vol} (S^{q}) }{M_D^q}  \left[ \frac{1}{f(\psi)} + \frac{(q+3)}{f(\psi)^2 (q+2)} \left( \frac{df}{d\psi}\right)^2  \right] g^{\mu \nu} \partial_{\mu} \psi \partial_{\nu} \psi - V (R, \psi)  \right),
\end{eqnarray}
where
\begin{equation}\label{eq:inflationpotential}
V(R, \psi) = \frac{M_4^2 M_D^2}{2 (M_D R)^q} \left[- \frac{q (q-1)}{(M_D R)^2}+ f(\psi)^{ - \frac{q+4}{q+2}} \frac{2 \Lambda}{M_D^2}+ \frac{h(\psi)}{f(\psi)^{\frac{4-q}{q+2}}} \frac{Q^2}{2 (M_D R)^{2 q}} \right] + f(\psi)^{ - \frac{q+4}{q+2}} \frac{{\rm Vol} (S^{q})  v(\psi)}{M_D^q (M_D R)^q}.
\end{equation}
For the moment, we will find it convenient to work with
the field $R$
instead of the canonically normalized radion field $\phi$.

Both models are specified by:
\begin{equation}
 f (\psi) = 1+ \xi_1 \frac{\psi^2}{M_D^2}, \ \ \ v(\psi) = m \psi^2 + \lambda \psi^n, \ \ \  h(\psi) = 1 + \xi_2 \frac{\psi^2}{M_D^2},
\end{equation}
where $m$ has mass dimension $q+2$ and $\lambda$ has mass dimension $q+4-n$. Assuming an exact shift symmetry $\psi \rightarrow - \psi$, this corresponds to the case of minimal coupling of the $\psi$ field to gravity and flux.

The key feature we will exploit is the $R$-dependent mass of the
$\psi$ field. Examining the
second derivative of the potential with
respect to $\psi$ at $\psi=0$ and fixed $R$, we obtain
\begin{equation}\label{eq:2ndderivpsi}
\frac{d^2 V}{d\psi^2} = 2 M_4^2 (M_D R)^{-q} \left[ \tilde{m} - \xi_1 \frac{q+4}{q+2} \frac{2 \Lambda}{M_D^2} + \left( \xi_2 - \xi_1 \frac{4-q}{q+2} \right) \frac{Q^2}{2 (M_D R)^{2 q} } \right],
\end{equation}
where $\tilde{m} = m / M_D^{q+2}$. The $\psi$ field is stabilized at large $R$ as long as $\tilde{m} >  \xi_1 \frac{q+4}{q+2} \frac{2 \Lambda}{M_D^2}$. However, if $\xi_2 - \xi_1 \frac{4-q}{q+2} < 0$, then $\psi$ becomes destabilized at small $R$.  The region where $\psi$ becomes tachyonic allows for the possibility of an altered vacuum structure where $\psi$ can obtain a VEV. We assume that the second term in the potential $V (\psi)$ does not affect the properties of the potential in the vicinity of $\psi = 0$, but it will be important in establishing the vacuum structure.

In order for the solutions described in Sec.~\ref{sec:FRWLg0} for
the radion field to exist, we
require in both models that the sign
change in the second derivative Eq.~\ref{eq:2ndderivpsi} occurs
somewhere between the maximum and minimum of the potential along
$\psi = 0$:
\begin{equation}\label{eq:signchangeloc}
\frac{q+4}{q+2} \frac{2 \Lambda}{M_D^2} - \left( \frac{\xi_2}{\xi_1} -  \frac{4-q}{q+2} \right) \frac{Q^2}{2 (M_D R_{-})^{2 q} } < \frac{\tilde{m}}{\xi_1} <  \frac{q+4}{q+2} \frac{2 \Lambda}{M_D^2} - \left( \frac{\xi_2}{\xi_1} -  \frac{4-q}{q+2} \right) \frac{Q^2}{2 (M_D R_{+})^{2 q} }.
\end{equation}
From the discussion in Sec.~\ref{sec:radionpotential}, if the
minimum along
$\psi = 0$ is positive, then $Q \sim Q_{max} \sim
(M_D^2 / \Lambda)^{(q-1)/2}$ and $R_{\pm} \sim \Lambda^{-1/2}$. The
ratio $\frac{\tilde{m}}{\xi_1}$ must therefore be of order
$\Lambda/M_D^2$ for the sign change in the second derivative to
occur in the desired position.

Consider first potentials of the type shown in the left panel
of
Fig.~\ref{fig:inflation}, where $\psi$ plays the role of the
inflaton. For potentials of this type, the interpolating solution
exists when the condition on the location of the second derivative
Eq.~\ref{eq:signchangeloc} is satisfied. The field configuration on
the big-bang surface of the $4$-dimensional open FRW universe inside
of the interpolating solution is set by the value of $\phi \sim
\phi_-$ at the stationary point of the interpolating solution and
the stabilized value $\psi \simeq 0$. Near the minimum of $\phi$,
Eq.~\ref{eq:2ndderivpsi} becomes negative and $\psi$ is
destabilized and begins to slowly roll.

These are essentially ``hilltop" models~\cite{Boubekeur:2005zm};
near
the minimum in $\phi$, the potential is very close to quadratic over
a range of $\psi$ of order $M_4$, after which the potential
steepens, and inflation ends. As $\phi$ is near its minimum, we can
analyze this as a single-field model of inflation, where $\psi$ is
the slowly rolling field. Inflation occurs near the maximum of the
potential in the $\psi$ direction which this guarantees that the
first slow-roll parameter
\begin{equation}
\epsilon = \frac{M_4^2}{2} \left( \frac{V'}{V} \right)^2,
\end{equation}
is small, but not the second slow-roll parameter $\eta$
\begin{equation}
 \eta = M_4^2 \frac{V''}{V}.
\end{equation}
When both are small, we will find models with greater than
60
$e$-folds of inflation. If $\epsilon$ is negligible, the spectral
index is
\begin{equation}
n_s - 1 \simeq - 2 |\eta|.
\end{equation}
Using the WMAP 5-year~\cite{Komatsu:2008hk} central
value of $n_s =
.96$, we require $|\eta| = .02$. We must also ensure that the scalar
power
\begin{equation}
\mathcal{P}_{\mathcal{R}} = \frac{1}{24 \pi^2 \epsilon} \frac{V}{M_4^4},
\end{equation}
is of the correct magnitude $\mathcal{P}_{\mathcal{R}} \simeq 2.5
\times 10^{-9}$. This,
together with the absence of tensor
perturbations in the CMB, constrains the scale of inflation to be
$V^{1/4} < 2 \times 10^{16}$ GeV.

Satisfying the constraint on the spectral index requires
$\tilde{m}$, $\xi_1$, an
 $\xi_2$ to be somewhat less than one so
that the second derivative Eq.~\ref{eq:2ndderivpsi} is sufficiently
small. This represents a true tuning, and corrections will introduce
the so-called ``$\eta$-problem"  (corrections drive $\eta \sim 1$).
The correct scalar power can be obtained by a combination of initial
conditions and either taking $\Lambda / M_D^2 \ll 1$ (scaling the
entire potential) or choosing an appropriate value of $Q$ such that
the minimum along $\psi = 0$ is of the appropriate height. One model
that satisfies all of the observational constraints has the
parameters $\{Q = 142, \Lambda = 5 \times 10^{-5}, \tilde{m} = 3.24
\times 10^{-7}, \xi_1 = 1.63 \times 10^{-3}, \xi_2 = - 9 \times
10^{-5} \}$. An additional tuning of $\lambda$ is necessary in order to
end inflation in a vacuum with very small cosmological constant. In
this example, it is necessary to take $n=6$ and $\lambda = 3.3
\times 10^{-12}$.

\begin{figure*}
\begin{center}
\includegraphics[width=18cm]{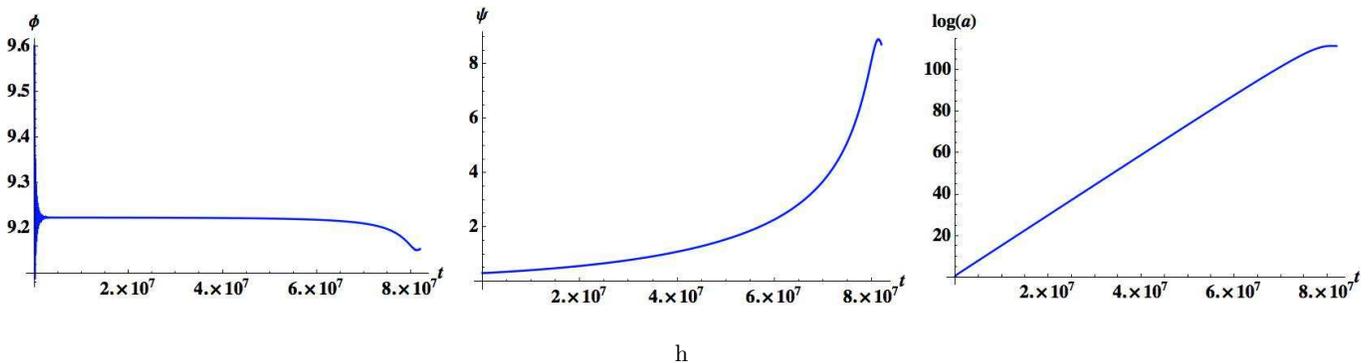}
\end{center}
h\caption{Numerical evolution in the presence of the potential in
the left panel of Fig.~\ref{fig:inflation}. The parameters
specifying the potential are $\{q=2, n = 6, Q = 142, \Lambda = 5
\times 10^{-5}, \tilde{m} = 3.24 \times 10^{-7}, \xi_1 = 1.63 \times
10^{-3}, \xi_2 = - 9 \times 10^{-5}, \lambda = 3.3 \times 10^{-12}
\}$. The stationary point of the interpolating solution sets the
initial condition for $\psi$ to $\phi_0 = 9.6$. We choose $\psi =
.3$ to obtain enough inflation (for this initial condition, there
are approximately 110 $e$-folds).} \label{fig:inflationmodel1}
\end{figure*}

The numerical evolution of the fields in this model is shown in
Fig.~\ref{fig:inflationmodel1},
where we have transformed from $R$ to
the canonical radion field $\phi$. The initial condition for the
$\phi$ field is set by the stationary point of the interpolating
solution which is approximately at $\phi \simeq 9.6$. Because the
$\psi$ field is stabilized in the $D$-dimensional de Sitter space,
it is deposited at $\psi \sim 0$ modulo fluctuations. We might
expect the size of typical fluctuations to be of order $\Delta \psi
\sim H \sim 10^{-6}$ in this example. The universe begins dominated
by curvature, and because the gradient in the $\psi$ direction is
quite small, the field moves in the $\phi$ direction, and begins
damped oscillations about the minimum in $\phi$. This begins an
epoch of inflation, as the field slowly rolls in the $\psi$
direction, as can be seen in Fig.~\ref{fig:inflationmodel1}. The
total number of $e$-folds depends on the initial condition for
$\psi$, which in this example is taken to be $\psi_0 = .3$. This is
much larger than the expected size of fluctuations, and so the total
number of $e$-folds will typically be vastly larger than the
required $60$. In fact, the potential in the $\psi$ direction is
sufficiently flat that an epoch of slow-roll eternal inflation can
be triggered inside of the interpolating solution. As long as there are enough $e$-folds (around 60, which is not surprisingly the number typically needed to solve the flatness problem), the original curvature of the FRW region in the interpolating solution is diluted below the current bound from observations of the CMB. This is identical to the dilution required in the epoch of  ``open inflation" that occurs inside of bubble universes, see e.g.~\cite{Freivogel:2005vv}.

We now move to the second type of inflationary potential,
shown in
the right panel of Fig.~\ref{fig:inflation}, where $\phi$ plays the
role of the inflaton. There is no interpolating solution for these
models, and the initial conditions are set by the compactification
solution. This places the field in the near-vicinity of the
potential maximum in the $\phi$ and close to the stabilized value
$\psi=0$. If the field starts close enough to the maximum, an epoch
of slow-roll eternal inflation will occur. Spacetime regions
fluctuate away from the maximum, beginning slow-roll inflation with
a variety of initial conditions.

\begin{figure*}
\begin{center}
\includegraphics[width=18cm]{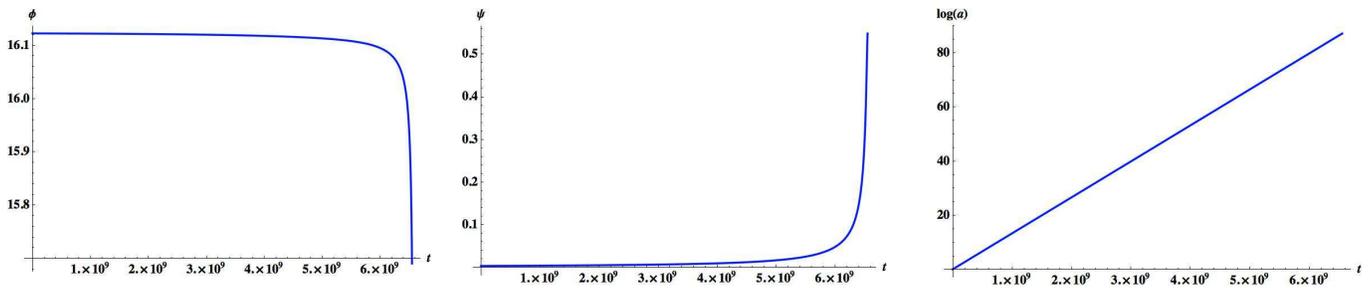}
\end{center}
\caption{Numerical evolution for the potential in the right panel of
Fig.~\ref{fig:inflation}.
The parameters specifying the potential are
$\{q=2, n = 6, Q = 3651.5, \Lambda =1  \times 10^{-7}, \tilde{m} =
1.32 \times 10^{-7}, \xi_1 = .199, \xi_2 = - 1 , \lambda = 3.6
\times 10^{-6} \}$. The Euclidean compactification solution places
the field near the maximum of the potential in the $\phi$-direction
near $\psi=0$. We choose the initial conditions to be $\{ \phi_0 =
16.122, \psi_0 = .003 \}$, which yields approximately $90$ $e$-folds
of inflation. } \label{fig:inflationmodel2}
\end{figure*}

This model is similar to hybrid inflation~\cite{Linde:1991km}; as
the field
slowly rolls in the $\phi$ direction, an instability
develops in the $\psi$ direction that ends inflation. In these
models $Q \sim Q_{max}$, and an approximate inflection point
develops in the $\phi$ direction. The scale of inflation is
therefore set exclusively by the overall height of the potential,
and it is necessary to take $\Lambda / M_D^2 \ll 1$ to satisfy the
constraint on the scalar power. However, because we no longer need
to tune the second derivative in the $\psi$ direction to be small,
we need only ensure that $\psi$ becomes destabilized at an
appropriate value of $\phi$. This requires $\tilde{m} \sim \Lambda /
M_D^2$ with $\xi_1$ and $\xi_2$ order unity. For example, a working
set of parameters is $\{Q = 3651.5, \Lambda =1  \times 10^{-7},
\tilde{m} = 1.32 \times 10^{-7}, \xi_1 = .199, \xi_2 = - 1 \}$.
Evolution from $\{ \phi_0 = 16.122, \psi = .003 \}$ yields the
trajectory shown in Fig.~\ref{fig:inflationmodel2}, and produces a
spectrum of fluctuations consistent with observation. Again, an
additional tuning of $\lambda$ is necessary to end inflation in a
minimum with small vacuum energy. In this example, we must take $n =
10$ and $\lambda = 3.6 \times 10^{-7}$.

\section{Conclusions}\label{sec:conclusions}

Despite the enormous amount of research into theories of extra
dimensions, very little is known about the possible ways to make a
transition from a higher-dimensional to lower-dimensional universe.
In this work we have explored how such a transition might occur in
one particular setting, and furthermore explored the landscape of
possible vacua of different dimensions and vacuum energy that would
emerge.

The starting point for this dynamical compactification scenario
was $D=p+q+2$-dimensional Einstein gravity in the presence of a
positive cosmological constant and a set of $q$-form field
strengths. In this theory, the $D$-dimensional de~Sitter vacuum
solution is semi-classically unstable to the nucleation non-singular interpolating and
compactification solutions that contain a region in which $q$ of the
dimensions are compact and there is a $p+2$-dimensional cosmological
spacetime. This region evolves towards either a big-crunch
singularity or a $p+2$-dimensional universe with zero or positive
vacuum energy, perhaps undergoing an epoch of slow-roll inflation in
the process. In this way, it is plausible that our universe arose
from such a dynamical compactification of some number of extra
dimensions.

The properties of the $p+2$-dimensional vacua depend not only on
fixed quantities such as the higher dimensional cosmological
constant and total dimensionality, but also on variable quantities
such as the charge and properties of other matter fields. Since the
charge can take a number of values, and there can be multiple kinds
of flux, a landscape of vacua with different vacuum energy and
effective dimensionality emerges. Each of these vacua is populated
by the dynamical compactification process described above. This
mechanism of  landscape population is fundamentally different from
false vacuum eternal inflation, which operates entirely within the
realm of the lower-dimensional effective theory.~\footnote{In the
landscape of Type IIB string theory, the volume modulus can provide
an obstacle to populating vacua through eternal
inflation~\cite{Johnson:2008vn}, and it is interesting to
identify an alternative mechanism.} Since the interpolating
solutions are close cousins of non-extremal $p$-branes, their semi-classical production is more analogous to the
nucleation of charged black holes in de Sitter space.

In any theory where a number of vacua are realized in
different spatiotemporal regions, the prediction of the parameters in a
low-energy effective theory becomes statistical at best. Such a
prediction will necessarily involve both dynamics and selection
effects. Before tackling such complicated questions, a zeroth order
requirement is for the theory to contain at least one vacuum with
the observed properties of our universe. In our toy model, we can
ask if there are vacua with the correct dimensionality and the
observed value of the cosmological constant. Finding vacua of the
correct dimensionality requires that there exists a $(D-4)$-form
field strength. Naturally obtaining the observed value of the
cosmological constant  requires that there is 1) tuning of the
higher dimensional cosmological constant, 2) tuning of the
fundamental unit of charge associated with the flux, 3) multiple
types of $q$-form flux with incommensurate charges, 4) many extra
dimensions, or some combination thereof. This is very similar to the
story in the string theory landscape.

To gain some understanding of the dynamics, we have
explored the
phenomenology of nucleation rates. Generically, the rates are
suppressed by some appreciable fraction of the higher dimensional de
Sitter action, and are higher for smaller values of the lower
dimensional vacuum energy (the highest rate is to the most negative
minimum). Another interesting behavior arises when comparing the
rate to vacua of different dimensionality, which at fixed normalized
charge $Q / Q_{max}$ is maximized for a particular $p$ or range of
$p$ depending on the total dimensionality. Comparing the rate into a
$p+2$-dimensional vacuum to the rate for decompactification back to
$D$-dimensions, we find in agreement with previous studies that
Minkowski vacua are completely stable and the rate into a
$p+2$-dimensional vacuum with a small positive vacuum energy is much
higher than the rate out. These rates can be used to examine the
structure of the higher-dimensional multiverse, for example
calculating the distribution of comoving volume.

Although we have presented a relatively complete picture of the solutions and their nucleation, we have neglected a number of potentially important effects. Perhaps the most pressing is the question of stability. It is well known that black branes in flat space are unstable~\cite{Gregory:1993vy}, which may have implications for the stability of the interpolating solutions. The stability of the $dS_{p+2} \times S^q$  compactification solutions was examined in Ref.~\cite{Bousso:2002fi} (see also~\cite{DeWolfe:2001nz,Contaldi:2004hr}), where an instability was discovered for $q \geq 4$. This is perhaps troubling since the existence of more than 3 extra dimensions is theoretically well motivated from string theory.

Recently this problem was revisited in Ref.~\cite{Kinoshita:2009hh},
who found stable warped solutions for $q \geq 4$, and proposed that
such solutions are the endpoint of evolution from the unstable
solutions without warping. In this case, the existence of an
instability would not affect the qualitative aspects of our
analysis in any number of dimensions, but addressing this question in more detail is clearly
necessary.

In addition, there are many are many directions in which one could
proceed to make more detailed and quantitative predictions in this
multiverse model. This necessarily involves determining an
appropriate measure over the various solutions, but there may be
other more direct signatures. For example, in false vacuum eternal
inflation, it is known that the inevitable collisions between bubble
universes might lead to direct signatures of other bubble
universes~\cite{Aguirre:2007an}. In complete analogy, we might
expect that some of the interpolating solutions overlap, leading to
potentially observable inhomogeneities in the $p+2$-dimensional
cosmologies. This situation is also analogous to multi-centered
black hole and black brane solutions.

It is hoped that this model will provide different insights
into issues surrounding extra dimensions and theories with multiple vacua
than those obtained by studying eternal inflation in the string
theory landscape.

\begin{acknowledgments}
The authors wish to thank Rob Myers and Matt Kleban for helpful conversations. Partial support for this research was provided by the U.S. Department of Energy and the Gordon and Betty Moore Foundation. L.R. thanks the California Institute of Technology, the Moore Fellowship Program, and NYU for their hospitality while this work was completed. L.R. is supported by NSF grant PHY-0556111.
\end{acknowledgments}

\begin{appendix}
\section{Homogenous but anisotropic solutions}\label{sec:bianchisolutions}

A more general  $p+2$-dimensional metric ansatz is to assume homogeneity but not isotropy. In four dimensions, these are the Bianchi-type cosmologies. Here, we detail the construction of solutions without spatial curvature, although it is possible to generalize our methods to all homogenous solutions in any number of dimensions. For flat spatial sections, the most general homogenous metric can be written as
\begin{equation}\label{eq:bianchimetric}
ds^2 = -d\tau^2 + \sum_{i=1}^{p+1} a_i (\tau)^2 dx_i^2,
\end{equation}
where the $x_i$ are cartesian coordinates that generally range between $-\infty < x_i < \infty$. There are $p+2$ (dimensionless) scale factors $a_i (\tau)$ which are all arbitrary functions of $\tau$.

It is possible to decouple the Einstein equations by finding equations of motion for the field and the product of metric functions
\begin{equation}
\zeta = \prod_{i=1}^{p+1} a_i.
\end{equation}
Making this substitution yields
\begin{equation}\label{eq:bianchiF}
\left( \frac{\dot{\zeta}}{\zeta} \right)^2 = \frac{2 (p+1)}{M_{p+2}^2 p} \left( \frac{\dot{\phi}^2 }{2} + M_{p+2}^{2-p} V \right) + \frac{\Sigma^2}{\zeta^2},
\end{equation}
\begin{equation}\label{eq:bianchizdd}
\frac{\ddot{\zeta}}{\zeta} = \frac{2 (p+1)}{p} \frac{V}{M_{p+2}^{p}},
\end{equation}
\begin{equation}\label{eq:bianchifield}
\ddot{\phi} + \frac{\dot{\zeta}}{\zeta} \dot{\phi} = - M_{p+2}^{2-p} V'.
\end{equation}
The constant $\Sigma$ parametrizes the degree of anisotropy, and is related to the metric functions $a_i (\tau)$ as we outline below.

The Ricci scalar is again given by Eq.~\ref{eq:FRWricci} (as can be explicitly checked using the equations of motion), and so as long as the field and its derivatives remain finite, the $\zeta = 0$ surface is completely non-singular. From the field equation Eq.~\ref{eq:bianchifield}, this requires that $\dot{\phi} (\tau = 0) = 0$. The $\zeta=0$ surfaces correspond to event horizons in the $D$-dimensional geometry, just as they did in the black hole example of the Introduction. From Eq.~\ref{eq:bianchiF}, if the field and its derivatives stay finite, then the anisotropy term always dominates at small $\zeta$. This leads to the universal behavior $\zeta = \Sigma \tau$ as $\tau \rightarrow 0$, which can be recognized as the Kasner solution.

The value of $\Sigma$ and the time dependence of the individual $a_i (\tau)$ can be found from the remaining Einstein equations. Near $\tau = 0$, the metric is always in the form of the Kasner solution, where the scale factors are a power law in $\tau$
\begin{equation}
a_i (\tau) = a^{(0)}_i \tau^{n_i}.
\end{equation}
There are two Kasner conditions
\begin{equation}
\sum_{i=1}^{p+1} n_i = 1, \ \ \ \sum_{i=1}^{p+1} n_i^2 = 1,
\end{equation}
that constrain the exponents $n_i$. It can be shown that unless there is a single non-zero $n_i$, at least one of the $n_i$ must be negative (see e.g. Ref.~\cite{Misner:1974qy}). This implies that if there is more than one non-zero $n_i$, at least one scale factor diverges at $\tau = 0$, while the others shrink to zero. If it is possible to continue across the $\tau = 0$ surface to a region where $\tau$ is spacelike, only one of the metric functions can go to zero (otherwise the analytic continuation becomes ambiguous). Therefore, in the vicinity of $\tau = 0$, only one of the Kasner exponents can be positive. From the Kasner conditions, in order to obtain $\zeta = 0$ with a single positive Kasner exponent, all of the other exponents must be zero.

From the equations of motion, the anisotropy parameter $\Sigma$ is given by
\begin{equation}
\Sigma = \prod_{i=1}^{p+1} a^{(0)}_i.
\end{equation}

The full time dependence of the scale factors can be determined by identifying a number of constants of the motion. For example, specializing to $p+2 = 4$, and following Ref.~\cite{Demianski:1992tu}, there are three constants of the motion
\begin{eqnarray}
\Pi_1 & = & a_1 \left( a_2 \dot{a}_3 - a_3 \dot{a}_2 \right), \nonumber \\
\Pi_2 & = & a_2 \left( a_3 \dot{a}_1 - a_1 \dot{a}_3 \right),  \nonumber \\
\Pi_3 & = & a_3  \left( a_1 \dot{a}_2 - a_2 \dot{a}_1 \right).
\end{eqnarray}
One then obtains equations for $a_1 (\tau)$ and $a_2 (\tau)$
\begin{eqnarray}\label{eq:qeqn}
\frac{\dot{a}_1}{a_1} &=& \frac{1}{3} \frac{\dot{\zeta}}{\zeta} -\frac{1}{3 \zeta} \left( \Pi_2 + 2 \Pi_1 \right), \\
\frac{\dot{a}_2}{a_2} &=& \frac{1}{3} \frac{\dot{\zeta}}{\zeta} + \frac{1}{3 \zeta} \left( \Pi_1 + 2 \Pi_2 \right).
\end{eqnarray}
The anisotropy parameter $\Sigma$ is given in terms of the constants of motion by
\begin{equation}
\Sigma^2 = \Pi_1^2 + \Pi_2^2 + \Pi_1 \Pi_2.
\end{equation}
The time-dependence of $a_3$ can then be determined from $\zeta$. This procedure can be applied in an arbitrary number of dimensions by again identifying the appropriate constants of the motion.

\subsection{$\Lambda = 0$}

We now construct anisotropic solutions in the absence of a $D$-dimensional cosmological constant. Before analyzing the dimensionally reduced theory, we briefly discuss some analytic solutions that can be generated from an anisotropic metric ansatz. Included in this set of solutions are the extremal and non-extremal $p$-brane solutions of Ref.~\cite{Horowitz:1991cd}. The metric for such solutions in D-dimensions is given by~\cite{Gibbons:1994vm}
\begin{eqnarray}\label{eq:nonextremalpbrane}
d \tilde{s}^2 = &-& \left[ 1 - \left( \frac{R_+}{R} \right)^{q-1} \right] \left[ 1 - \left( \frac{R_-}{R} \right)^{q-1} \right]^{\frac{1-p}{1+p}} dt^2 + \left[ 1 -  \left( \frac{R_-}{R} \right)^{q-1} \right]^{\frac{2}{p+1}} d \vec{x} \cdot d \vec{x} \nonumber \\ &+& \left[ 1 - \left( \frac{R_+}{R} \right)^{q-1} \right]^{-1} \left[ 1 - \left( \frac{R_-}{R} \right)^{q-1} \right]^{-1} dR^2 + R^2 d\Omega_q^2,
\end{eqnarray}
where $R_+$ and $R_-$ are the two event horizons related to the charge by
\begin{equation}
Q^2 = \frac{(q+p) (q-1)}{2 (p+1)} \left( R_+ R_- \right)^{q-1}.
\end{equation}
There is a curvature singularity at $R = R_-$. When $R_+ = R_-$, the event horizons become degenerate, and the solution is said to be extremal. These are none other than the extremal solutions we constructed in Sec.~\ref{sec:solutionsL0}. It is also possible to have super-extremal solutions, in which there is a naked singularity (these are similar to the timelike singular solutions of Sec.~\ref{sec:solutionsL0}).

In the region between the horizons $R_- < R < R_+$, $R$ is a timelike variable. Near the horizon at $R_+$, we now show that the metric is of the Kasner form. This can be done analytically for $p = 2$ by dimensionally reducing to 4-dimensions and performing a conformal transformation to the 4-dimensional Einstein frame. The metric Eq.~\ref{eq:nonextremalpbrane} becomes
\begin{eqnarray}
ds^2 =  &-& (M_D R)^{q} \left[ \left( \frac{R_+}{R} \right)^{q-1} -1 \right]^{-1} \left[ 1 - \left( \frac{R_-}{R} \right)^{q-1} \right]^{-1} dR^2 + (M_D R^{q}) \left[ 1 -  \left( \frac{R_-}{R} \right)^{q-1} \right]^{\frac{2}{3}}  d \vec{x} \cdot d \vec{x} \nonumber \\ &+& (M_D R)^{q} \left[ \left( \frac{R_+}{R} \right)^{q-1} - 1 \right] \left[ 1 - \left( \frac{R_-}{R} \right)^{q-1} \right]^{-\frac{1}{3}} dt^2.
\end{eqnarray}
Changing coordinates to
\begin{equation}
L \equiv R_+ - R,
\end{equation}
and expanding to lowest order around the horizon at $L=0$
\begin{eqnarray}
ds^2 = &-&  \frac{ \left( M_D R_+\right)^q R_+ \left[ 1 - \left( \frac{R_-}{R_+} \right)^{q-1} \right]}{(q-1) }  \frac{dL^2}{L} + \left( M_D R_+\right)^q \left[ 1 -  \left( \frac{R_-}{R_+} \right)^{q-1} \right]^{\frac{2}{3}}  d \vec{x} \cdot d \vec{x} \nonumber \\ &+& \frac{(q-1) \left( M_D R_+\right)^q }{R_+ \left[ 1 - \left( \frac{R_-}{R_+} \right)^{q-1} \right]^{1/3}} L dt^2.
\end{eqnarray}
Re-scaling $L$, $\vec{x}$, and $t$ by the constant factors, we obtain
\begin{equation}
ds^2 = - \frac{r_+}{L} dL^2 + d \vec{x} \cdot d \vec{x} + \frac{L}{r_+} dt^2.
\end{equation}
Finally, defining the proper time variable $\tau = 2 \sqrt{r_+ L}$, we recover the metric
\begin{equation}
ds^2 = - d\tau^2 + d \vec{x} \cdot d \vec{x} + \frac{\tau^2}{r_+^2} dt^2.
\end{equation}
This is of the Kasner form, with all but one of the exponents zero as expected.

The non-extremal $p$-brane solutions discussed above can also be obtained from the dimensionally reduced theory by evolving in the presence of the radion potential, just as in the analysis of Sections~\ref{sec:FRWsolutions},~\ref{sec:solutionsL0}, and~\ref{sec:FRWLg0}. The non-extremal, extremal, and super-extremal $p$-brane solutions correspond to different initial conditions for the radion field. In fact, these are the only solutions that possess an asymptotically flat $D$-dimensional region, and the radion and scale factor profiles will be qualitatively similar to the extremal and singular (timelike and spacelike) solutions found in Sec.~\ref{sec:solutionsL0}.

Our method of constructing solutions in the dimensionally reduced theory will be identical to the procedure outlined in Sec.~\ref{sec:FRWsolutions}, where spacetime regions with timelike and spacelike $\tau$ are matched across event horizons. We will discuss the evolution of the radion field in each case separately.

Returning to the equations of motion for $\zeta$, it can be seen from Eq.~\ref{eq:bianchizdd} that the sign of the second time derivative of $\zeta$ is fixed by the sign of $V$. For $\Lambda = 0$, the potential for $\phi$ is negative in the vicinity of the potential minimum and approaches zero from below as $\phi \rightarrow \infty$. Evolution to $\phi \rightarrow \infty$ is therefore in a spacetime region where $\tau$ is a spacelike variable and $V \rightarrow -V$. Just as for the maximally symmetric case, there are attractor solutions approached from generic initial conditions. The analysis is similar to that of Sec.~\ref{sec:FRWsolutions}.

We now look for solutions in regions where $\tau$ is a timelike variable as in Eq.~\ref{eq:bianchimetric}. Concentrating on motion in the vicinity of the potential minimum, we can apply the methods used in Sec.~\ref{sec:FRWsolutions}. Since the minimum is negative, the scale factor is bounded, with the frequency of motion given from Eq.~\ref{eq:bianchizdd} by
\begin{equation}\label{eq:omegaabianchi}
\omega_\zeta^2 = \frac{2 (p+1)}{p} |V(\phi_{-})|.
\end{equation}
Substituting into the field equation Eq.~\ref{eq:bianchifield}, the solutions for $\phi$ can be written in terms of Gegenbauer polynomials, with the index determined by
\begin{equation}
\sigma^2 + \sigma = \frac{|V''(\phi_{-})|}{\omega_\zeta^2} = \frac{p}{2 (p+1)} \frac{|V''(\phi_{-})|}{|V(\phi_{-})|},
\end{equation}
which, substituting with Eq.~\ref{eq:VppoverV}, yields $\sigma = 1$ independent of $p$. Therefore, for solutions with stationary points infinitely close to the minimum, there are solutions that interpolate across the minimum exactly once, and the oscillatory solutions found in Sec.~\ref{sec:FRWLg0} cannot not exist. For an initial condition displaced slightly from the minimum, the frequency increases, causing the solutions to go singular in the vicinity of the would-be second stationary point.

Putting the full solution together, if there is a non-singular stationary point on the right side of the potential minimum, then it is possible to connect the asymptotic region $\phi \rightarrow \infty$ with $\tau$ spacelike across an event horizon to a region in the basin of attraction of the minimum with $\tau$ timelike that contains a singularity. This is exactly the expected behavior for the non-extremal $p$-branes discussed previously. The causal structure will resemble the singular (spacelike) solutions of Sec.~\ref{sec:FRWLg0}. If there is no non-singular stationary point, then the geometry will contain a timelike naked singularity. The causal structure in this case will be identical to the singular (timelike) solution described in Sec.~\ref{sec:FRWLg0}. The extremal solution will be identical to the solution obtained in Sec.~\ref{sec:solutionsL0}, since the regime where the anisotropy dominates the energy density is pushed to $\tau \rightarrow - \infty$.

\subsection{$\Lambda > 0$}

When $\Lambda > 0$, the potential approaches zero from above, implying that $\tau$ is a timelike variable in the asymptotically $D$-dimensional region of the geometry. There are attractor solutions where $\tau$ is timelike interpolating between a non-singular stationary point and $\phi \rightarrow \infty$ of the type discussed in Sec.~\ref{sec:FRWLg0}. Continuing across the stationary point to a region where $\tau$ is spacelike, we can look for motion in the vicinity of the inverted potential maximum. The solution for the scale factor is oscillatory and the solution for the field is again a Gegenbauer polynomial with index given by the solution to
\begin{equation}
\sigma^2 + \sigma = \frac{p}{2 (p+1)} \frac{|V''(\phi_{+})|}{|V(\phi_{+})|}.
\end{equation}
The solution for $\sigma$ is always less than one, implying that there are no non-singular trajectories crossing the maximum of the potential -- the field cannot evolve quickly enough to traverse the maximum before the scale factor hits its second zero. Thus, in contrast to the solutions in Sec.~\ref{sec:solutionsL0} for an FRW metric ansatz, there can be no interpolating solutions connecting the asymptotic $\phi \rightarrow \infty$ region to the potential minimum. The geometries will therefore have a causal structure similar to the singular solutions of Sec.~\ref{sec:solutionsL0}. Moving the stationary point in the motion of the radion field to the potential maximum, the anisotropy becomes unimportant, and it will be possible to find the Nariai solution discussed in Sec.~\ref{sec:solutionsL0}

\end{appendix}

\bibliography{dynamicalcompactification}

\end{document}